\documentclass[twoside,twocolumn,9pt]{article}
\usepackage{extsizes}
\usepackage[super,sort&compress,comma]{natbib} 
\usepackage[version=3]{mhchem}
\usepackage[left=1.5cm, right=1.5cm, top=1.785cm, bottom=2.0cm]{geometry}
\usepackage{balance}
\usepackage{mathptmx}
\usepackage{sectsty}
\usepackage{graphicx}
\usepackage[utf8]{inputenc}
\usepackage{wrapfig}
\usepackage{lastpage}
\usepackage[format=plain,justification=justified,singlelinecheck=false,font={stretch=1.125,small,sf},labelfont=bf,labelsep=space]{caption}
\usepackage{float}
\usepackage{fancyhdr}
\usepackage{fnpos}
\usepackage[english]{babel}
\addto{\captionsenglish}{%
  
}
\usepackage{array}
\usepackage{droidsans}
\usepackage{charter}
\usepackage[T1]{fontenc}
\usepackage[usenames,dvipsnames]{xcolor}
\usepackage{setspace}
\usepackage[compact]{titlesec}
\usepackage{hyperref}
\usepackage{import}

\usepackage{epstopdf}

\definecolor{cream}{RGB}{222,217,201}

\begin{document}

\pagestyle{fancy}
\thispagestyle{plain}
\fancypagestyle{plain}{
\renewcommand{\headrulewidth}{0pt}
}

\makeFNbottom
\makeatletter
\renewcommand\LARGE{\@setfontsize\LARGE{15pt}{17}}
\renewcommand\Large{\@setfontsize\Large{12pt}{14}}
\renewcommand\large{\@setfontsize\large{10pt}{12}}
\renewcommand\footnotesize{\@setfontsize\footnotesize{7pt}{10}}
\makeatother

\renewcommand{\thefootnote}{\fnsymbol{footnote}}
\renewcommand\footnoterule{\vspace*{1pt}%
\color{cream}\hrule width 3.5in height 0.4pt \color{black}\vspace*{5pt}} 
\setcounter{secnumdepth}{5}

\makeatletter 
\renewcommand\@biblabel[1]{#1}            
\renewcommand\@makefntext[1]%
{\noindent\makebox[0pt][r]{\@thefnmark\,}#1}
\makeatother 
\renewcommand{\figurename}{\small{Fig.}~}
\sectionfont{\sffamily\Large}
\subsectionfont{\normalsize}
\subsubsectionfont{\bf}
\setstretch{1.125} 
\setlength{\skip\footins}{0.8cm}
\setlength{\footnotesep}{0.25cm}
\setlength{\jot}{10pt}
\titlespacing*{\section}{0pt}{4pt}{4pt}
\titlespacing*{\subsection}{0pt}{15pt}{1pt}

\fancyfoot{}
\fancyfoot[LO,RE]{\vspace{-7.1pt}\includegraphics[height=9pt]{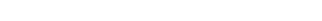}}
\fancyfoot[CO]{\vspace{-7.1pt}\hspace{13.2cm}\includegraphics{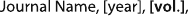}}
\fancyfoot[CE]{\vspace{-7.2pt}\hspace{-14.2cm}\includegraphics{head_foot/RF}}
\fancyfoot[RO]{\footnotesize{\sffamily{1--\pageref{LastPage} ~\textbar  \hspace{2pt}\thepage}}}
\fancyfoot[LE]{\footnotesize{\sffamily{\thepage~\textbar\hspace{3.45cm} 1--\pageref{LastPage}}}}
\fancyhead{}
\renewcommand{\headrulewidth}{0pt} 
\renewcommand{\footrulewidth}{0pt}
\setlength{\arrayrulewidth}{1pt}
\setlength{\columnsep}{6.5mm}
\setlength\bibsep{1pt}

\makeatletter 
\newlength{\figrulesep} 
\setlength{\figrulesep}{0.5\textfloatsep} 

\newcommand{\topfigrule}{\vspace*{-1pt}%
\noindent{\color{cream}\rule[-\figrulesep]{\columnwidth}{1.5pt}} }

\newcommand{\botfigrule}{\vspace*{-2pt}%
\noindent{\color{cream}\rule[\figrulesep]{\cohttps://www.overleaf.com/project/634d752cf2e722257ef91e17lumnwidth}{1.5pt}} }

\newcommand{\dblfigrule}{\vspace*{-1pt}%
\noindent{\color{cream}\rule[-\figrulesep]{\textwidth}{1.5pt}} }

\makeatother

\twocolumn[
  \begin{@twocolumnfalse}
{\includegraphics[height=30pt]{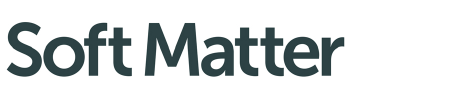}\hfill\raisebox{0pt}[0pt][0pt]{\includegraphics[height=55pt]{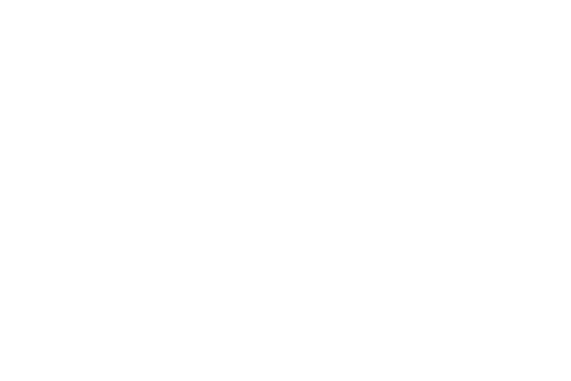}}\\[1ex]
\includegraphics[width=18.5cm]{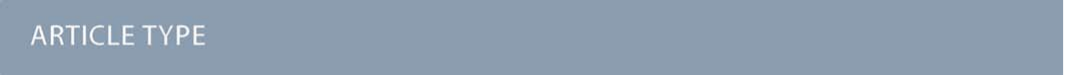}}\par
\vspace{1em}
\sffamily
\begin{tabular}{m{4.5cm} p{13.5cm} }

\includegraphics{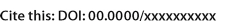} & \noindent\LARGE{\textbf{Soft matter physics of the ground beneath our feet}} \\
\vspace{0.3cm} & \vspace{0.3cm} \\

 & \noindent\large{Anne Voigtl\"ander\textit{$^{a\dag}$}, 
 Morgane Houssais\textit{$^{b\dag}$}, 
 Karol A. Bacik\textit{$^{c}$}, 
 Ian C. Bourg\textit{$^{d}$}, 
 Justin C. Burton\textit{$^{e}$}, 
 Karen E. Daniels\textit{$^{f}$}, 
 Sujit S. Datta\textit{$^{g}$}, 
 Emanuela Del Gado\textit{$^{h}$}, 
 Nakul S. Deshpande\textit{$^{f}$}, 
 Olivier Devauchelle\textit{$^{i}$}, 
 Behrooz Ferdowsi\textit{$^{j}$}, 
 Rachel Glade\textit{$^{k}$}, 
 Lucas Goehring\textit{$^{l}$},  
 Ian J. Hewitt\textit{$^{m}$}, 
 Douglas Jerolmack\textit{$^{n}$}, 
 Ruben Juanes\textit{$^{o}$},  
 Arshad Kudrolli\textit{$^{b}$}, 
 Ching-Yao Lai\textit{$^{p}$}, 
 Wei Li\textit{$^{o,q}$}, 
 Claire Masteller\textit{$^{r}$},
 Kavinda Nissanka\textit{$^{e}$}, 
 Allan M. Rubin\textit{$^{s}$}, 
 Howard A. Stone\textit{$^{t}$}, 
 Jenny Suckale\textit{$^{u}$}, 
 Nathalie M. Vriend\textit{$^{v}$}, 
 John S. Wettlaufer\textit{$^{w}$}, 
 Judy Q. Yang\textit{$^{x}$}} 
 \\

\includegraphics{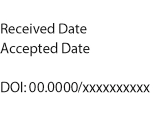} & \noindent\normalsize
{Inspired by presentations by the authors during a workshop organized at the Princeton Center for Theoretical Science (PCTS) in January 2022, we present a perspective on some of the outstanding questions related to the "physics of the ground beneath our feet."
These identified challenges are intrinsically shared with the field of Soft Matter but also have unique aspects when the natural environment is studied.}

\end{tabular}

\end{@twocolumnfalse} \vspace{0.6cm}

]


\renewcommand*\rmdefault{bch}\normalfont\upshape
\rmfamily
\section*{}
\vspace{-1cm}

\footnotetext{\dag~These authors contributed equally to this work. Following authors in alphabetical order.}
\footnotetext{\textit{$^{a}$~German Research Centre for Geosciences (GFZ), Geomorphology, Telegrafenberg, 14473 Potsdam, Germany; now at: Lawrence Berkeley National Laboratory (LBNL), Energy Geosciences Division, 1 Cyclotron Rd, Berkeley, CA 94720, USA. E-mail: anne.voigtlaender@gfz-potsdam.de}}
\footnotetext{\textit{$^{b}$~Clark University, Department of Physics, 950 Main St, Worcester, MA 01610, USA.}}
\footnotetext{\textit{$^{c}$~University of Bath, Department of Life Sciences and Department of Mathematical Sciences, Bath, BA2 7AY, United Kingdom.}}
\footnotetext{\textit{$^{d}$~Princeton University, Civil and Environmental Engineering (CEE) and High Meadows Environmental Institute (HMEI), E208 EQuad, Princeton, NJ 08540, USA.}}
\footnotetext{\textit{$^{e}$~Emory University, Department of Physics, 400 Dowman Dr., Atlanta, GA 30033, USA.}}
\footnotetext{\textit{$^{f}$~North Carolina State University, 2401 Stinson Dr., Raleigh, NC 27607, USA.}}
\footnotetext{\textit{$^{g}$~Princeton University, Department of Chemical and Biological Engineering, Princeton, NJ 08544, USA.}}
\footnotetext{\textit{$^{h}$~Georgetown University, Department of Physics, Institute for Soft Matter Synthesis and Metrology, Washington DC, USA.}}
\footnotetext{\textit{$^{i}$~Université Paris Cité, Institut de Physique du Globe de Paris, 1 rue Jussieu, CNRS, F-75005 Paris, France.}}
\footnotetext{\textit{$^{j}$~University of Houston, Department of Civil and Environmental Engineering, Houston, TX 77204, USA.}}
\footnotetext{\textit{$^{k}$~University of Rochester, Earth \& Environmental Sciences Department and Mechanical Engineering Department, 227 Hutchison Hall, P.O. Box 270221, Rochester, NY 14627.}}
\footnotetext{\textit{$^{l}$~Nottingham Trent University, School of Science and Technology, Nottingham NG11 8NS, United Kingdom.}}
\footnotetext{\textit{$^{m}$~University of Oxford, Mathematical Institute, Woodstock Road, Oxford OX2 6GG, United Kingdom.}}
\footnotetext{\textit{$^{n}$~University of Pennsylvania, Department of Earth \& Environmental Science, Philadelphia, PA 19104, USA.}}
\footnotetext{\textit{$^{o}$~Massachusetts Institute of Technology, Department of Civil and Environmental Engineering, 77~Massachusetts Avenue, Cambridge, MA 02139, USA.}}
\footnotetext{\textit{$^{p}$~Stanford University, Department of Geophysics, Stanford, CA 94305, USA.}}
\footnotetext{\textit{$^{q}$~now at: Stony Brook University, Department of Civil Engineering, Stony Brook, NY 11794, USA.}}
\footnotetext{\textit{$^{r}$~Washington University in St. Louis, Department of Earth and Planetary Sciences, St. Louis, MO, USA.}}
\break
\footnotetext{\textit{$^{s}$~Princeton University, Department of Geosciences, Princeton, NJ 08544, USA.}}
\footnotetext{\textit{$^{t}$~Princeton University, Department of Mechanical and Aerospace Engineering, Princeton, NJ 08544, USA.}}
\footnotetext{\textit{$^{u}$~Stanford University, Computational and Mathematical Engineering, and Environmental Engineering, Stanford, CA 94305, USA.}}
\footnotetext{\textit{$^{v}$~University of Colorado at Boulder, Department of Mechanical Engineering, Boulder, CO 80309, USA.}}
\footnotetext{\textit{$^{w}$~Yale University, Departments of Earth \& Planetary Sciences, Mathematics and Physics, New Haven, CT 06520, USA, and Nordic Institute for Theoretical Physics, 106 91, Stockholm, Sweden.}}
\footnotetext{\textit{$^{x}$~University of Minnesota, Saint Anthony Falls Laboratory and Department of Civil, Environmental, and Geo-Engineering, Minneapolis, MN, USA.}}




The ground beneath our feet, plain and static as it might seem, holds the key to many of the most pressing global issues. More than ever, in a world that is experiencing global warming and changing precipitation patterns, we need to better understand, predict, and feasibly control processes within the ground that can cause natural hazards \citep{Dauxois2021}, shape future landscapes \citep{Sharp1982, Dietrich2003, Houssais2017, JerolmackDaniels2019}, impact the health of our agricultural lands \citep{Jansson2020}, and sequester carbon. 

\begin{figure*}[ht]
\centering
\includegraphics[width= \textwidth]{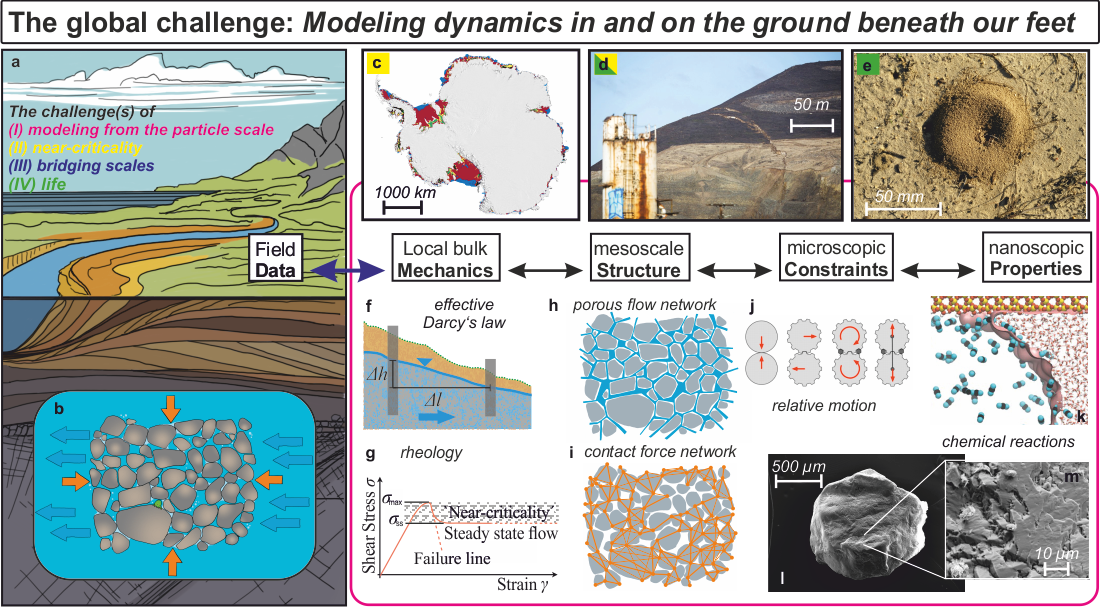}
\caption{(a) Overview sketch of the four challenges of modeling the soft matter physics of the ground identified in this paper: (I) modeling processes from the grain scale; (II) measuring and capturing the ground dynamics near critical states; (III) connecting laboratory and theory results to the field-scale observations; (IV) understanding and taking into account the many effects of life. (b) In a given element of the ground, subjected to normal stresses (orange arrows) and a mean groundwater flow (blue arrows); the soft matter physics of the ground encompasses simultaneously multiple phases, processes, dimensions, and scales, which can have various expressions at the Earth's surface, e.g., (c) an Antarctica map showing ice-shelf areas vulnerable to hydro-fracture (marked in red) in a warming world by Lai \textit{et al.} \citep{Lai2020vulnerability}; (d) a photograph of a crack that appeared on Jan 3 2018 in Rattlesnake Ridge, near Union Gap, WA, April 2018 (200 km from the 2014 Oso landslide with industrial infrastructure in the foreground, Photo credit: Shawn Gust, Yakima Herald-Republic via AP); and (e) a photograph of a mound of grains built by ants (imaggeo/EGU). To model the multi-dynamics of the ground, diverse methods, concepts, and approaches are used to link the ground's properties, constraints, structures, mechanics, and observational data. For example,  models of (f) effective groundwater flow, or (g) rheological behavior in experiments and simulations, rely on assumptions relative to (h) porous flow  and (i) contact force networks. (j) Individual motions in these networks are constrained by the properties and mechanisms of the phases involved \citep{Singh2022}. Often a phase can both be a constituent of the bulk system and, also, define an interface where chemical reactions occur, like a $CO_2$ bubble in contact with a water meniscus in a silica nanopore (k, courtesy of I.C. Bourg). An expression of these interactions, for example, a mineral particle such as a natural quartz sand grain (l) can display a chemo-mechanically altered surface topography (m), as seen in the scanning electron microscopy images (courtesy of A. Voigtl\"ander). Such nanoscale phenomena can in turn affect the effective rheology of the ground and groundwater flow.}
\label{fig: IntroFig}
\end{figure*}

What we are calling the ground is all the complex material at the surface of the Earth that is not essentially a single-phase natural fluid, such as air and water. It is either a composite solid (e.g., crystalline or sedimentary rocks, or ice, of various compositions, age, and deformation history) or \textendash ~and most commonly \textendash ~an assembly of grains of different types (e.g., sand grains, clay platelets, shells, ice crystals, boulders, bacteria). The ground evolves because it is constantly being  processed by reactive solutions, biological activity, and capillary and thermal stresses and sheared by gravitational forces and fluid flows. 
As a result, locally, and at a given time $t$, the ground exhibits normal and shear ($\gamma (t)$) strain deformations, an effective solid \textendash ~or packing \textendash ~fraction $\phi (t)$, and associated porosity $s(t) = 1 - \phi(t)$ occupied by a water volume fraction $w (t)$. Over time, the ground surface water content can vary from $0~\%$ (dry) to $100~\%$ (fully saturated, and sometimes overflowing), through intermediate $w$ values (partially saturated). Studies of the physical and chemical dynamics of such a system \textendash ~coarsely, a heterogeneous and fragile ``sponge'' or pile of sticky grains \textendash  ~give rise to a multitude of Soft Matter problems. 

Echoing the richness of the ground, a multitude of disciplines investigate the thin layer making the Earth's surface, including the Earth and environmental sciences, engineering, physics, chemistry, and ecology. Different disciplines tend to develop different cultures and approaches, as well as different motivations, constraints, and even notations. As a result, a wide and complementary range of foci regarding temporal and spatial scales, concepts, laboratory, and field study methods coexist. Nevertheless, scientific events for active exchanges between these different communities studying active processes related to the Earth's surface remain rare. Consequently, a workshop was organized at the Princeton Center for Theoretical Science (PCTS) in January 2022. The participants considered concepts and challenges in and out of well-controlled laboratory experiments and models and the fantastic messiness one can confront, and be enriched by, while studying the natural environment in the field (see Fig.~\ref{fig: IntroFig}). The authors contributed presentations of recent and diverse results, which represent only a small fraction of the many outstanding questions related to the field of ``physics of the ground''. 

From conversations amongst the participants there emerged useful clarification on the mechanics of various systems over different length and time scales, field site location and conditions, process interactions, and feedbacks. Fundamental questions were at the heart of the discussion, such as:  How can the complexity of nature’s dynamics and mechanics be simplified, measured, down-scaled, and structured to fit into a model, or are such simplifications not possible? How do we treat intermittency in natural phenomena? What does near-criticality look like in nature, and over what scales should we consider it? How can the development of complex material rheology frameworks be of future use? 

Our intention with this survey is to be both broad and specific, highlighting some of the ongoing fronts of the scientific research on the ground beneath our feet, to inspire future new and collaborative works. The paper is organized in four sections, corresponding to  four major scientific challenges (see Fig.~\ref{fig: IntroFig}). These challenges are shared intrinsically with the field of Soft Matter, yet they also have unique aspects when one studies the natural environment. Each contribution within the distinct sections \textendash ~corresponding to material presented during the workshop \textendash ~introduces  a selection of specific outstanding questions and ongoing efforts to tackle them.

\textit{I - The challenge of modeling from the grain scale.}\\ The ground is essentially a range of partially-wet particulate (hence porous) systems where chemical, physical, and biological processes occur (see right side of Fig.~\ref{fig: IntroFig}). Modeling each of these processes from the smaller scale is both fundamental and essential to building larger-scale and multi-parameter, multi-physics models. 

\textit{II - The challenge of near-criticality}.\\ The Earth surface constantly evolves and sets itself near its material yield criterion (typically, a critical shear stress $\sigma_c$). As a result, we mainly live in and observe a quasi-static environment \textendash ~an apparently stable ground \textendash ~but which is often about to fail (see Fig.~\ref{fig: IntroFig}c, d).  Surveying and predicting the ground's mechanical behavior, in particular the rare times it deforms plastically (e.g., during floods and landslides), requires good understanding and modeling of its near-critical behavior. This challenge demands advancements in both fundamental physics understanding and development of new methods of quantitative observations.

\textit{III - The challenge of bridging scales}. \\ All Earth (near-) surface processes occur at given length and time scales. Some mechanisms are universal across a wide range of the spatial and temporal scales, some structures are hierarchical and emergent, some material properties are bounded with typical magnitudes, lengths, and characteristic times or rates. Knowledge of how and when to bridge scales (see Fig.~\ref{fig: IntroFig}), mathematically, numerically, and methodologically from the laboratory to the field sites (and vice versa) is a difficult key to forge, but one that can unlock advanced predictive capabilities.  

\textit{IV - The challenge(s) of life.} \\ Living matter has the unique property of reproducing itself, and grows or decays over time. The self-propelled motion of many organisms is another specific impact of life on the environment. In the end, many organisms contribute to the constant alteration of their surroundings, while also depending on it (see Fig.~\ref{fig: IntroFig}d, e). The interconnections between all forms of life \textendash ~including human life \textendash ~and the dynamics of the ground, although obeying the laws of physics, bring additional complexity and carry a large number of original and pressing questions.

\section{The challenge of modeling from the grain scale}
\label{sec:(I)}

Understanding the time dynamics of a process from the scale of a single grain provides clear definitions of transients, steady state dynamics (e.g., rheology), fundamental causes for instability growth, and how interactions with other processes can be tackled.
Studying and modeling a single process in the ground at the grain scale requires identifying, visualizing or parameterizing, and quantifying the relevant elementary phenomena happening at that scale. This section presents a few examples of ongoing efforts in this domain based on experimental, numerical, and analytical approaches (see Fig.~\ref{fig: IntroFig}). 

In particular, Section \ref{sec:Datta} highlights some technological advancements that allow in-situ observation and quantification of dynamics in  porous materials. Section \ref{sec:Juanes} presents recent technical advances in analog models to visualize granular interactions, including emergent force chains, and feedbacks in poroelastic granular materials. The authors in section \ref{sec:RubinAndFerdowsi} show that grain-scale models of granular flow can produce transients in frictional behavior similar to those observed in laboratory rock and gouge friction experiments, behavior for which a first-principles understanding is currently lacking. Section \ref{sec:Bourg} presents novel ways to model the complex behavior of natural colloidal materials such as clay using simulations coupling porous flow, chemistry and mechanics. A different approach to grasp the dynamics and interactions from particles to bulk behavior is non-linear statistical physics, as presented in section \ref{sec:Wettlaufer} in a model of the dynamics of grains trapped within ice sheets. 

\subsection{In situ visualization of soft matter dynamics in granular and porous media - \textit{S. S. Datta}}
\label{sec:Datta}
\begin{figure}[ht]
 \centering
 \includegraphics[width=\columnwidth]{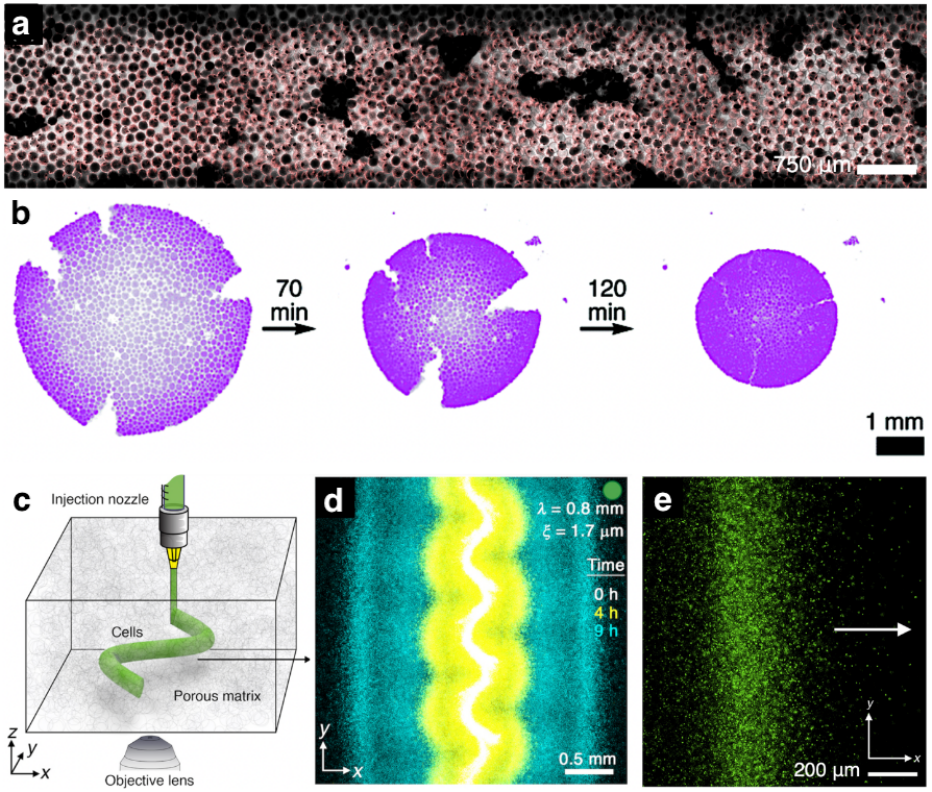}
 \caption{In situ visualization of soft matter dynamics in granular and porous media. (a) Large-scale confocal micrograph taken inside a 3D porous medium (section through solid matrix shown by black circles), showing trapped oil (additional black) and deposited colloidal particles (red) in the pore space (from~\cite{schneider2021using}). (b) Self-healing of a cracked packing of hydrogel beads; color shows fluorescence due to an excited dye that has diffused within the hydrogel beads, with an intensity that increases with bead shrinkage (from~\cite{cho2019crack}). (c) Schematic showing 3D printing of bacteria inside a porous granular hydrogel matrix. (d) Superimposed experimental confocal micrographs (different colors show different times) of bacteria spreading collectively from a 3D-printed population with an undulatory initial structure; the spreading cells smooth out these morphological perturbations. $(\lambda)$ and $(\xi)$ refer to the undulation wavelength and hydrogel matrix mean pore size, respectively (from ~\cite{bhattacharjee2022chemotactic}). (e) Magnified view of a front of bacteria spreading by chemotaxis in a crowded, porous, granular hydrogel matrix (from~\cite{bhattacharjee2021chemotactic}).}
 \label{fig:DattaFig}
\end{figure}

Advancements in our understanding of ``Soft Earth Geosciences'' are, in many ways, being driven by experimental advances in visualizing grain-scale processes. In particular, tricks from physical chemistry and colloidal science, coupled with developments in microscopy and imaging science, have yielded unprecedented ability to visualize the dynamics of soft materials in models of complex and crowded environments akin to the porous soils, sediments, aquifers, and reservoirs in the ground beneath our feet. 

In these cases, the environment alters the material, the material itself alters the environment \textendash ~and these coupled dynamics give rise to new behaviors that challenge current understanding. For example, despite its importance in energy, environmental, manufacturing, separations, industrial, and microfluidic processes, prediction and control of complex fluid transport in porous media is challenging and often operates by trial and error. Even basic prediction of where injected fluid distributes through a porous medium, and what the associated macroscopic resistance to flow is, remains elusive: how physicochemical interactions dictate fluid microstructure and transport is poorly understood due to their time-dependent and multi-scale nature. 

Disentangling these interactions through in situ visualization is therefore an exciting frontier of research. While even basic characterization has traditionally been difficult due to the opacity and complexity of most three-dimensional (3D) environments, confocal microscopy of refractive index-matched fluids and model solid media now enables researchers to directly visualize soft matter dynamics in 3D porous media with controlled pore structures and chemistries~\cite{anbari2018microfluidic}. 

Below, we summarize selected projects from our lab that used this approach to develop insights in five areas. A common feature is the ability to simultaneously probe pore space topology, dynamic changes in fluid microstructure, multi-scale flow patterns, and macroscopic transport \textendash ~which provides a way to directly connect phenomena across multiple scales.

(i) \emph{Polymer solution flow instabilities}. As they squeeze through tight pores, polymers can deform, resist deformation due to their elasticity, and alter subsequent transport~\cite{larson1990,shaqfeh1996,mckinley1996,pakdel1996,burghelea2004chaotic,rodd2007,afonso2010purely,zilz2012,galindo2012,ribeiro2014,clarke2016,machado2016extra,kawale2017a,qin2019flow,sousa2018purely,browne2020pore,browne2020bistability,walkama2020disorder,haward2021stagnation}. Direct visualization revealed that these coupled effects lead to unexpected spatial and temporal fluctuations in the transport of polymer solutions that are often applied in processes such as groundwater remediation and oil recovery~\cite{browne2020pore,browne2020bistability,datta2022perspectives}.  Moreover, analysis of the measured pore-scale flow fields quantitatively established that the pore-scale onset of these chaotic fluctuations generates a strong anomalous increase in the macroscopic flow resistance~\cite{Browne2021} \textendash ~a phenomenon that is well-documented, but that had eluded explanation for >50 years~\cite{marshall1967flow,james1975laminar,durst1981,dursthaas1981,ChauveteauMoan,kauser,hawardodell,odellhaward,zamani2015effect,clarke2016,skauge2018polymer,ibezim2021viscoelastic}. Given that the macroscopic flow resistance is one of the most fundamental descriptors of fluid flow, these findings not only help deepen understanding of polymer solution flows but also provide quantitative guidelines to inform their geophysical applications at large scales (e.g.,~\cite{browne2023homogenizing}). 

(ii) \emph{Particulate transport}. Particulate transport underlies a wide array of processes in porous media that affect our everyday lives, ranging from the beneficial \textendash ~e.g., groundwater remediation~\cite{phenrat2009particle,zhao2016overview,kanel2006arsenic} \textendash ~to the harmful, such as the migration of microplastics, contaminants, and pathogens in the environment~\cite{schijven2003bacteriophages,zhong2017transport,harvey1991use}. As colloidal particles navigate a tortuous porous medium, they can alter the medium in turn by depositing onto its solid matrix, making prediction of macroscopic particle distributions challenging~\cite{bizmark2019transport,DressaireEmilie2016Coms,sahimi1991hydrodynamics,zeman2017microfiltration,linkhorst2016microfluidic,molnar2015predicting,gerber2019self,kusaka2010morphology,lin2016examining,auset2006pore,wyss2006mechanism,de2016dynamics,lin2017characterizing,mays2011static,roth2015colloid,li2006role,gerber2018prl}. Direct visualization enabled identification of the fundamental mechanisms by which particles are distributed throughout a porous medium, demonstrating that the interplay between hydrodynamic and colloidal interaction forces controls this process~\cite{bizmark2020multiscale}.  Moreover, it enabled characterization of how interactions between particles and trapped non-aqueous fluids influence subsequent transport (Fig.~\ref{fig:DattaFig}a)~\cite{schneider2021using}. These results help shed light on the multi-scale interactions between fluids, particles, and porous media that have traditionally been represented in black-box models using ``lumped'' empirical parameters \textendash ~guiding the development of more accurate and generalizable models that could be applied in diverse geophysical settings. 

(iii) \emph{Water-absorbent hydrogels}. Hydrogels are elastic networks of hydrophilic polymers that can absorb large quantities of water. They, therefore, hold tremendous promise as water reservoirs for plant roots in dry soils, potentially reducing the burden of irrigation in agriculture~\cite{braun2021spotlight,said2018environmentally,wei2016using,abedi2008evaluation,hemvichian2014synthesis,nascimento2021temperature,guilherme2015superabsorbent,bandak2021effects,souza2016water,banedjschafie2015water,garbowski2020soil,sojka1998water,wei2013effect,cejas2014kinetics,wei2014rain,wei2014morphology,woodhouse1991effect,bai2010effects,frantz2005actual,hejduk2012evaluation} \textendash ~which is critically important given the increasing threat of water scarcity for a growing world population. However, this application requires hydrogel water absorption and swelling to be predictable and controllable. Using direct visualization of hydrogel swelling in granular media akin to soil, we demonstrated that confinement in a granular medium can dramatically hinder the ability of hydrogels to absorb water~\cite{louf2021under}. These studies now help to elucidate how environmental factors such as soil structure, chemistry, water saturation, and mechanics influence hydrogel water absorption, providing physical principles to ultimately guide the design of hydrogels whose swelling and mechanics are optimized for use in a given environment~\cite{misiewicz2022characteristics}. 

In related research, hydrogel packings themselves acted as models of shrinkable granular media e.g., soft clay-rich soils, whose deformations influence the integrity of built structures and barriers for waste isolation~\cite{goehring2015desiccation,doi:10.2113/gselements.5.2.105,Espinoza:2012iq}. Deforming such a soft porous material alters fluid transport through its pores, which in turn further deforms the material. Using direct visualization of this coupling between fluid transport and solid deformations, we have shown how material physicochemical properties that regulate fluid permeability and mechanical deformations, as well as interactions with external boundaries, together control how these materials swell/shrink and deform~\cite{cho2019crack}, fracture~\cite{cho2019scaling}, and potentially even self-heal~\cite{cho2019crack} (Fig.~\ref{fig:DattaFig}b) \textendash ~providing new insights into the desiccation of soft earth materials.  

(iv) \emph{Immiscible fluid displacement}. The displacement of a fluid from a porous medium by another immiscible fluid underlies groundwater contamination and remediation, subsurface carbon sequestration, oil and gas migration and recovery, and moisture infiltration in/drying of soil and wood~\cite{morrow2001recovery, mattax1962imbibition, li2000characterization,nicolaides2015impact,bennion2006supercritical,bennion2010drainage, hatiboglu2008pore,celia2015status,schaefer2000experimental,esposito2011remediation,penvern2020,zhou2019,weisbrod2002imbibition,tesoro2007relative,hassanein2006investigation}. While these processes are well-studied in homogeneous porous media with randomly-distributed pores of different sizes and similar surface chemistries~\cite{lenormand1984role, chang2009experimental, hatiboglu2008pore, sun2016micro, hughes2000pore, lenormand1988numerical, lenormand1990liquids, sahimi1993flow, alava2004imbibition, stokes1986interfacial, weitz1987dynamic, hultmark2011influence,zhao2016wettability,tanino2018oil,odier2017forced}, many naturally-occurring media have structural and chemical heterogeneities, such as pore size gradients, strata of different permeabilities, and regions of differing surface chemistry~\cite{bear2013dynamics,galloway2012terrigenous,gasda2005upscaling,king2018microstructural}. Direct visualization in 2D and 3D porous media revealed that such heterogeneities fundamentally alter fluid displacement pathways and dynamics. In particular, a pore size or surface energy gradient can either suppress or exacerbate both capillary fingering~\cite{lu2019controlling} and viscous fingering~\cite{rabbani2018suppressing}, distinct interfacial instabilities that typically arise in homogeneous media.  Furthermore, for the case of stratified media, visualization revealed that immiscible fluid displacement is spatially heterogeneous, with different strata being invaded at different rates~\cite{lu2020forced}, leading to differing amounts of fluid removal \textendash ~phenomena that are not predicted by typically-used, spatially-averaged models of fluid flow, but are captured by new theoretical models inspired by the experiments~\cite{lu2021forced}.

(v) \emph{Bacterial communities}. Bacterial communities in the ground beneath our feet critically impact our everyday lives: they degrade contaminants, fix nitrogen, help sustain plant growth, and decompose organic matter~\cite{dechesne,souza,turnbull,watt,babalola,roseanne18,Adadevoh:2016,ford07,wang08,raina2019role}. However, despite their ubiquity and importance, how such bacterial communities spread, self-organize and stably function, and interact with their surroundings is poorly understood. Laboratory studies typically focus on bacteria in liquid culture or at flat interfaces; however, these do not exhibit many of the features of environmental communities that inhabit soils or sediments in terrestrial environments. Thus, we have developed “transparent soils” using e.g., granular hydrogels that enable the behavior of bacteria to be probed in 3D granular media over length scales ranging from the single-cell to the community scale~\cite{bhattacharjee2019confinement}~(Fig.~\ref{fig:DattaFig}c,d,e). This capability revealed that current understanding of bacterial motility \textendash ~which is based on studies performed in bulk liquid \textendash ~is incomplete: for example, confinement in a crowded medium fundamentally alters how bacteria move, both at the single cell~\cite{bhattacharjee2019bacterial} and population scales~\cite{bhattacharjee2021chemotactic}, in previously unknown ways. Ultimately, these results could guide the development of new theoretical models~\cite{amchin2022influence} to more accurately predict the motion and growth of bacterial populations in complex environments akin to those in the ground beneath our feet \textendash ~potentially helping to provide quantitative guidelines for the control of these dynamics in processes ranging from bioremediation to agriculture.

These examples highlight the utility of direct visualization of soft matter dynamics in model porous media in shedding new light on problems in soft earth geosciences. Moving forward, it will be important for researchers to continue to develop new imaging approaches to access, e.g., 3D fluid flow fields. In addition, a useful direction for future research will be to examine soft matter dynamics in granular and porous media with additional complexities such as deformability and rearrangements of the granular matrix, and different grain shapes, sizes, surface chemistries, and packing geometries. While a great deal of empirical evidence indicates that these factors strongly alter soft matter dynamics in complex environments, unifying principles that describe how remain lacking; this research could provide a way to discover these. Not only will this work deepen fundamental understanding of soft matter dynamics in geoscientific settings, but it will also enable researchers to develop guidelines for the application of existing soft materials and complex fluids, as well as principles for the formulation of new materials and fluids, e.g., in controlling solute transport and transport-limited chemical reactions in environmental remediation, as well as other industrial and environmental processes more broadly.

\subsection{Force chains underpin emergent poromechanical behavior in granular media - \textit{W. Li \& R. Juanes}}
\label{sec:Juanes}
\begin{figure*}[ht]
 \centering
  \includegraphics[width=\textwidth]{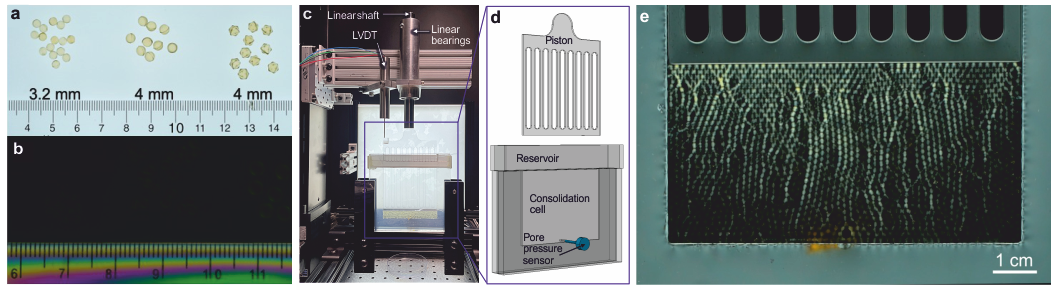}
 \caption{Consolidation test using photo-poromechanics. (a)~Millimeter-size photoelastic particles in two different shapes (spheres and icosahedra) under white light. (b)~Photoelastic particles under a circular polariscope. The polystyrene ruler, having residual stress, shows color stripes. However, the particles, being residual-stress-free, are hardly visible (from \citet{li2021}). (c)~Experimental setup for the 1D consolidation experiment. A granular pack of fluid-saturated photoelastic spheres is loaded suddenly with a constant weight, while the video, deformation, and excess pore pressure are recorded. (d)~Detailed schematic of the consolidation cell. Two glass plates are glued with a 2~mm thick U-shaped spacer where the beads are inserted to form a monolayer pack. The excess pore pressure is measured at the bottom of the cell with a pressure sensor. The pore fluid fills the cell to provide a constant-pressure boundary condition. A piston made of a 1.8~mm acrylic plate (with slots cut out to reduce resistance to fluid flow) allows for the fluid to seep out of the cell (from \citet{li2021}). (e)~A snapshot of the photoelastic response of the granular pack during the consolidation test. The force chains \textendash ~which quantify the Terzaghi stress in the granular pack \textendash ~develop from the top boundary, then progress downwards through the pack as the pore-fluid pressure diffuses upwards. }
 \label{fig:JuanesFig1}
\end{figure*}

Photoelasticity has a long history as a technique to quantify internal stresses in solid bodies\citep{frocht1941}, but it has been traditionally applied to granular media consisting of cylindrical (usually circular) disks\citep{AbedZadeh2019, Daniels2017}. This particle geometry has the advantage of allowing for precise quantification of stresses\citep{majmudar2005}, but the disadvantage that it prevents connectivity within the pore space, thus restricting severely its purpose as an analogue of permeable porous media, where fluid flow and mechanical deformation are often strongly coupled\citep{Biot1941}. This is because it is effective stress \textendash ~the fraction of the total stress that is transmitted through the solid skeleton \textendash ~that controls the mechanical behavior of porous media, from land subsidence due to groundwater pumping to the cohesion of sand in sandcastles\citep{juanes2020}. Karl von Terzaghi, father of soil mechanics, introduced the concept of effective stress a century ago\citep{terzaghi1925, terzaghi1943}. Until recently, however, this physical quantity could only be calculated by subtracting pore pressure from the normal total stress, or inferred from its ``effect'', typically the solid skeleton deformation. 

For a proper analogy of a porous medium in terms of pore geometry, connectivity and morphology, a pack of 3D particles, such as spheres, should be used. Extending photoelasticity to such systems, however, requires developing a method to manufacture residual-stress-free photoelastic particles, and obtaining quantitative information on the forces acting on these 3D particles. During the workshop we presented a fabrication process, similar to ``squeeze casting'', to produce millimeter-scale residual-stress-free photoelastic particles (spheres and other shapes, such as icosahedra) with high geometric accuracy (Fig.~\ref{fig:JuanesFig1}a,b). The combined photoelastic response from light intensity and light color permits a rough quantification of forces acting on the particles over a wide range of forces. We then presented a first application of the new technique, coined photoporomechanics\citep{li2021}, to illustrate the evolution of effective stress during vertical consolidation (Fig.~\ref{fig:JuanesFig1}c,d): a process by which the stresses caused by a sudden load are gradually transmitted through a fluid-filled granular pack as the fluid drains and excess pore pressures dissipate. We show that compaction of the granular pack is concomitant with the emergence of particle-particle force networks, which originate at the top boundary (where the pore fluid seeps out) and propagate downwards through the pack as the pore pressure gradually dissipates (Fig.~\ref{fig:JuanesFig1}e).
 
These examples highlight the utility of direct visualization of soft-matter dynamics in model porous media in shedding new light on problems in soft-earth geosciences. Moving forward, it will be important for researchers to continue to develop new imaging approaches to access, e.g., 3D fluid flow fields. In addition, a useful direction for future research will be to examine soft matter dynamics in granular and porous media with additional complexities such as deformability and rearrangements of the granular matrix, and different grain shapes, sizes, surface chemistries, and packing geometries. While a great deal of empirical evidence indicates that these factors strongly alter soft matter dynamics in complex environments, unifying principles that describe how remain lacking; this research could provide a way to discover these. Not only will this work deepen fundamental understanding of soft matter dynamics in geoscientific settings, but it could also enable researchers to develop guidelines for the application of existing soft materials and complex fluids, as well as the formulation of new materials.

This extension of photoelasticity to 3D particles provides a powerful experimental model system to study the strong coupling of solid and fluid in granular media that take place in geoscience processes like landslides\citep{palmer2017}, gas vents from ocean sediments \citep{skarke2014}, and injection-induced seismicity\citep{guglielmi2015}. This is especially attractive in three dimensions, where (while long-standing issues related to the interpretation of light transmission in fully-3D stress fields\citep{frocht1952} still need to be resolved) the method can form the basis for force-chain tomography\citep{li2023}.

\subsection{The role of granular flow in fault friction - \textit{A. M. Rubin \& B. Ferdowsi}}
\label{sec:RubinAndFerdowsi}
\begin{figure*}[ht]
 \centering
 \includegraphics[width=\textwidth]{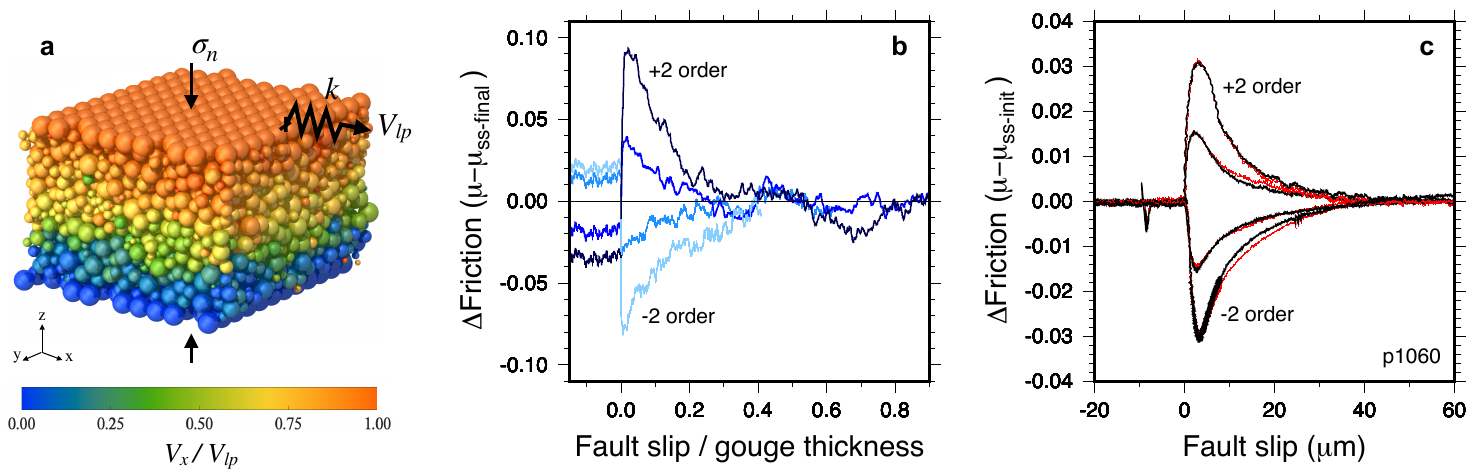}
 \caption{(a) Snapshot from a DEM simulation of a sheared granular layer\cite{FerdowsiRubin2020}. Grains are spherical, polydisperse, and have elastic properties appropriate for glass beads. Colors indicate grain velocity in the x direction, relative to the load-point velocity, averaged over an upper-plate sliding displacement of 1 mean grain diameter. (b) The friction signal, relative to the future steady-state value, following simulated velocity steps of $\pm$ 1 and 2 orders of magnitude from an initial velocity of 0.01~$m/s$\cite{FerdowsiRubin2020}.  Slip distance is defined to be zero at the time of the step. (c) Solid black lines indicate the friction signals, relative to the prior steady-state value, following velocity steps of $\pm$ 1 and 2 orders of magnitude at sliding speeds from 1 to 100~$\mu/s$, from experiments conducted in the Penn State Rock and Sediment Mechanics Lab\cite{Bhattacharya2015}. The starting material is a 3-mm-thick layer of synthetic quartz gouge, with particles ranging from 50-150~$\mu m$ in diameter (shear ultimately localizes to a narrower zone where particles have been comminuted, a process not modeled in the DEM).  The synthetic quartz gouge is nearly steady-state velocity neutral, whereas the DEM is steady-state velocity strengthening. The red lines are a fit to the data using the empirical “Slip” version of the rate-state friction equations (“direct effect” coefficient a=0.0073; “state evolution effect” coefficient b=0.0075; e-folding slip distance $D_c$=12.2~$\mu m$).}
 \label{fig:RubinFig}
\end{figure*}

Faults in the Earth are invariably filled with granular material (gouge) derived from wear of the surrounding rock \cite{sammis1989fractals,   
  marone1998laboratory}.  How the properties of this heterogeneous gouge layer control fault slip at both elastodynamic and quasi-static sliding speeds is not understood \cite{bhattacharya2022evolution,daub2010friction,rice2006heating,Rempel2006,kato2001composite}.  

Numerical models of fault slip require a constitutive law for fault friction.  The current state-of-the-art, originally conceived for two rough surfaces in contact but observed to apply to sheared gouge layers as well, falls under the heading of ``rate- and state-dependent'' friction. In this formalism, the friction coefficient (the ratio of shear to normal stress during sliding) depends upon the fault sliding rate (or strain rate), and a more nebulous property termed ``state''. State is conventionally thought to reflect a combination of the true contact area and the intrinsic strength of those contacts. Also conventionally, the state dependence is thought to be due to time-dependent plastic flow or chemical bonding at those contacts, although the opaque nature of rock makes the origin of state evolution difficult to decipher. How state evolves for surfaces not at steady-state sliding is parameterized by ``state evolution laws'' that are largely empirical, yet still do not adequately describe all the relevant features of laboratory experiments. 

Numerical simulations of faults obeying rate-state friction show that the precise description of state evolution, or the transient friction between sliding surfaces when not at steady state, significantly influences processes that Earth scientists care about (e.g., earthquake nucleation\cite{AmpueroRubin2008}). The lack of an accurate or physics-based description of state evolution thus severely hampers our ability to extrapolate the results of numerical models of fault slip to the Earth.  Recent Discrete Element Method (DEM) simulations of a granular gouge layer show that much of the phenomenology of transient rock and gouge friction seen in laboratory experiments (both the rate-dependence and the state-dependence) can be reproduced by numerical models in which this dependence arises only from momentum transfer between the grains, with no chemical reactions or time-dependent plasticity at grain/grain contacts\cite{FerdowsiRubin2020}. 

Panel (a) of Figure~\ref{fig:RubinFig} shows a snapshot from a granular DEM simulation designed to mimic laboratory rock friction experiments.  A 2D layer, periodic in the $x$ and $y$ directions, is sheared between two rigid parallel plates. A specified velocity history in the $x$ direction is applied to a very stiff spring attached to the upper plate, while a constant normal stress of 5~MPa is maintained in the $z$ direction.  A constant sliding friction acts at grain/grain contacts.  Panel (b) shows the friction signal, relative to the future steady-state value, following simulated velocity steps of $\pm$ 1 and 2 orders of magnitude. At the time of the velocity step, there is an abrupt change in stress of the same sign as that of the step (the “direct velocity effect” of rate-state friction), followed by an exponential decay to a new steady-state value (the “state evolution effect”). The magnitudes of the direct and evolution effects are approximately proportional to the logarithm of the velocity jump, with an e-folding strain for friction evolution of $\sim$0.13.  

These results are similar to those from laboratory friction experiments on rock and many other materials.  The solid black lines in Figure~\ref{fig:RubinFig}c indicate the friction signals, relative to the prior steady state value, following velocity steps of $\pm$ 1 and 2 orders of magnitude from experiments on synthetic quartz gouge\cite{Bhattacharya2015}. The magnitude of the logarithmic rate- and state-dependence in the DEM and lab experiments are similar to within a factor of $\sim$2 (there is some rounding and diminishing of the peaks in panel (c) not present in panel (b) because the elastic stiffness of the lab system is smaller).  The red lines in Figure~\ref{fig:RubinFig}c are a fit to the data using the empirical “Slip” version of the rate-state friction equations \cite{marone1998laboratory,Bhattacharya2015,ruina1983slip}, using a single set of parameter values for all 4 steps.  

The source of the rate- and state-dependence in the DEM, which lacks time-dependence at the contact scale, remains an area of active investigation.  It appears to be possible to understand the direct strain-rate dependence semi-quantitatively in terms of an Arrhenius process, with the kinetic energy of the grains playing the role of the molecular kinetic energy in the classical understanding of rate-state friction\cite{rice2001rate}, as grains hop from one potential well to another\cite{FerdowsiRubin2020}.
Although the nature of granular friction has been studied extensively in the physics and engineering literature, most of this work concerns friction during steady flow \cite{kim2020power,jop2015rheological,henann2013predictive,forterre2008flows,lois2005numerical, silbert2001granular,thompson1991granular}. The transient frictional properties of granular flow thus represent a rich and underexplored field of interest to Earth scientists, physicists, and engineers. 

\subsection{Coupled flow and mechanics of clays and muds - \textit{I. C. Bourg}}
\label{sec:Bourg}
A recurrent theme in efforts to predict the physics of the ground is the existence of complex couplings between hydraulic, mechanical, and chemical phenomena in fine-grained natural materials often referred to as clays or muds \citep{Ghezzehei2001,Grabowski2011,Carrillo2019}. These materials include a variety of phases including phyllosilicate minerals, nanocrystalline Al and Fe oxides, soil organic matter, and biofilms \citep{Kleber2021}. A common feature of these materials is that they consist of assemblages of colloidal particles with dimensions $\sim10^3$ times smaller than the sand or silt grains discussed in previous sections. Despite vast chemical and microstructural differences, these phases exhibit common properties \textendash ~low permeability, cohesion, chemo-mechanical couplings, a significant yield stress \textendash ~suggesting that these features are associated with particle dimensions on the order of nanometers and with interactions across thin fluid films.

An important manifestation of the impact of clays and muds is that for most sediments and sedimentary rocks, a 30$\%$ increase in clay content causes a 10$^6$-fold decrease in permeability \citep{Bourg2017}. This large impact on permeability can transform the ground from a material where fluids readily flow to a material where they do not, drastically modifying subsurface hydrology and poromechanics \citep{Neuzil2019}. For example, the ability of mud flows to carry suspended debris is sensitive to the effective stress (section \ref{sec:Juanes}) on the mud solid framework, which depends on the extent to which external stresses cause fluid expulsion vs a fluid pressure increase \citep{Iverson1997,Kaitna2016}.

A second illustration of the complex properties and impacts of clays and muds is that these materials can transition from non-cohesive to cohesive mechanics depending on conditions. Whereas interparticle interactions in coarser-grained materials predominantly consist of repulsive grain contact forces, potentially supplemented by attraction due to capillary fluid menisci \citep{Koos2011}, interparticle interactions in clayey materials involve a variety of attractive and repulsive interactions across thin water films—including osmotic, electrostatic, van der Waals, hydration, and excluded volume effects—with different length scales and sensitivities to particle shape, surface charge, and solution chemistry \citep{Underwood2020,Shen2021}. These complex water-mediated interactions play key roles in the emergence of complex dynamic properties in clayey media, including cohesion, yield stress, and thixotropy (with implications, for example, for the impact of clay in fault slip, section \ref{sec:RubinAndFerdowsi}, and in debris flow, section \ref{sec:JerolmackAndDeshpante}), yet they remain insufficiently understood to predict these dynamics \citep{Ancey2007,Seiphoori2020}.

Finally, a third illustration of the impact of clays is their inhibition of the transition from ductile to brittle mechanics during soil freezing or sediment diagenesis. Because of their high specific surface area and hygroscopic nature, clays can hold relatively large amounts of water (on the order of half of their volume or more) with most water molecules located in direct contact with the nearest surface \citep{Underwood2020}. This gives rise to distinct aqueous chemistries, including inhibition of freezing (section \ref{sec:Wettlaufer}) and complex impacts on cementation that can cause a persistence of ductile mechanics despite exposure to below-freezing temperatures \citep{Dash2006,Jin2020} or deep geologic burial \citep{Gratier2013,Bourg2015}.

A key challenge in efforts to resolve the processes outlined above from the grain scale is that they involve a large-scale separation between the grain scale discussed in previous sections \textendash ~associated with sand grains, grain contacts, force chains, and microbial processes with characteristic scales on the order of 10$^{-6}$ to 10$^{-4}$~m \textendash ~and the scale of clay colloidal interactions discussed above, on the order of 10$^{-9}$ to 10$^{-7}$ m. With the exception of idealized subsurface materials such as pure sand or clay, efforts to understand the physics of the ground at the grain scale are inherently multiscale because of the ubiquitous co-existence of clay or mud and coarser-grained material with more than three orders of magnitude separation in grain size \citep{Bourg2017,Carrillo2019,Seiphoori2020}.

\subsection{Polar nonequilibrium statistical physics - \textit{J. S. Wettlaufer}}
\label{sec:Wettlaufer}
\begin{figure} [ht]
 \centering
 \includegraphics[width= \linewidth]{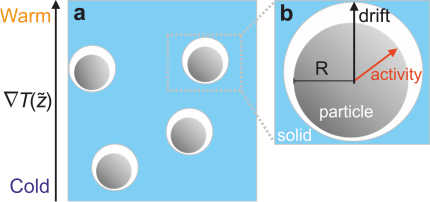}
 \caption{The interface between ice and inert or living particles is separated by a so-called premelted water film below the bulk melting point. (a) Perspective view of few active particles embedded inside ice against which they premelt and experience an external temperature gradient $\nabla T$, which creates a
 thermomolecular pressure gradient driving the flow of liquid from high to low temperatures, so that particles translate from low to high temperatures. (b) An expanded view of one active particle inside the solid. The radius of the particle is $R$, the black arrow shows the drift velocity induced by the temperature gradient, and the red arrow denotes the activity given by an active force (from \cite{Vachier_Wettlaufer_2022}).}
\label{fig:WettlauferFig}
\end{figure}

All phases of matter are separated by interfaces. Nonetheless, it is tempting, and indeed common, to ignore their energies and the associated implications. This, despite the fact that interfaces can control the large-scale phase behavior of all materials, from multiphase flows to the deformation of polycrystals. Interfacial influences on melting and freezing control the phase behavior of all materials; modern research shows that a layer of water exists on the surface of ice, even at temperatures well below freezing. These unfrozen films can influence everything from the slipperiness of glaciers to the electrification of thunderclouds\cite{Wettlaufer2019}.

In cold climates, roads are salted in winter, harnessing the freezing point depression of impurities. Each salt crystal, however, abuts an ice surface where the phase change occurs. Less commonly thought of, but equally important, are other mechanisms that can extend the equilibrium domain of a liquid phase into the solid region of the normal phase diagram. The causes of this ``premelting'', which, in addition to impurities \cite{Navaneeth_Wettlaufer_2020}, include surface melting, interface curvature and substrate disorder, allow for the persistence of water at interfaces well below the bulk melting point. The thickness of the liquid film depends on the temperature, impurities, material properties (intermolecular forces) and geometry. A temperature gradient is accompanied by a thermomolecular pressure gradient that drives the unfrozen interfacial liquid from high to low temperatures and hence particles in ice, as shown in Figure \ref{fig:WettlauferFig}, migrate from low to high temperatures. Such premelting dynamics are operative in a wide range of settings, from the heaving of frozen ground and planetary regolith, to the scavenging of atmospheric trace gases by snow and the redistribution of climate proxies in ice sheets, to the collisional processes in protoplanetary disks. Moreover, the unfrozen films act both as a refuge for biota and a transport mechanism for nutrients, waste and the biota themselves.  

New research considers such processes in the framework of active matter, wherein particles are endowed with intrinsic mobility mimicking life, and addresses the interplay between a wide range of problems, from extremophiles of both terrestrial and exobiological relevance to ecological dynamics in Earth's cryosphere. For example, biota are found in glaciers, ice sheets, and permafrost, evolving in a complex mobile environment facilitated or hindered by a range of bulk and surface interactions.  Survival strategies, such as producing exopolymeric substances and antifreeze glycoproteins, that enhance the interfacial water also facilitate bio-mobility.  Such phenomena can be cast in the stochastic framework of active Ornstein-Uhlenbeck dynamics and chemotaxis \cite{Vachier_Wettlaufer_2022}, to find that for an attractive (repulsive) nutrient source, that thermomolecular motion is enhanced (suppressed) by biolocomotion. This phenomenon is essential for understanding the persistence of life at low temperatures.

\section{The challenge of near-criticality}
The study of the near-critical behavior of material has a long history and encompasses problems such as glass transition and deformation,  shear-thickening behavior of suspensions, particulate material jamming, solid creep, and fracture dynamics. Most of these systems exhibit behaviors that are non-linear functions (e.g., see Fig.~\ref{fig: IntroFig}) of the system's temperature, applied stresses, and density of grains or atoms. Material failure in the environment  shares these common features and presents specific challenges.
Section \ref{sec:JerolmackAndDeshpante} lists the most fundamental and concrete soft matter problems embedded into the pressing question of ``how does a hill material turn into a debris flow?" It also reviews some recent experimental results on hillslope creep dynamics.

The Earth surface, just as any soft condensed matter near one of their failure criteria, is generally far from equilibrium and from presenting isotropicity. Finding insightful measurements of such system responses, and how to use them practically to predict material failure, has been a crucial scientific endeavor. Section \ref{sec:Daniels} presents recent experimental results on this front and their implications for further developing failure prediction in the environment. Different types of  environments  exhibit near-critical behavior, and they have been recorded and analyzed in different ways: for example, section \ref{sec:Burton} presents a recent highly-resolved spatial and temporal recording of iceberg collective dynamics along the coast of Greenland; section \ref{sec:Masteller} presents the challenge of observing and modeling river bed dynamics from flood to flood. In both natural systems, as in experiments presented in section \ref{sec:Daniels}, the stress \textendash ~or energy \textendash ~landscape in the system appears significantly changed after a failure event, leading to hysteretic behavior. Finally, section \ref{sec:Suckale} presents some of the major questions in modeling volcanic processes from the fundamental scale of mineral crystals and gas bubbles. 
\label{sec:(II)}

\subsection{Landscapes of glass - \textit{D. Jerolmack and N. Deshpande}}
\label{sec:JerolmackAndDeshpante}
Students of Physics have long been attracted to other fields of study, while the frontiers of Physics can also be advanced by challenges arising from other fields. Biophysics is now a recognized discipline within Physics where the essential elements of life \textendash ~far-from-equilibrium dynamics, self-organization, soft materials and complex fluids \textendash ~have required novel developments in theory and experiments from Physics \cite{glaserBiophysicsIntroduction2012}. Geoscience also presents novel problems, patterns, materials, and extreme conditions that have drawn the attention of some physicists. But, the field of ``Geophysics'' is not endemic in physics departments and education.

The traditional field of Geophysics can be roughly split into two areas. The first, ``Solid Earth Geophysics'' is concerned with the application of solid mechanics to measuring and modeling rocks and ice over a variety of length and time scales \cite{guptaEncyclopediaSolidEarth2011}. The second, ``Geophysical Fluid Dynamics'', is deeply rooted in applied mathematics and its application to atmospheric and ocean dynamics \cite{pedloskyGeophysicalFluidDynamics2013}. 
We recognize that there is a large gap between these two areas that we tentatively call ``Soft Earth Geophysics''. This field is concerned with the behaviors of Earth's granular materials \textendash ~from clays to boulders \textendash ~and their interactions with each other, biological materials, and fluids \cite{jerolmackViewingEarthSurface2019}. These materials transit many regimes: thermal to athermal and reactive to inert (clay to sand), jammed and creeping to dilute and flowing, and soft and porous to hard and dense. Often these regimes are mixed, and/or systems repeatedly transition among them. Certainly, there are many scientists and engineers at work on these problems, in fields such as geotechnical engineering, hydrology, and others. But physicists with expertise in soft matter, statistical mechanics, and nonlinear dynamics have not organized themselves to attack these problems with the formality associated with fields such as Biophysics. 

Consider the formation of post-wildfire debris flows, which is an increasingly frequent and deadly hazard. Debris flows are highly concentrated slurries of soil and water that form on steep hillslopes \cite{regmiChapter11Review2015}. 
Predicting the conditions that will trigger debris flows, and assessing the hazard associated with their runout, still relies largely on empirical relations derived from observations of previous flows. 
The challenges for understanding the failure and dynamics of debris flows represent frontier challenges in soft matter science:

(i) Some wildfires are known to leave behind a hydrophobic layer beneath the surface, which may help to confine rainfall to a shallow surface layer of soil that accelerates saturation and failure \cite{hubbertPostFireSoilWater2012,gabetPostfireThinDebris2003,alessioPostWildfireGenerationDebrisFlow2021}. 
Rapid wetting of surface soils may also create strong capillary pressure gradients that regulate soil failure and erosion style. 
Progress in this problem will require understanding of wetting under extreme conditions, and the influence of interfacial soil properties on infiltration and capillarity.

(ii) Debris flows may form by an unjamming transition in which soil experiences a sudden loss of rigidity associated with a decrease in volume fraction; i.e., a landslide \cite{shenEDDAIntegratedSimulation2018}. However, they may also form by progressive soil entrainment that increases volume fraction until it reaches close to the jamming point \cite{cannonProcessFirerelatedDebris2001,alessioPostWildfireGenerationDebrisFlow2021}. The conditions that lead to one or the other mechanism are not known. 

(iii) The rheology of debris flows is certainly non-Newtonian; generally, debris flows appear to be yield stress materials with some degree of shear thinning \cite{schippaEffectsSedimentSize2018,sosioRheologyConcentratedGranular2009}. However, rheology appears to be extraordinarily sensitive to the concentrations of clay and sand \cite{parsonsExperimentalStudyGrain2001}. Concepts of jamming and lubrication are just beginning to be applied to heterogeneous debris-flow materials, and offer some hope to explain and even collapse the variability observed in disparate studies \cite{kostynickRheologyDebrisFlow2022}. 

(iv) Debris flows entrain large boulders that migrate to the front of the flow and act as a battering ram \cite{bakerParticleSizeSegregationSpontaneous2016,keanInundationFlowDynamics2019}. Whether this is the result of granular segregation like the Brazil nut effect or a consequence of phase separation of  granular (boulder) and liquid (mud) materials, is unknown.

A major limitation in understanding failure and yield of Earth materials is that models still rely exclusively on a Mohr-Coulomb criterion \cite{terzaghi1943}. We know from observations, however, that sub-yield creep is pervasive in hillslope soils \cite{eylesSoilCreepHumid1970,flemingRatesSeasonalCreep1975,auzetSoilCreepDynamics1996,matsuokaRelationshipFrostHeave1998}. Gravity-driven creep of (athermal) granular materials has not been examined in Physics until recently \cite{houssaisOnsetSedimentTransport2015,Houssais2021,cunez_strain_2022,Ferdowsi-PNAS2018}. Surprisingly, it has been shown that even an undisturbed sandpile creeps; even more surprising, relaxation by creep is very similar to aging in a glass following application of a stress \cite{Deshpande2021}. These dynamics can be modulated by disturbances: tapping the pile hardens the bed and accelerates aging of the granular material, while heating it can reverse aging and sustain high creep rates. The dynamics suggest that mechanical noise in granular materials may play a role akin to thermal fluctuations in glasses. This is a frontier topic in the physics of amorphous solids, and gives rise to a host of new questions. Patterns of localized strain in the relaxing granular pile are similar to numerical simulations of metallic glasses, and also to large-scale strain fields around a ruptured fault following an earthquake \cite{jonssonPostearthquakeGroundMovements2003}. What is the role of material properties in setting the length and time scales of localized strain? Earth materials are subject to a wide spectrum of forcing \textendash ~hydrologic, seismic, biogenic, etc. Which kinds of forcing harden granular landscapes, and which ones set them up for failure? Finally, we observe in the field that in some locations soil creep is widely distributed within the bulk, while in other locations strain is highly localized along failure planes to create ``earthflows'' \cite{murphyVadoseZoneThickness2022}. What determines whether creep is spatially localized or diffuse, and can we understand earthflow dynamics as creep along a shear band?

\subsection{Rigidity, nonlocality, and acoustics in dense granular materials - \textit{K. Daniels}}
\label{sec:Daniels}
\begin{figure}[h]
 \centering
 \includegraphics[width=\linewidth]{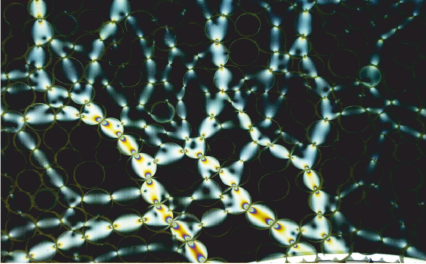}
 \caption{An image of photoelastic disks resisting a shear force applied by the roughened boundary visible at the lower right, imaged with a darkfield polariscope  \cite{AbedZadeh2019,Daniels2017}. These methods allow for the quantitative determination of the vector contact forces between particles when performed in monochromatic light. In this image, the brighter particles are those carrying more force, while the darker particles carry little force. Under increased shear, the chains of forces buckle and rearrange (Source: E. Berthier, F. Fazelpour, C. Kirberger, NC State Physics)}
 \label{fig:DanielsFig}
\end{figure}

Forecasting when Earth's critical zone will flow \textendash ~whether through creep \cite{Deshpande2021}, flow \cite{kostynickRheologyDebrisFlow2022}, or catastrophic failure \citep{Cascini2022} \textendash ~underlies many of the problems presented in this paper. Within the soft matter physics community, these questions have been addressed as questions of rigidity: how resistance to flow arises from the particle-scale to the meso-scale and to the system scale. Within a granular or amorphous material, internal stresses are transmitted by a heterogeneous network of forces known as force chains, as shown in Figure~\ref{fig:DanielsFig}. This network provides the material with its global rigidity, and several techniques exist for probing the spatio-temporal evolution of rigidity at various scales. Physicists have constructed models based on nearly-perfect particles residing within an energy landscape of valid states \cite{Wyart2005}, as well as simplified models comparing the number of constraints to the number of degrees of freedom\cite{Mao2018}. 

For simplified laboratory systems, in spite of their dissipative nature and the difficulty of defining a frictional failure criterion, it now appears that the energy- and constraint-based frameworks both predict the same regions as being rigid or not \cite{Liu2021}. When passively listening to acoustic emissions transmitted through the material, the statistical distribution of the resulting vibrational modes subtly shifts as a laboratory granular material approaches its point of failure \cite{Brzinski2018}, as would be predicted for model materials developing low-frequency vibrational modes as they approach a state with zero rigidity \cite{Ohern2003}. Finally, for models of disordered solids \textendash ~networks manufactured to have a disordered network of thin beams \textendash ~it is possible to forecast the most likely failure locations using only the meso-scale topology of the network's  connectivity, without including any mechanical information\cite{Berthier2019}.  

It remains an open question whether these frameworks can translate to the rough, heterogeneous, anisotropic particles and wet environments necessary to understand geophysical dynamics. For instance, is it possible to measure a quantity like the density of vibrational modes\cite{Owens2013} using seismometers or strain sensors? When a hillslope or glacier progresses towards a point of failure, do similar hallmarks forecast likely failure locations and times? Already, network science has been successfully used to evaluate kinematic data obtained from ground-based radar, interpreted in light of the underlying micro-mechanics of granular failure, to successfully  forecast the location and time of granular failure\citep{Tordesillas2018}.

\subsection{Floating granular materials - \textit{J. Burton and K. Nissanka}}
\label{sec:Burton}
\begin{figure}[p]
 \centering
 \includegraphics[width=\linewidth]{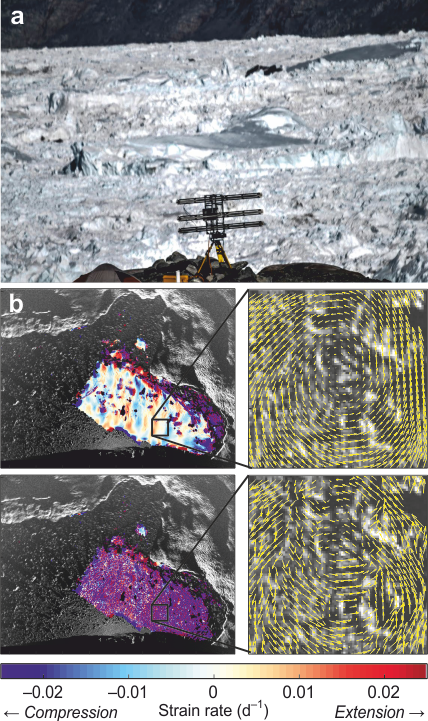}
 \caption {Adapted from Cassotto et al. \cite{cassotto2021granular}. (a) View of ice m\'{e}lange at Jakobshavn Isbræ on the Western coast of Greenland. The ground-based radar visible in the center of the image is a few meters in size and is perched on the rocky cliffs above the fjord. (b) Top-left, divergence of velocity field under steady flow. Red areas represent extension of the flow, and blue areas represent compression. Overall, the field is smoothly varying. Top-right: variation of velocity field in the black rectangle after subtracting the mean of the underlying steady flow. Bottom-left, divergence of velocity field less than 1 hour before a calving event (the fracture and discharge of a cubic-kilometer-sized iceberg from the glacier into the ice m\'{e}lange). The divergence field is rapidly-varying and noisy. Bottom-right: variation of velocity field before calving, showing heterogeneous flow patterns. }
 \label{fig:BurtonFig}
\end{figure}

A single drop of soap can drastically affect the fluid dynamics of a liquid interface. Surfactant molecules spread rapidly, drive flows, and provide stability and elasticity. Although the Earth’s oceans, seas, and rivers cover immense length scales, granular collections of ice, trees, organisms, and pollutants can shield the air/water interface. Such floating granular materials often jam in converging flows or narrowing geometries, creating hazards or ephemeral perturbations to the dynamics of Earth’s aquatic interfaces. Examples include logjams \cite{deshpande2019logjams}, river ice \cite{ettema1990jam}, sea ice \cite{herman2013numerical}, and volcanic pumice \cite{jutzeler2014fate}. In biological systems, granular rafts can be formed intentionally to survive flooding, as in the case of fire ants \cite{mlot2011fire}. Although the fractional coverage of Earth’s water bodies with floating granular materials is small, they can be exceedingly important, since crucial veins of transport can become quickly jammed with buoyant terrestrial debris. 

During the workshop, we showcased an outsized example of this behavior: ice m\'{e}lange, a buoyant agglomeration of icebergs and sea ice that forms in the narrow fjords of Greenland (Fig.~\ref{fig:BurtonFig}a). Ice m\'{e}lange is perhaps the world’s largest granular material \cite{burton2018quantifying,robel2017thinning,amundson2018quasi}, with individual clasts ranging from 10s to 100s of meters in size. As ice m\'{e}lange is slowly pushed through fjords that are many kilometers wide, it jams, buckles, and breaks as friction from the rocky walls transmits stress to the buoyant interior.  Importantly, ice m\'{e}lange has recently been shown to affect ice-sheet mass losses by inhibiting iceberg calving \cite{cassotto2021granular}. During calving, cubic-kilometer-sized icebergs are fractured from the main glacier and discharged into the ice m\'{e}lange. Surprisingly, centimeter-scale iceberg displacements can be measured with ground-based radar every 3 minutes. The measurements revealed that a period of incoherent granular flow preceded iceberg calving events (Fig.~\ref{fig:BurtonFig}b), representing an important first step towards real-time detection of failure in geophysical granular flows. Within the context of floating granular materials, there remain a few key challenges. These materials are very sensitive to particle shape and confinement, both of which are essential for their ability to jam and transmit stress. Also, these materials can interact with the water, e.g., melting ice drives stratified flows from below. Finally, laboratory studies combined with continuum modeling using granular rheologies are needed to provide a larger-scale picture of how floating granular materials shape and respond to their dynamic environment.

\subsection{How rivers remember - \textit{C. Masteller}}
\label{sec:Masteller}
Erosion and morphological change in gravel-bed rivers require bedload transport, or the transport of sediment by rolling, sliding, or saltating close to the riverbed.  Almost all existing model predictions of bedload transport rates are underpinned by the degree to which flow conditions exceed some threshold value in dimensionless shear stress \cite{engelund_sediment_1976, luque_erosion_1976, Meyer-Peter1948, wong_reanalysis_2006} representing the initiation of motion of sediment particles, e.g., \cite{shields_application_1936,wiberg_calculations_1987}, or some “reference” transport rate, e.g., \cite{parker_self-formed_1978, wilcock_surface-based_2003}. For gravel-bed rivers, the bulk of sediment transport occurs close to these thresholds \cite{parker_self-formed_1978, phillips_self-organization_2016, Phillips2022}, making estimates of the threshold for particle motion critical for accurate predictions of fluvial transport rates. 

Further, there is a strong link between the width of a river during bankfull flow, or a flow that fills a channel to the top of its banks, and the entrainment threshold of bed and bank sediments. This model of channel stability, referred to here as “near-threshold channel” theory (NTC), was formalized by Parker\cite{parker_self-formed_1978} and has been validated within coarse-grained alluvial and bedrock influenced rivers \cite{metivier_laboratory_2017, mueller_variation_2005,phillips_self-organization_2016,Phillips2022}  
Most bedload transport models that use dimensionless shear stress, also termed critical Shields stress, ${\sigma^*_c}$, as the threshold parameter, have assumed a constant value near ${\sigma^*_c}$ = 0.045. However, Buffington and Montgomery  \cite{buffington_systematic_1997}  demonstrated that ${\sigma^*_c}$ varied systematically with the ratio of the median sediment grain size to flow depth, and argued that a universal threshold should be applied with caution. 

More recent work has explored how flow and grain interactions lead to inherent variability in the threshold for particle motion. Reid et al.\cite{reid_incidence_1985} first suggested the influence of antecedent flows based on field-based bedload transport monitoring, hypothesizing that longer inter-flood durations led to increases in ${\sigma^*_c}$ and reduced sediment transport rates. 

We present experiments that confirm that the magnitude and duration of inter-event flows affect ${\sigma^*_c}$ evolution. We show that with little to no active sediment transport, grain-scale changes in interlocking, and subtle surface reorganization increase particle resistance to motion \cite{masteller_interplay_2017}. We suggest that the increase in particle resistance under inter-event flows is akin to granular creep and compaction of granular materials under low to moderate shear rates. In response to higher magnitude flows (e.g., floods) surface reorganization of the bed leads to a decrease in particle entrainment thresholds via an increase in surface roughness, akin to dilation of granular materials under high shear rates.
We show that the magnitude of antecedent flows was the dominant control on the evolution of ${\sigma^*_c}$ in the Erlenbach torrent in Switzerland, with a secondary, short-lived duration effect\cite{Masteller2019}. Consistent with experiments, these direct measurements of the onset of motion showed increases in critical Shields stress with increasing inter-event flow magnitude. The study also explored responses of ${\sigma^*_c}$ to higher magnitude, sediment transporting flows. As with below-threshold flow, strengthening effects were also observed following low to intermediate-magnitude bedload-transporting floods. However, following high-magnitude flows, the threshold for motion decreased. Masteller et al.\cite{Masteller2019} hypothesized that the transition from bed strengthening to bed weakening was associated with a transition from local rearrangement of particles to more intense transport, which acts to significantly disrupt bed structure via particle collisions or long-distance particle transport. 
To capture these variations in particle erosion thresholds, we develop a flow history-dependent model in which ${\sigma^*_c}$ evolves through time as a function of bed shear stress \cite{2020AGUFMEP008..04M}.  We calibrate the model to a 23-year record of flow and bedload transport at the Erlenbach.  We demonstrate that the model predicts field-based ${\sigma^*_c}$ values more accurately than the assumption of a constant ${\sigma^*_c}$. We suggest that this model may be more generally applied to capture time-varying erosion thresholds in gravel-bed rivers. 

There remain outstanding challenges:

(i)	What constraints does a deformable boundary (channel container) place on variations in bed shear stress, and by extension, evolution of particle entrainment thresholds?
Work by Cuñez et al.\cite{cunez_strain_2022} demonstrates that dilation occurs at shear stresses well above those commonly observed to result in channel widening – suggesting that bed disruption or weakening may be infrequent or buffered by channel width adjustments. How do adjustments in particle entrainment thresholds impact thresholds for channel widening? 

(ii) Definitive links between granular processes observed in physical experiments and field observations in gravel-bed rivers are precluded by measurement limitations related to natural grains including complex shape (e.g.,\cite{deal_grain_2023}), observational of grain arrangement limited to surface topography, and limits of fine-scale measurement capabilities in the field, as dynamic changes in grain scale topography or erosion thresholds are relatively small and thus challenging to measure.  
There is potential for the application of geophysical methods, including environmental seismology (e.g.,\cite{burtin_continuous_2013, Cook2022, Larose2015, Schmandt2017}) and distributed acoustic systems \cite{2022AGUFMEP33A..01R}, but these methods are still largely unexplored in a fluvial context.

\break
\subsection{Why do persistently degassing volcanoes erupt? - \textit{J. Suckale}}
\label{sec:Suckale}
\begin{figure}[ht]
 \centering
 \includegraphics[width=\linewidth]{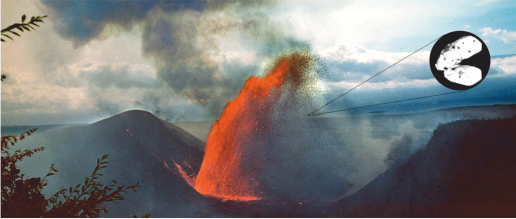}
 \caption{Photo of a lava fountain during the 1959 Kilauea Iki eruption (courtesy of USGS) in the background and a photo of a crystal cluster later identified in erupted samples by Schwindinger and Anderson\citep{schwindinger1989synneusis}.}
 \label{fig:SuckaleFig}
\end{figure}

Not all volcanic eruptions are rare. According to the Volcano Watch by the United States Geological Survey, dozens of volcanoes erupt every day and often the same ones. These volcanoes are commonly referred to as persistently active. Persistently active volcanoes are able to maintain an open connection between the magma storage regions and the surface vent, creating a dynamic system that erupts frequently. Their volcanic activity spans a wide spectrum from continuous passive degassing to intermittent explosive or effusive eruptions with more violent, paroxysmal eruptions emerging with little or no clear precursory activity\citep{albert2016years, passarelli2012correlation}. The transitions between different eruptive regimes are sudden and unpredictable, creating large uncertainty in risk assessments\citep{ripepe2017forecasting}. 

An example is K{\=\i}lauea volcano, Hawaii, where the continual and slow oozing-out of magma during year-long eruptions is disrupted episodically by dramatic, day-long lava fountains that propel an explosive spray of molten clasts and ash hundreds of meters into the air\citep{richter1970chronological,swanson1976february,wolfe1987puu,heliker2003first}. Current classifications of these eruptive regimes are largely phenomenological. To make progress towards predicting changes in eruptive behavior, we need to underpin this existing typology with an improved understanding of the physical processes that could cause a regime transition. In fact, the National Academies declared the development of multi-scale models that capture critical processes and can be tested against field data as one of the three grand challenges in modern volcanology\citep{NAS2017}. 

At the heart of the question about eruptive regime transitions is the challenge of near-criticality. Most of the time, persistently active volcanoes are not erupting and still emitting copious quantities of gas and thermal energy\cite{StoiberandWilliams1986, ALLARD1994, KAZAHAYA1994, Palmaetal2008, OPPENHEIMER2009, woitischek2020strombolian}; why not always? Near-criticality could provide a valuable framework for understanding why seemingly small increases in gas flux, pressure or crystallinity could lead to a sudden and dramatic change in behavior: For example, petrological data shows that the uppermost few hundred meters of the plumbing system at Stromboli volcano, Italy, are composed of a highly crystalline magmatic mush with a solid fraction of 45-60\%\citep{metrich2001crystallization,francalanci2004volcanic,francalanci2005intra}. This mush is prone to tensile failure beneath the observed vent locations driven by gas overpressure and the tectonic stress field, suggesting that Strombolian eruptions could be related to a transition from flow to failure\citep{suckale2016flow}.

A transition from distributed flow to localized failure can also occur in flow configurations with low crystallinity. One process that could trigger this transition even in a largely fluid system is the hydrodynamic interaction between individual crystals\citep{qin2020flow}. The hydrodynamic interactions between crystals are amplified by the high viscosity of magmatic melts, roughly five to twelve orders of magnitude higher than water, because it implies that individual crystals interact hydrodynamically over spatial distances many orders of magnitude larger than their size. These long-scale hydrodynamic interactions between individual crystals favor self-organizing behavior, reflected in spatially variable crystal distributions. This self-organization depends on the ambient, pure-fluid flow field and at the same time modifies it \citep{suckale2012crystals,qin2020direct,qin2020flow}. 

The diversity of physical processes that can disrupt conduit flow may be reflected in the diversity of observed eruptive regimes, but we will never know unless we can test different models directly against data. Some of the most precious clues may emerge from the smallest scales, namely individual crystals or bubbles, often at the micrometer scale. While large-scale data sets like seismicity, crustal deformation or heat and gas flux measurements at volcanoes provide impressive testimony of syn-eruptive processes, crystal-scale data may record at least some pre-eruptive processes directly\citep{dibenedetto2020crystal}. Lending a helping hand in preserving this information is the glass transition, another aspect of near-criticality in natural systems. Once the eruption starts, the melt in the conduit quenches to a glass, freezing-in the crystals and bubbles it contains. 

Figure~\ref{fig:SuckaleFig} shows an example of a lava fountain during the 1959 eruption at K{\=\i}lauea Iki, Hawaii, and a close-up photo of the crystal clusters later found in erupted samples\citep{schwindinger1989synneusis}. A detailed analysis of the crystal cluster showed that the cluster formed by two crystals drifting together during flow and intergrowing over time\citep{schwindinger1989synneusis}. The puzzling aspect of these crystal clusters is the abundance of relatively large misalignment angles separating the two crystals\citep{wieser2019sink}. A smaller angle would be hydrodynamically more favorable, but is only observed in a surprisingly small percentage of clusters. Reanalyzing this data, we showed that the angles between these clustered crystals could be produced by exposure of the individual crystals to a traveling wave in the conduit prior to eruption\citep{dibenedetto2020crystal}.

In linear-shear flows, crystals tumble along in Jeffery orbits \citep{jeffery1922}, but wavy flows align crystals \citep{DiBenedetto2018b, DiBenedetto2019} onto a preferential angle that depends on both the flow conditions and the crystal geometry. From the crystal geometry, it is possible to infer that the observed high percentage of large misalignment angles is indicative of a downward propagating wave in a volcanic conduit with low crystallinity\citep{dibenedetto2020crystal}. The inferred crystallinity is consistent with the lower range of observed crystallinities\citep{richter1966petrography} and consistent with the possibility that a spatially heterogeneous arrangement of crystals inside the volcanic conduit could trigger a transition from flow to sliding \citep{qin2020flow}.

Many questions regarding the eruptive behavior of persistently active volcanoes remain. Making progress on these is a challenge that benefits from the input of fields outside of classical volcanology, including but not limited to multi-phase flow, non-linear system dynamics, thermodynamics, and numerical analysis. It is also a challenge that touches on several themes discussed later in this paper, particularly the challenge of modeling the grain scale and the challenge of bridging scales. As many other natural systems, volcanic systems span an enormous range of physical conditions and scales from microns to hundreds of kilometers.  


\section{The challenge of bridging scales}
Scaling analysis tries to identify and match relevant physical regimes at the different spatial and temporal scales, and is at the heart of Earth science. One can upscale model results to predict field observations, or downscale geophysical results in explaining them by essential processes studied in the laboratory (Fig.~\ref{fig: IntroFig}). Identifying and scaling the essential physics for each pattern is challenging at times, because the wide range of spatial and temporal scales involved gives rise to several regimes, and the same patterns can result from different mechanisms. To test the validity of models across scales, analyzing the characteristics of patterns, in the time series of, or shapes produced by, landscape mechanical behavior is often key. Mathematical modeling can offer pathways to bridge laboratory observation to field observation scales in analyzing the dynamics of regimes and emergence of the associated patterns.

Section \ref{sec:Lai} highlights that mechanisms of viscous and elastic deformations might differ in the temporal and spatial domains, in the specific case of Antarctic ice dynamics. It is shown that brittle fracture dominates in local and momentary observations, but ice sheets appear to flow viscously when observed for years over large areas. 
Section \ref{sec:Devauchelle} tests if rainfall time signals, modulated in space via groundwater flow, can be approximated by averaging over the domains. Such results bring subtle questions to a common approach of bridging scales by averaging over a grid point.

Section \ref{sec:Hewitt} considers the complex fluid dynamics at the bottom of glaciers, and how to reproduce the evolution of subglacial channel systems, coupling models of sediment transport and ice melting. 
Section \ref{sec:Goehring} focuses on understanding how physicochemical mechanisms in the ground result in the striking formation of surficial salt patterns. Section \ref{sec:Glade} presents a scaling analysis and remote sensing measurements of periglacial soil patterns and investigates their relation to fluid-flow instabilities.  Finally, the dynamics of dunes \textendash ~fragile but perpetual forms in deserts \textendash ~remains challenging to predict; section \ref{sec:Vriend} uses laboratory-scale experiments to show how dunes persist and set a length scale in landscapes by interacting, attracting, and repelling each other. 
\label{sec:(III)}

\subsection{Ice cracks in a warming climate - \textit{Y. Lai}}
\label{sec:Lai}
\begin{figure*}[ht]
 \centering
 \includegraphics[width=\textwidth]{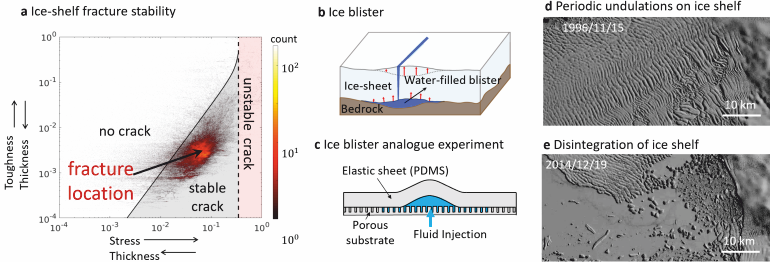}
 \caption{(a) Fracture stability diagram for Antarctic ice-shelf fractures. Most ice-shelf fractures identified by a neural network on Antarctic satellite imagery (red dots) lie in the stable-fracture regime\cite{Lai2020vulnerability}. (b) Formation of water-filled “blister” at the bottom of the ice sheet after a lake drains\cite{lai2021hydraulic}. (c) Analogue laboratory experiment mimicking a water-filled blister beneath an ice sheet relaxing on a porous water network\cite{chase2021relaxation}. (d) Satellite image showing undulation patterns on ice shelves due to fracture formations on the Thwaites Ice Shelf in 1996. (e) Same region as (d), in 2014 when the ice-shelf was broken into icebergs ((d, e) are from Landsat image). }
 \label{fig:LaiFig}
\end{figure*}

Interactions between fluids \cite{fowler1986sliding,schoof2010ice,schoof2013ice,hewitt2012flotation,schoof2007marine,pegler2018suppression,sayag2019instability,rallabandi2017wind,wettlaufer2006premelting}, elasticity \cite{vella2007finger,vella2008explaining,sayag2013elastic,wagner2016role,wagner2014footloose}, sediments \cite{fannon2017numerical,zoet2020slip,warburton2023shear}, granular flows \cite{burton2018quantifying}, and porous flows \cite{rempel2004premelting,meyer2017continuum,moure2023thermodynamic} are ubiquitous in the polar regions. Ice sheets and ice shelves are viscous gravity currents spreading above bedrock and ocean \cite{schoof2013ice}, respectively. Ice  flows as a viscous fluid (e.g., glaciers) at longer timescales but breaks as a solid at shorter timescales (e.g., iceberg calving \cite{benn2007calving,bassis2013diverse}). Because the mass loss of ice sheets contributes to the rising sea levels, it is important to understand the fate of ice sheets in a changing climate. In this workshop, we highlighted a few processes involving interactions between fluids and solids with important implications for ice-sheet dynamics. We also identified some open questions to be addressed.

A major open question affecting future sea levels is whether meltwater-driven fracturing of ice shelves will significantly impact the future loss of the Antarctic Ice Sheet. Atmospheric warming threatens to accelerate the retreat of the Antarctic Ice Sheet by increasing surface melting and facilitating hydrofracturing \cite{scambos2000link}, where meltwater flows into and enlarges fractures on ice shelves \cite{weertman1973can,van1998fracture}, potentially triggering ice-shelf collapse \cite{scambos2000link,banwell2013breakup,robel2019speed} and acceleration of sea-level rise \cite{deconto2016contribution}. In this context, we combined theory and deep-learning to develop a stability diagram for Antarctic fractures\cite{Lai2020vulnerability}. Figure~\ref{fig:LaiFig}a illustrates a theoretical prediction of the stability of Antarctic fractures depending on the ice thickness, ice toughness, and glaciological stresses on ice shelves. To compare observations with theory, we trained a deep convolutional neural network to detect continent-wide fracture features on ice shelves\cite{Lai2020vulnerability}. We find that most ice-shelf locations that the deep neural network detects as fractures, shown as points in Figure~\ref{fig:LaiFig}a, lie in the parameter regime where our theory predicts stable fractures (gray triangle), and are consistent with the fracture theory. We find that if climate warms and causes the Antarctic ice surface to melt, large portions of Antarctic ice shelves will likely collapse due to hydrofracture\cite{Lai2020vulnerability}.

Besides theory and field observations, in this workshop, we also explored the unique contribution of analogue experiments to the understanding of ice-sheet processes. The benefit of analogue experiments is that the parameters can be well controlled. We connected the findings in analogue experiments with the large-scale geophysical observations by matching the relevant nondimensional parameters. We discussed the use of an analogue experiment to mimic the formation and relaxation of a water-filled ``blister'' (Fig.~\ref{fig:LaiFig}b) beneath an ice sheet due to the injection of meltwater \cite{tsai2010model,lai2021hydraulic}. The analogue experiment \cite{chase2021relaxation} (Fig.~\ref{fig:LaiFig}c) validated a mathematical model describing meltwater leaking from a pressurized “blister” into the surrounding water network (modeled as a porous substrate) beneath the ice sheet (modeled as an elastic sheet). The mathematical model has been used to constrain the hydrological property of the water network beneath the ice sheets, which is otherwise difficult to measure\cite{lai2021hydraulic}.

Many unanswered questions are to be explored, such as the processes governing the catastrophic collapse of ice shelves, including the mechanisms responsible for the periodic undulations observed in satellite imagery (Fig.~\ref{fig:LaiFig}d). The surface periodic undulations are highly correlated with locations of basal crevasses\cite{luckman2012basal, mcgrath2012basal, buck2021flexural}. Importantly, the undulation spacing is relevant to the size of the icebergs (Fig.~\ref{fig:LaiFig}e). The types of mechanical instabilities \cite{coffey2022enigmatic} that give rise to these periodic spacing are still poorly understood. How the complex rheology impacts the mechanical instabilities, the disintegration of ice shelves, and the dynamics of ice sheets, is still to be investigated.

\subsection{Can we average the rainfall signal? - \textit{O. Devauchelle}}
\label{sec:Devauchelle}
\begin{figure}[h]
 \centering
 \includegraphics[width=\linewidth]{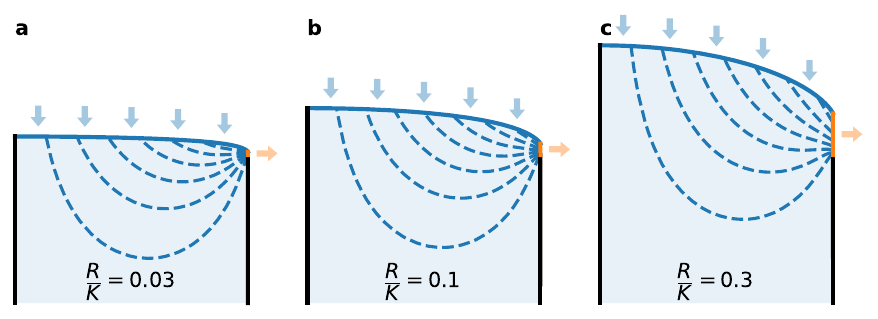}
 \caption{Flow in unconfined aquifers, of different volumes/heights (increasing from (a) to (c)) of hydraulic conductivity $K$ recharged by a constant rainfall $R$ (analytical solution\cite{jules2020couplage}). Rainwater infiltrates into the ground (blue arrows), and joins the water table (solid blue line). From there, it follows the groundwater flow lines (dashed blue lines) until it reaches the outlet (orange line), where it seeps into a river. The river flows towards the reader. The solid black lines are impervious (left: groundwater divide, right: axis of symmetry). All lengths are made dimensionless with the distance that separates the river from the divide. There remains only one dimensionless parameter in this problem: $R/K$.}
 \label{fig:DevauchelleFig}
\end{figure}

Rainwater infiltrates into the ground, until it reaches the water table, where the porous matrix is saturated with groundwater (Figure~\ref{fig:DevauchelleFig}). There, it begins the slow underground travel that will eventually bring it back to the surface, where it will join a stream, and run to the sea. How long does the underground part of this travel take? Clearly, the residence time of water in an aquifer is $\tau = \dfrac{V}{R A}$, where $V$ is the groundwater volume, $A$ is the area of the catchment, and $R$ is the rainfall rate\citep{jules2021flow} (typically expressed in mm$\,$year$^{-1}$). Residence time is thus tantamount to storage.
It is also a prime control on the biological and chemical reactions that weather the porous matrix\citep{maher2011role, harman2019low}, and a good estimate of the time it takes for groundwater to recover from pollution.

As a first approximation, we can average the rainfall signal over the years, and treat its mean $\langle R \rangle$ as a steady forcing of the groundwater flow. The resulting Darcy problem is then stationary, and amenable to classic fluid mechanics. For illustration, Figure~\ref{fig:DevauchelleFig} shows the stationary flow of groundwater through deep, unconfined aquifers that discharges into neighboring streams\citep{jules2021flow}. As the rainfall rate increases, the water table rises. The domain over which the flow equations need to be solved thus expands, and this makes the problem non-linear. Even in steady-state, therefore, the residence time of water in an aquifer is not just inversely proportional to the rainfall rate, because the volume of groundwater needs to accommodate the flux it carries ($V$ is a function of $R$). 

In reality, of course, the rainfall signal is intermittent, and so is the groundwater flow it induces\citep{Guerin2019}. Since this  problem is non-linear, we cannot expect that the time average will gracefully propagate through the equations, as it would in a linear system. There is no reason to believe, therefore, that the steady flow of Figure~\ref{fig:DevauchelleFig} is the average of the actual groundwater flow. To find the latter, we generally need to solve the non-stationary problem, and average the result over time \textendash ~a procedure far more costly than solving the steady-state problem. In short, finding the average groundwater flow is a difficult problem, because it is not just the solution of the average equations.

This issue, which might seem anecdotal, is in fact ubiquitous. One simply needs to consider a non-linear system driven by some fluctuating forcing \textendash ~rainfall or temperature. In cold climates, for instance, the soil cycles through freezing and thawing (Glade, Sec.~\ref{sec:Glade}), which obviously modulates its rheology, and therefore its downward creep (Jerolmack and Desphante, Sec.~\ref{sec:JerolmackAndDeshpante})\citep{Ferdowsi-PNAS2018}. Could this parametric forcing explain why some soil patterns appear only in the Arctic, as suggested by J.~Burton during the 2022 PCTS workshop? 

In other words, fluctuations do not always average out. In Earth sciences, this might be the rule rather than the exception.

\subsection{Subglacial plumbing systems - \textit{I. Hewitt}}
\label{sec:Hewitt}
Increased glacier and ice-sheet melting is an obvious consequence of climate warming, with significant impacts for sea-level rise and for water resources in mountainous regions.  Vast quantities of meltwater are transported beneath the ice, along the interface between ice and the underlying bedrock or till, driven out towards the ocean by the overlying weight of the ice.  With little opportunity for direct observations, various conceptual theories for how to envisage the subglacial drainage system have been developed.  There are similarities, and some important differences, to subaerial water flow and stream formation.  Open questions abound about the relevant physics, and how it can be modeled.  In particular, these include the role of erosion, deposition, ice-melting, and ice creep in enlarging and contracting the space available for water flow.  There are potentially useful analogies with other ``deformable'' or ``reactive'' porous media, and for an increased role for analogue laboratory experiments.

An important aspect of these systems is their temporal evolution \textendash ~it is inferred from tracing experiments that there is a massive expansion of the drainage system during the summer melt season (due to dissipation-driven melting of the basal ice), but that this subsequently collapses (due to the viscous ``creep'' of the ice) during the winter\citep{chandler2013evolution}.  The system is believed to transition between a relatively low permeability system in which water moves through porous sediments or ``linked cavities'', and a more efficient river-like network of channels\citep{werder2013modeling}.  The channels can be incised both upwards into the ice\citep{nye1976water} and downwards into the sediments\citep{Damsgaard2017}.

\begin{figure}[h]
 \centering
 \includegraphics[width= \linewidth]{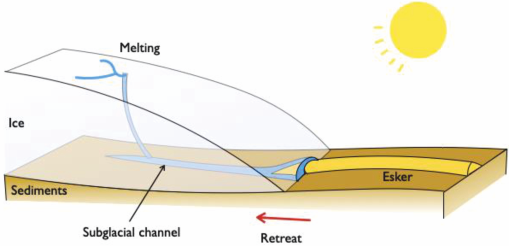}
 \caption{Suggested formation mechanism for an esker. Eskers are long ridges of sediment, found particularly in areas of Canada and Scandinavia, which were deposited as the ice-sheets retreated at the end of the last glacial period. Sediments are deposited as the flow velocity in a water-filled subglacial channel decreases near the retreating ice margin. With a better understanding of this formation mechanism, they could tell us more about the plumbing system under present-day ice sheets.}
 \label{fig:HewittFig}
\end{figure}

One aspect of these channels that provoked some discussion at the workshop is that they are believed to be responsible for depositing eskers (long, sinuous ridges of sand and gravel) during the last deglaciation\citep{hewitt2019model}. A model was presented in which these form through continual deposition at the widening mouth of the channel as the ice-sheet margin retreats (see Fig.~\ref{fig:HewittFig}), with various suggestions made for how the sediment size distributions within the eskers might be used to test this hypothesis. 

\subsection{Patterns in dry salt lakes - \textit{L. Goehring}}
\label{sec:Goehring}
\begin{figure}[ht]
 \centering
 \includegraphics[width=\linewidth]{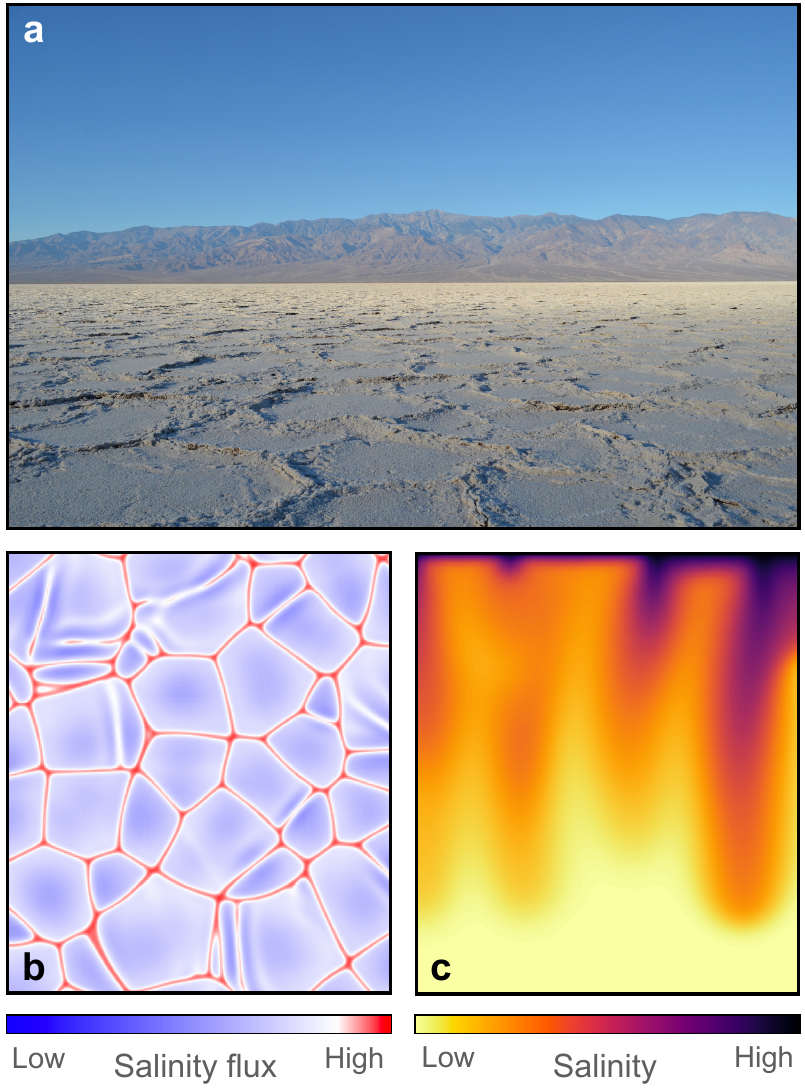}
 \caption{A convective model of salt polygons in dry salt lakes.  (a) The dry lake surface at e.g. Badwater Basin, Death Valley (CA, USA), is covered by a pattern of ridges in an approximately 10 cm thick crust lying over moist sandy soil.  Here, the polygonal features are typically about 1.5 meters across.  The model of buoyancy-driven flows used to simulate the emergent length scales and time scales of pattern formation at such sites predicts (b) salt flux into the crust (and hence crust growth rates) and (c) salinity profiles in the soil beneath. Panels (b, c) courtesy of and copyright Matthew Threadgold. }
 \label{fig:GoehringFig}
\end{figure}

Dry salt lakes, playas and salt pans represent some of the most extreme environments on Earth. They form in dry terminal basins where groundwater collects just beneath the surface and where evaporation dominates over precipitation\cite{Lowenstein1985}.  The otherworldly landscapes that result are ones of beautifully ordered polygonal patterns in a surface salt crust, and an inspiration to fantastic settings like Star Wars’ planet Crait.  Found worldwide, some noteworthy dry salt lakes include Badwater Basin in Death Valley (CA, USA, see Fig.~\ref{fig:GoehringFig}a), Salar de Uyuni (Bolivia), Dasht-e Kavir (Iran) and Sua Pan (Botswana).  The example of Qaidam Basin, China, has also been studied as an analogue for strikingly similar features found on Mars\cite{Dang2018}.  Although fracture\cite{Christiansen1963} and buckling\cite{Krinsley1970} of the surface crust are associated with these features, until recently, no clear mechanism has been able to accurately explain the emergent spatial and temporal scales of the salt crust patterns.  The main challenge to any such explanation involves identifying a mechanism specific to salt lake environments that can account for the consistent growth of 1-3 m wide closed polygonal features in the crust~\cite{Nield2015,Lasser2020}, over timescales of a few months \cite{Lokier2012,Nield2015}, and in a way that is insensitive to the exact salt chemistry and soil composition of any particular lake site.  

In order to predict the formation and dynamics of salt crust patterns, an intimate link between these dynamics and the convection of salty water beneath the soil has been proposed, where convection cells template the crust pattern\cite{Lasser2021,Lasser2022}.  Convection in porous media is itself well-studied, with a variety of approaches and applications summarized in a recent, extensive review\cite{Hewitt2020}. In the context of salt crusts, the connection is made to the particular problem of convection in the presence of a through-flow of fluid.  This problem was originally raised in the context of geysers\cite{Wooding1960}, but has since been developed to explain the subsurface flows observed at playas or dry salt lakes\cite{Wooding1997} and sabkhas, which are evaporate pans near tidal flats\cite{Stevens2009}.  

Briefly, the resulting model considers the Darcy flow of water in the porous sand or sediment of a dry lake, where the water table remains close to the surface\cite{Lasser2021,Lasser2022}.  The water contains salt, which accumulates at the evaporating surface.  The salt moves advectively with the water, and diffusively along any concentration gradients.  It adds to the density of the water, providing buoyancy forces that can drive additional flows.   As boundary conditions, there is a continuous loss of water at the surface, caused by evaporation, and the groundwater is recharged from below by some distant reservoir.  This leads to the accumulation of salt-rich, denser water near the surface, which can be unstable to convection.  The convective dynamics are captured by a single dimensionless group, the Rayleigh number, which describes the ratio of convective to diffusive effects.  Essentially, this group describes the vigor of any convection\cite{Lasser2021}, as it also characterizes the speed of the convective flows, relative to the background flows caused by the surface evaporation.  

Building on a body of recent field observations\cite{Nield2015,Lasser2020,Lasser2022}, this model of salt playa allows for the dynamic evolution of convection cells, which then modulate the salt flux into the surface crust\cite{Lasser2022}. As confirmed by direct field data of crust growth rates\cite{Nield2015} and subsurface flow patterns\cite{Lasser2022}, it predicts that surface salts will accumulate fastest above down-welling flows that spontaneously arrange into a polygonal network (see Fig.~\ref{fig:GoehringFig}b,c) when simulated in large, three-dimensional domains\cite{Lasser2022}.  When the model parameters are constrained by relevant field data, it also accurately accounts for the observed growth rates of the polygons, and their remarkably consistent size, which arises naturally from a balance between evaporation rates and salt diffusivity. Further development of these ideas would require considering more carefully the feedback between the crust and the groundwater flows and accounting for how the development of differences in crust thickness will, in turn, influence local evaporation rates\cite{Groeneveld2010,Eloukabi2013,Lasser2021}. This effort would not only contribute to our understanding of these patterns but also to environmental work.  For example, Owens Lake has been the focus of a decades-long remediation effort to reduce dust formation off of dry lake surfaces, which is linked to the dynamics of the salt crusts\cite{Groeneveld2010}.  However, even without further elaboration, the convective model demonstrates how the self-organization of flows beneath our feet can naturally explain the emergent length scales and time scales of salt polygons in nature.

\subsection{Large-scale arctic soil patterns analogous to small-scale fluid flow instabilities - \textit{R. Glade}}
\label{sec:Glade}
\begin{figure}[h]
\centering
\includegraphics[width= \linewidth]{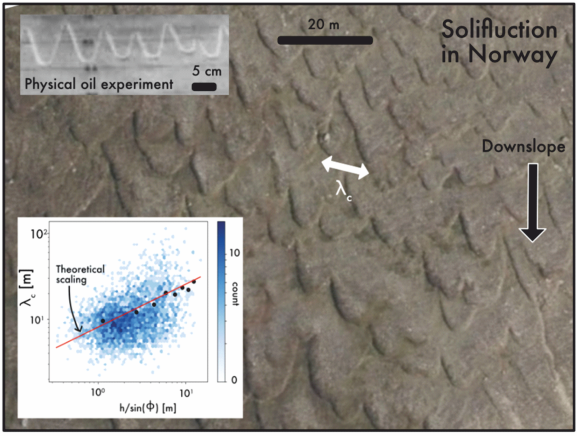}
\caption{Background: Aerial image of solifluction lobes with wavelength $\lambda_c$ ~ tens of meters in Norway (Image credit: The Norwegian Mapping Authority).
Top left inset: fluid contact line instability in a laboratory experiment with wavelength of centimeters \cite{huppert1982flow}. 
Bottom left inset: $\lambda_c$ plotted against lobe thickness divided by topographic slope, $h/\sin{\theta}$, measured from a remote-sensing derived digital elevation model.
Blue color indicates the number of data points in each hexagonal bin. Black dots represent average wavelengths binned by $h\sin{\theta}$. 
Red line is theoretical prediction from fluid-inspired scaling analysis, $\lambda_c \propto \sqrt{h \gamma/\sin{\theta}}$. Because we have no constraints on the cohesive length scale, $\gamma$, we assume it is constant in our plot. Figure modified from \cite{glade2021arctic}.}
\label{fig:GladeFig}
\end{figure}

A key challenge in linking soft matter physics with Earth science lies in dealing with the high degree of heterogeneity in natural landscapes (e.g., \cite{furbish2021rarefied, haff199614}). In cold landscapes, icy soil composed of an ever-changing mixture of heterogeneous sediment grains, liquid water, and ice demonstrates the complexity of Earth materials \cite{JerolmackDaniels2019}. Soil in these settings moves downhill at slow rates of millimeters to centimeters per year due to freeze-thaw processes (e.g.,\cite{matsuoka2001solifluction}); over time, the soil self-organizes into distinct finger-like patterns known as solifluction lobes, with wavelengths of tens to hundreds of meters (Fig.~\ref{fig:GladeFig}). Inspired by contact line instabilities in thin film fluids (e.g., paint dripping down a wall) that form due to competition between viscous forces and surface tension (e.g.,\cite{huppert1982flow}), we develop a theoretical prediction for the solifluction lobe wavelength that aims to connect grain-scale cohesion and fluid-like motion of the soil to large-scale pattern development while acknowledging the importance of natural heterogeneity. We find that similar to surface-tension dominated flows, competition between body forces and resisting forces (here in the form of enhanced soil cohesion at raised soil fronts) may drive pattern formation. Allowing for a hydrostatic component to account for large scale topographic roughness not present in thin films, we find a new scaling relation that predicts the cross-slope wavelength ($\lambda_c$) varies as a function of soil thickness ($h$), topographic slope ($\sin{\theta}$), and a length scale characterizing spatial variations in cohesion ($\gamma$):  $\lambda_c  \propto \sqrt{h \gamma/\sin{\theta}}$.

Using remote sensing data of thousands of solifluction lobes across Norway, we find that though solifluction wavelength data contains a large amount of scatter typical of field data, our theoretical prediction is able to predict average wavelengths\cite{glade2021arctic}. Using a long term climate dataset, we also find that average lobe heights and wavelengths increase with elevation, pointing toward a broad climate control on solifluction patterns and illustrating the possible importance of external driving factors in addition to smaller-scale soil dynamics. 
This work demonstrates that even granular material in the non-inertial regime can exhibit instabilities fundamentally similar to those found in small scale systems, at time and length scales orders of magnitude larger than previously observed. The presence of these patterns only in cold landscapes suggests that the exceptionally large amount of heterogeneity found in icy landscapes may allow for the development of sub-critical instabilities in non-inertial flows. 

Our findings point toward the need to address key knowledge gaps that impact our ability to understand landscape dynamics through a soft matter lens. First, we lack adequate rheological models that can account for (i)) the non-inertial regime \cite{deshpandePerpetualFragilityCreeping2021, Houssais2021}, (ii) heterogeneity in grain size and material properties \cite{fazelpour2022effect},  (iii) cohesion between grains \cite{mandal2020insights}, and (iv) the presence of liquid water and ice \cite{harris2003gelifluction}. While our highly simplified treatment of soil creep as a viscous fluid works surprisingly well to explain average pattern wavelengths, without a more accurate representation of soil rheology we are severely limited in our predictive capabilities. Second, field observations of soil transport processes are difficult to obtain because they operate over such long timescales, though recent advances show promise for obtaining high-resolution surface deformation of slow-moving solifluction lobes \cite{harkema2023monitoring}. This illustrates the necessity for a collaborative, holistic approach that incorporates theory, laboratory experiments, numerical modeling,  and field observations to better bridge grain-scale dynamics with landscape-scale processes and patterns.

\subsection{Bedform dynamics: interaction, attraction and repulsion of dunes - \textit{N. M. Vriend and K. A. Bacik}}
\label{sec:Vriend}
In desert landscapes, we observe individual sand dunes of different sizes, with a characteristic length scale of up to kilometers, which seamlessly interact with each other and their environment~\cite{Bagnold41, Lorenz14, Livingstone07}. 
As migrating sand dunes notoriously bury human-made infrastructure and lead to degradation of arable land, this interaction has important practical implications too~\cite{Dauxois2021, Berte10}. 
From a physical point of view, the evolution of a sedimentary surface is a result of an intricate coupling between the turbulent flow and the granular bed~\cite{Andreotti13,Kroy02}. 
Relevant scales of motion in this system span several orders of magnitude, from sediment transport, through dune migration, to large-scale organization of a dune field~\cite{Charru2013}. 

\begin{figure}[h]
 \centering
 \includegraphics[width= \linewidth]{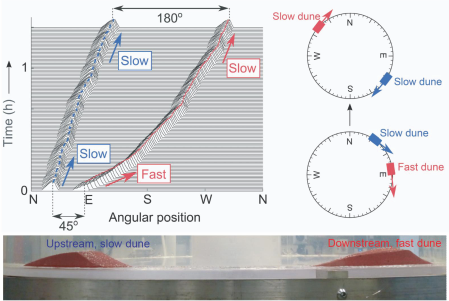}
 \caption{Two equal size miniature dunes, placed $45^{\circ}$ apart within the periodic annular channel, due to the drag imposed by the water current, start migrating, and over long times converge to a symmetric antipodal configuration~\cite{Bacik2020}.}
 \label{fig:VriendFig}
\end{figure}

During the workshop, we focused on the system-level dynamics, which in the field occur over decades and thus are difficult to investigate in detail. 
However, in the lab, by using appropriately scaled miniature subaqueous dunes, we can investigate the key physical processes in a matter of hours.
Specifically, we discussed three research questions, recently investigated within a new laboratory experiment (Fig.~\ref{fig:VriendFig}), which is uniquely suited for investigating long-term dynamics due to its circular quasi-2D geometry~\cite{Bacik2020}. 

First, we explored the pairwise interaction between two dunes, leading to either coalescence (merging), ejection (sediment exchange)~\cite{Jarvis2022a}, or wake-induced repulsion of bedforms~\cite{Bacik2020} and presented a phase-space diagram outlining the possible interaction outcomes derived from experiments and cellular automaton simulations~\cite{Jarvis2022b}. 
Second, as a first step towards understanding the system-level dynamics of a dune field, we studied the long-term behavior of a periodic two-dune system~\cite{Bacik2021a}: is it always attracted to a symmetric state with two identical equi-spaced dunes, or are there conditions where the symmetry is spontaneously broken? 
We found that the key to understanding the dynamics is “turbulence”: for flows with a relatively low turbulence intensity, the dunes will display fast-slow dynamics before settling at a stable equilibrium, but for high levels of turbulence an asymmetric attractor appears. 
This indicates that, at least in theory, the hydrodynamic coupling between neighboring dunes can either promote or inhibit regular organization of a dune field. 
Lastly, by placing idealized objects in the path of our model dunes, we directly addressed the engineering challenge of dune-obstacle interaction. 
We observed that both object size and shape of the obstacle determine whether the dune is blocked or able to overcome an obstacle and reform on the other side~\cite{Bacik2021b}, and once again, we discovered the importance of turbulent flow structures. 
Indeed, we showed that the outcome of the dune-obstacle interaction can be predicted with a simple data-driven tool based on the modal decomposition of the flow field around the obstacle (without sediment or dunes present).

In summary, we revealed surprising connections between the rapid small-scale processes (such as turbulent fluctuations interacting with a granular interface) and the slow large-scale evolution of sedimentary landscapes. 
Remarkably, by taking advantage of scaling laws, we have managed to investigate such processes in a controlled laboratory experiment, but we are hopeful that with the advance of remote sensing, one day we will be able to validate our predictions with observational data.

\section{The challenge(s) of life}
Life has an impact on the ground dynamics and is impacted by the physicochemical evolution of the ground. Over time, plants, fungi, and burrowing animals alter the ground composition, constitutive structure, and consequently, its mechanical properties (e.g., Fig.\ref{fig: IntroFig}e). Sometimes the presence of life makes the ground more cohesive and enhances its resistance to erosion, sometimes its own dynamics pull apart or alter the ground such that it is destabilized. Such puzzling observations illustrate why it is crucial to isolate and study the biophysical processes happening in the ground. We present here two examples of such phenomena: in section \ref{sec:Kudrolli}, recent results and outlooks on the dynamics and impact of invertebrates (such as worms) in the ground are presented, while section \ref{sec:StonesAndYiang} highlights a newly recognized effect bacteria can have on fluid dynamics in porous media.
Finally, section \ref{sec:DelGado} reminds us that human beings, via the building of our infrastructure, are also a major part of the life disturbance of the Earth's ground and atmosphere. It presents recent results to support a fundamentally different usage of the ground as a human resource.
\label{sec:(IV)}

\subsection{Intruder dynamics in granular sediments - \textit{A. Kudrolli}}
\label{sec:Kudrolli}
The ground beneath us is constantly shaped by animal and human activities that can further impact movement of fluids and erosion~\cite{Hole1981}. Exopedonic and endopedonic activities leading to creation of mounds, voids, and burrows in loose sediments, besides anthropogenic activities leading to desertification, and trawling for resources on the ocean floor are problems of great importance in ecosystem management. To overcome the opacity of granular matter, where much of these activities occur, X-ray imaging and index-matching techniques have been employed to understand locomotion strategies from undulatory to peristaltic motion in situ~\cite{Hosoi2015,Kudrolli2019}. Besides water jets and fluidization, various strategies from plastic grain rearrangements to sand fluidization and burrow extension by fracture have been uncovered depending on depth, compaction, and grain size~\cite{Dorgan2013}. Considerable work is underway to understand the observed locomotion speeds to the observed gaits employed based on the rheology of the medium and in developing models starting from resistive force theory and slender-body theory introduced in the context of viscously dominated fluid dynamics~\cite{Maladen2009,Hewitt2022}. Further work is required to extend rheological models developed under uniform flow and shear conditions to time-dependent and unsteady flows encountered in such dynamics.  

These considerations have motivated studies of the drag encountered by spherical and cylindrical solid intruders moving across a granular bed~\cite{Panaitescu2017,Jewel2018,Allen2019,Pal2021,Chang2022}. These studies have found that the non-dimensional Inertial Number and Viscous Number used to characterize the properties of granular matter and granular suspensions introduced under uniform shear rate conditions can be extended to unsteady cases by using an effective shear rate set by the length scale and speed of the intruder. Visualization of the flow of the granular medium have revealed that flow is more narrowly confined around the intruder with far greater slip at the solid-medium interface compared with a viscous fluid~\cite{Jewel2018,Chang2022}. Granular flow around an intruder was found to result in far greater drag anisotropy compared to viscous flow, which is important for drag-assisted propulsion, with still greater anisotropy while considering grains with negligible surface friction~\cite{Pal2021}. The experiments robustly support the increase of drag with overburden pressure in a granular bed and scaling of drag with projected cross-sectional area in the case of simple intruder shapes such as spheres and solids. However, wakes generated by more complex or multi-component shapes were found to lead to non-additive drag. In particular, two rods moving in tandem~\cite{Chang2022}, are observed to present a drag as a function of separation distance that is different compared to that in viscous fluids. While drag acting on the leading and following rod in viscous fluids at low Reynolds numbers are essentially the same, in a granular material the drag acting on the leading rod exceeds that acting on the following rod even in the quasi-static limit~\cite{Chang2022}. These studies both point to the complexity and nuanced nature of granular matter encountered while moving or digging in them. 

While the above discussion has focused on the limit where the intruder is strong enough to move the material, a complementary limit is where the intruder cannot move the material, but is restricted to moving within the pore spaces. Thigmotactic behavior, as in motion along the edges of surfaces due to sensory feedback and environmental cues, can play an important role in determining transport~\cite{Park2003}. The importance of body shape, stroke, and topology is receiving attention in determining the dynamics of bacteria as discussed in the following section by Stone and Yang. The second part of the presentation focused on the dynamics of centimeter-scale oligochaeta {\it Lumbriculus variegatus} in model porous media, where its natural strokes were hindered by the tight passages between idealized grains. A persistent random walk model along boundaries was found to capture the observed time-distributions to escape the dynamical traps posed by the pore-throats~\cite{Biswas2023}. Active polymer models where simple steric interactions are emphasized have a significant role to play in determining general principles of transport in this limit~\cite{Bechinger2016,Mokhtari2019,Kurzthaler2021}.   

\subsection{Bacteria-mediated processes in porous media - \textit{H. A. Stone and J. Q. Yang}}
\label{sec:StonesAndYiang}
The ground under our feet is home to a wide range of life that forms the basis for the agriculture on which we all depend for food, for the plants and trees that regulate atmospheric carbon dioxide and oxygen, and for an enormous range of organisms, from bacteria, algae, and fungi to ants, worms, etc. Microorganisms are central to many of the underground processes. In recent years, the fate of carbon stored in soil has become an important frontier research area: can soils act as a “negative emission” technology, serving as a reservoir for carbon released from fossil fuel combustion, or will carbon, possibly long-stored in soils, be released as the climate warms~\cite{melillo2017long}? Of course, there are many types of soils and environmental conditions~\cite{doetterlSoilCarbon}. Models used to project future climate have wide variability for the contributions of soil carbon to projections of atmospheric CO2, even differing in the sign of the effect. Motivated by these questions, in Stone’s talk he described two laboratory studies, led by Yang, of bacteria-mediated processes in porous media, one bearing on the soil carbon storage question and the other identifying a previously unrecognized transport process for bacteria in partially saturated porous media. These kinds of problems have natural links to the topics discussed in challenge~\ref{sec:(I)} on modeling fundamental processes beginning at the particle scale, since several of these themes probe dynamics and transport in porous systems, e.g., the contributions of Bourg (\ref{sec:Bourg}), Datta (\ref{sec:Datta}), and Li and Juanes (\ref{sec:Juanes}). In addition, the bacteria-mediated processes discussed in this summary naturally involve properties of clay, as discussed in challenge~\ref{sec:(IV)} by del Gado (\ref{sec:DelGado}) and dynamics of particles in porous environments, which has links to the contribution of Kudrolli (\ref{sec:Kudrolli}). 

The first half of the presentation at the workshop summarized a microfluidic experimental study, which incorporated important elements of the soil carbon problem, including clay aggregates, different molecular weight carbon molecules, bacteria, and enzymes, embedded in flows of water~\cite{YangZhangBourgStone}. Confocal microscopy was used to document the space and time dependence of how molecules adsorb onto and diffuse into and out of transparent clay aggregates. Smaller molecular weight molecules displayed reversible transport while the larger molecular weight molecules were quasi-irreversibly adsorbed, i.e., diffusion into the aggregates occurred rapdily (within minutes), but most molecules remained adsorbed even after flushing for tens of hours with water. Bacteria were too large to fit into the nano-size pore space of the aggregates, so accumulated on the outside of clay aggregates, whereas enzymes were shown to effectively penetrate the small pores where they broke down and released the trapped large molecular weight sugars. A brief summary was given of typical models of soil carbon storage, and the experimental results were used to suggest improvements to the models.

In the second half of the presentation, we documented a previously unreported mechanism of transport of bacterial cells in unsaturated porous media. In particular, experiments were presented showing that surfactant-producing bacteria cause changes in the wettability (to a hydrophilic state) of an initial hydrophobic substrate, which, through a capillary pressure change, causes millimeter per hour fluid flow, comparable to other rapid bacterial swimming speeds, along corners of a model chamber~\cite{YangSanfilippoAbbasiGitaiBasslerStone}. Similar experimental observations of bacterial transport were then demonstrated in a porous medium of packed angular grains, which served as a model soil. The dynamics are controlled by quorum sensing, which regulates biosurfactant production. This transport process can also lead to movement of non-motile bacteria in the solution. The results suggest that this kind of surfactant-driven transport through changes in wettability, instead of the better-known Marangoni motion, may be relevant to natural porous environments. 

Subsurface life and transport processes are poorly understood, in part because they are difficult to visualize and monitor in space and time. They are important problems that will benefit from designing lab-scale experiments and can help provide insights about soil and surface processes relevant to agriculture, sustainability, water, energy, and Earth surface dynamics.  Microfluidic tools~\cite{StanleyGrossmannSolvasDeMello,AleklettKiersOhlssonShimizuCaldasHammer} and new experimental approaches can help to shed light on these questions. 
\break
\subsection{Nanoscale forces in hydrated clays and the physics of sustainable construction materials - \textit{E. Del Gado}}
\label{sec:DelGado}
\begin{figure}[h]
 \centering
  \includegraphics[width=\linewidth]{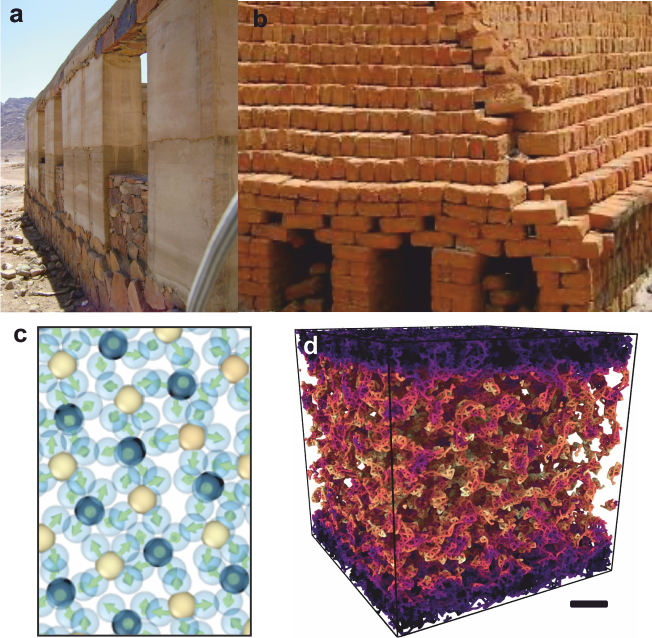}
 \caption{Earth construction materials, examples, and scientific questions: (a) rammed earth constructions, (b) rural brick kiln, (c) water-ion structures from semiatomistic simulations of Ca ions solutions confined between charged surfaces \cite{Dragulet2022ion}, and (d) mesoscale porous structures of cement hydrate gels from coarse-grained simulations, where particle sizes are on the order of 10-50~nm (see bar) \cite{Goyal2020}.}
 \label{fig:DelgadoFig}
\end{figure}

The ground beneath our feet is also a unique, and sustainable, source of construction materials (Fig.~\ref{fig:DelgadoFig}a,b). Clay soils and other Earth materials have been used for construction over the centuries \cite{BatirEnTerre,Guillaud2009}. Examples of Earth-based architecture, from the most modest to the most monumental ones, are available on all continents and in all climates. $15~\%$ of the architectural sites recognized as part of the UNESCO world heritage are entirely or partially built with soils and sediments, demonstrating the durability of these materials and construction techniques. Even in this century, around $50~\%$ of the world population lives in buildings made of raw soils and completely natural clays. Construction materials alone make up a sizeable portion of the GHG emissions of the entire construction sector, since, independently of building operations, they are currently responsible for close to $11~\%$ of the world's global $CO_2$ emissions \cite{globalstatus2017}. Most of the carbon footprint is in cement production for concrete, with the latter being the most used synthetic material on Earth, due to its centrality to construction technologies and the built infrastructure. 

Among immediately implementable strategies that would allow for substantial decarbonization of the cement industry, greener cement mixtures based on reduced cement content and partial substitution with clays and natural soils are probably the most interesting and valuable option \cite{Habert2020environmental, Gangotra2023}. Hence, the soil beneath our feet and the world's oldest construction material is also probably the most ecologically responsible and a potentially novel source of more sustainable construction technologies.   
Clay sensitivity to salinity, pH, moisture, and stresses, which is central to cohesive strength, stability of soil and building foundations, originates from the nanoscale physical chemistry and ionic composition of clay layers, from which larger-scale structures with complex pore networks and load bearing properties grow \cite{mud-2020,Bourg2017}. 
In hydrated clays, nanoscale surface forces develop from the accumulation and confinement of ions in solution between charged surfaces, a phenomenon which also controls the cohesion of hydrated cement, and it is well-known in soft matter, ranging from colloidal materials to biological systems \cite{israelachvili2015intermolecular}. 
However, the cohesive forces that develop during hydration of cement and clays are strongly ion-specific and dramatically depend on confinement and humidity conditions. Hence, they cannot be properly captured by existing mean-field theoretical descriptions used in other cases for surface forces in ionic solutions, raising a number of outstanding fundamental questions on the nanoscale physics of confined ion and water \cite{Goyal2021physics,Palaia2022likecharge,Dragulet2022ion}. 
Increasing confinement and surface charge densities promote ion-water structures, distinct from bulk ion hydration shells, that become strongly anisotropic, persistent, and self-organized into optimized nearly solid-like assemblies. 
Under these conditions, the dramatically reduced dielectric screening of water and the highly organized water-ion structures (Fig.~\ref{fig:DelgadoFig}c) lead to strongly attractive interactions between charged surfaces. 

Molecular simulations effectively fill the gap with the experimental characterization of cohesive forces in hydrated clays and cement, providing novel insight into the strong ionic correlations that govern them, to be used in continuum theories and larger scale studies \cite{Goyal2021physics,Palaia2022likecharge,Dragulet2022ion}. 
The nanoscale forces, in fact, together with the non-equilibrium environmental conditions, eventually determine the growth of microscale grain assemblies, layered meso-phases, and porous structures that can be obtained through coarse-grained simulations (Fig.~\ref{fig:DelgadoFig}d) and govern the rheology and mechanics of clay-based soil and construction materials, providing the missing link from the nanoscale physical chemistry to the meso-scale aggregation kinetics and morphological variability of soil and clay-based binders \cite{ioannidou2016crucial,Goyal2020,ioannidou2016mesoscale}. Understanding how mechanical and rheological behaviors emerge in porous clay matrices that gel and solidify starting from nanoscale forces, which are chemically specific and sensitive to non-equilibrium conditions and environmental reactivity, is an area rich with challenging scientific questions, where soft matter scientists are poised to contribute with key insights.

\section*{Discussion}
In each contribution associated with the four challenges identified above, a unique soft matter and geoscience research topic was addressed. The outstanding open questions raised by each author show commonalities that relate to the multi-dimensional, multi-scale, multi-phase, and multi-process character of the ground beneath our feet. Some of the most crucial shared questions are briefly summarized and discussed here. 

The opaque and time-dependent nature of the ground profoundly hinders our capacity to observe and understand it. In the laboratory, advancements in microscopic visualization and experimental techniques are, on the one hand, key to remediating this difficulty (see sections \ref{sec:Datta}, \ref{sec:StonesAndYiang}); yet illuminating all spatial dimensions, while allowing temporal dependencies, is a technical frontier. On the other hand, at the cost of simplifications, properties, stress and flow fields in 2D-3D granular porous media can be visualized and quantified (see sections \ref{sec:Juanes}, \ref{sec:RubinAndFerdowsi}, and \ref{sec:Daniels}). In the field, the time-dependent and temporally variable nature of surface and subsurface processes requires additional efforts for monitoring (see sections \ref{sec:Masteller}, \ref{sec:Hewitt}, and \ref{sec:Glade}). Extension and advancements in the use of geophysical techniques, such as seismology, hold promise in enabling tracking of the temporal evolution of e.g. transport processes \cite{Cook2022, Roth2017}. Simultaneously, temporally resolved remote sensing techniques are making details of large-scale phenomena, like ice shelf dynamics, observable (see sections \ref{sec:Burton}, and \ref{sec:Lai}).

Though the ground is becoming more observable, seemingly basic questions of when grains start moving, where particles jam, or where and when landslides rupture, are still wide open. Especially in natural systems, identifying the essential conditions (see sections \ref{sec:Bourg}, and \ref{sec:Wettlaufer}), material properties (see sections \ref{sec:Daniels}, \ref{sec:Suckale}, and \ref{sec:DelGado}), and processes (see sections \ref{sec:RubinAndFerdowsi}, \ref{sec:Burton}, and \ref{sec:Masteller}) to understand and model the bulk dynamics remains a research frontier. Beyond the relevant components, modeling approaches still struggle to represent key relations (e.g., how microscale processes change the bulk observations, or how to integrate over local conditions or time to obtain bulk mechanical behavior), feedbacks (how macrostructures affect the microscale conditions, or co-dependent processes and scales \cite{Houssais2021}), and temporal evolution. 
While computational advancements enable the modeling of complex systems, coupled dynamics require further efforts in developing the proper way to implement relationships governing elementary phenomena (see sections \ref{sec:Wettlaufer}, \ref{sec:Devauchelle}, \ref{sec:Goehring}, and \ref{sec:DelGado}), including life-related ones (see sections \ref{sec:Datta}, \ref{sec:Kudrolli}, and \ref{sec:StonesAndYiang}). 

Fundamental open questions, for many of the topics addressed here, emerge from the complex interplay of several phases. The dynamics of, for example, three-phase systems where interfaces of liquid-air-grains evolve over time, are key to a better understanding how natural rafts, such as ice mélanges, behave (see section \ref{sec:Burton}), how landslides start and stop (see section \ref{sec:JerolmackAndDeshpante}), and how groundwater evaporates into salt crusts (see section \ref{sec:Goehring}). Conceptually taming the interactions and feedbacks to understand more than two-phase systems challenges concepts, experiments, and observation throughout the Earth sciences (e.g., \cite{Lovejoy2008, Sadler2015, Hebert2022, Franzke2020, Houssais2021}).

Understanding and modeling contact lines, capillary stresses, and resulting cohesion in particulate materials is a vast and ongoing effort for the case of a single-phase liquid in contact with another fluid and solid particles \cite{Koos2011, Ha2020capillarity}. While, more often than not, flows in the ground are themselves suspensions of minerals (see sections \ref{sec:Datta}, and \ref{sec:Bourg}), or biological (active) particles (see sections \ref{sec:StonesAndYiang}, and \ref{sec:Kudrolli}); the solid part of the ground itself can be composed of vastly different materials in terms of wettability properties (see sections \ref{sec:Bourg}, and \ref{sec:StonesAndYiang}), density (see sections \ref{sec:Wettlaufer}, and \ref{sec:Suckale}), and rigidity (see section \ref{sec:Daniels}). 
Especially where life is involved and shapes its surroundings (see sections \ref{sec:Datta}, \ref{sec:Kudrolli}, and \ref{sec:StonesAndYiang}), or where landforms and patterns of different scales interact (see sections \ref{sec:Suckale}, \ref{sec:Goehring}, \ref{sec:Glade}, and \ref{sec:Vriend}), modeling the ground accurately requires solving among the most difficult questions of multi-phases systems.

Another recurrent question for many of the addressed topics is how to model transience in materials state and behavior. The ground is heterogeneous in composition and properties, therefore it presents diverse physicochemical behaviors and properties, such as rigidity (see section \ref{sec:Daniels}), plasticity (see section \ref{sec:Bourg}), or viscosity (see section \ref{sec:Suckale}). As the ground evolves over time under external physicochemical forcings, it also undergoes change in the material bulk dynamics. Such transients may result in instability growth and produce characteristic patterns. Direct observables of transience are often only the resulting expressions of dynamics hidden in the ground, like ordered patterns in salt lakes (see section \ref{sec:Goehring}), arctic soils (see section \ref{sec:Glade}), ice shelves (see sections \ref{sec:Lai}, and \ref{sec:Burton}), or glacial deposits (see section \ref{sec:Hewitt}). Linking these dynamics to climatic changes, geodynamics, as well as anthropogenic and biogenic activities, poses further outstanding open questions, including topics related to subsurface storage and transport of gases and fluids (see sections \ref{sec:Devauchelle}, and \ref{sec:Suckale}), or the anticipation of ruptures that can turn into devastating natural hazards (see sections \ref{sec:RubinAndFerdowsi}, and \ref{sec:JerolmackAndDeshpante}). 

\section*{Conclusion}
Most Earth surface materials can be categorized as Soft Materials. Studying their dynamics necessitates a diversity of approaches, thus understanding the ground is intrinsically a highly interdisciplinary field. Following a PCTS workshop, we addressed here a broad outline of the “physics of the ground beneath our feet”, considering the ground within the broad category of Soft Matter and all the variety of challenges one faces when trying to model its behavior. We identified four major challenges: (I) modeling from the grain scale, (II) near-criticality, (III) bridging scales, and (IV) life, which structure this paper (see Figure \ref{fig: IntroFig}). Within each section, contributions by the workshop participants presented examples of works tackling each challenge and open questions related to their specific topic. We hope for this collective effort to provide a new broad perspective on the field and invite more Soft Matter scientists to study the fascinating ground we live and build our future on.  

\section*{Author Contributions}
All authors were involved in the conceptualization of this review article. The writing was administered by AV and MH, including drafting the opening and closing sections. All authors were involved in both writing the original draft of their individual sections and reviewing and editing the entire article.



\section*{Conflicts of interest}
There are no conflicts to declare.

\section*{Acknowledgements}
AV and MH are indebted to Abhinindra Singh, for fruitful discussions on the early design of the manuscript.

SSD, ICB, C-Y Lai, and HAS thank the Princeton Center for Theoretical Science at Princeton University for their support of the 2022 workshop that stimulated the writing of this paper. In particular, we are grateful for the help provided by Charlene Borsack, both leading up to and during the workshop.

HAS and JQY thank the High Meadows Environmental Institute (Princeton University) and the NSF (grant MCB-1853602). These projects were led by Judy Yang (first a postdoc at Princeton and now a Professor at the University of Minnesota) and the wonderful collaborations included Niki Abbasi, Bonnie Bassler, Ian Bourg, Zemer Gitai, Joe Sanfilippo, and Xinning Zhang. 

SSD acknowledges funding from the High Meadows Environmental Institute and Andlinger Center for Energy and Environment (Princeton University), NSF Grant CBET-1941716,  a Camille Dreyfus Teacher-Scholar Award of the Camille and Henry Dreyfus Foundation, and the Princeton Center for Complex Materials, a Materials Research Science and Engineering Center supported by NSF grants DMR-1420541 and DMR-2011750.

OD is indebted to E.~Lajeunesse, V.~Jules, A.~Gu{\'e}rin and F.~M{\'e}tivier for sustained and enjoyable collaboration on groundwater flows.

ICB thanks NSF Grant EAR-2150797 and DOE Grant DE-SC0018419 for financial support.



\balance


\bibliography{PhysOfTheGround} 

\providecommand*{\mcitethebibliography}{\thebibliography}
\csname @ifundefined\endcsname{endmcitethebibliography}
{\let\endmcitethebibliography\endthebibliography}{}
\begin{mcitethebibliography}{418}
\providecommand*{\natexlab}[1]{#1}
\providecommand*{\mciteSetBstSublistMode}[1]{}
\providecommand*{\mciteSetBstMaxWidthForm}[2]{}
\providecommand*{\mciteBstWouldAddEndPuncttrue}
  {\def\EndOfBibitem{\unskip.}}
\providecommand*{\mciteBstWouldAddEndPunctfalse}
  {\let\EndOfBibitem\relax}
\providecommand*{\mciteSetBstMidEndSepPunct}[3]{}
\providecommand*{\mciteSetBstSublistLabelBeginEnd}[3]{}
\providecommand*{\EndOfBibitem}{}
\mciteSetBstSublistMode{f}
\mciteSetBstMaxWidthForm{subitem}
{(\emph{\alph{mcitesubitemcount}})}
\mciteSetBstSublistLabelBeginEnd{\mcitemaxwidthsubitemform\space}
{\relax}{\relax}

\bibitem[Dauxois \emph{et~al.}(2021)Dauxois, Peacock, Bauer, Caulfield,
  Cenedese, Gorl{\'e}, Haller, Ivey, Linden,
  Meiburg,\emph{et~al.}]{Dauxois2021}
T.~Dauxois, T.~Peacock, P.~Bauer, C.-C.~P. Caulfield, C.~Cenedese,
  C.~Gorl{\'e}, G.~Haller, G.~N. Ivey, P.~F. Linden, E.~Meiburg \emph{et~al.},
  \emph{Physical Review Fluids}, 2021, \textbf{6}, 020501\relax
\mciteBstWouldAddEndPuncttrue
\mciteSetBstMidEndSepPunct{\mcitedefaultmidpunct}
{\mcitedefaultendpunct}{\mcitedefaultseppunct}\relax
\EndOfBibitem
\bibitem[Sharp(1982)]{Sharp1982}
R.~P. Sharp, \emph{Proceedings of the National Academy of Sciences}, 1982,
  \textbf{79}, 4477--4486\relax
\mciteBstWouldAddEndPuncttrue
\mciteSetBstMidEndSepPunct{\mcitedefaultmidpunct}
{\mcitedefaultendpunct}{\mcitedefaultseppunct}\relax
\EndOfBibitem
\bibitem[Dietrich \emph{et~al.}(2003)Dietrich, Bellugi, Sklar, Stock, Heimsath,
  and Roering]{Dietrich2003}
W.~E. Dietrich, D.~G. Bellugi, L.~S. Sklar, J.~D. Stock, A.~M. Heimsath and
  J.~J. Roering, \emph{Prediction in geomorphology}, 2003,  103--132\relax
\mciteBstWouldAddEndPuncttrue
\mciteSetBstMidEndSepPunct{\mcitedefaultmidpunct}
{\mcitedefaultendpunct}{\mcitedefaultseppunct}\relax
\EndOfBibitem
\bibitem[Houssais and Jerolmack(2017)]{Houssais2017}
M.~Houssais and D.~J. Jerolmack, \emph{Geomorphology}, 2017, \textbf{277},
  251--264\relax
\mciteBstWouldAddEndPuncttrue
\mciteSetBstMidEndSepPunct{\mcitedefaultmidpunct}
{\mcitedefaultendpunct}{\mcitedefaultseppunct}\relax
\EndOfBibitem
\bibitem[Jerolmack and Daniels(2019)]{JerolmackDaniels2019}
D.~J. Jerolmack and K.~E. Daniels, \emph{Nature Reviews Physics}, 2019,
  \textbf{1}, 716--730\relax
\mciteBstWouldAddEndPuncttrue
\mciteSetBstMidEndSepPunct{\mcitedefaultmidpunct}
{\mcitedefaultendpunct}{\mcitedefaultseppunct}\relax
\EndOfBibitem
\bibitem[Jansson and Hofmockel(2020)]{Jansson2020}
J.~K. Jansson and K.~S. Hofmockel, \emph{Nature Reviews Microbiology}, 2020,
  \textbf{18}, 35--46\relax
\mciteBstWouldAddEndPuncttrue
\mciteSetBstMidEndSepPunct{\mcitedefaultmidpunct}
{\mcitedefaultendpunct}{\mcitedefaultseppunct}\relax
\EndOfBibitem
\bibitem[Lai \emph{et~al.}(2020)Lai, Kingslake, Wearing, Chen, Gentine, Li,
  Spergel, and van Wessem]{Lai2020vulnerability}
C.-Y. Lai, J.~Kingslake, M.~G. Wearing, P.-H.~C. Chen, P.~Gentine, H.~Li, J.~J.
  Spergel and J.~M. van Wessem, \emph{Nature}, 2020, \textbf{584},
  574--578\relax
\mciteBstWouldAddEndPuncttrue
\mciteSetBstMidEndSepPunct{\mcitedefaultmidpunct}
{\mcitedefaultendpunct}{\mcitedefaultseppunct}\relax
\EndOfBibitem
\bibitem[Singh \emph{et~al.}(2022)Singh, Jackson, van~der Naald, de~Pablo, and
  Jaeger]{Singh2022}
A.~Singh, G.~L. Jackson, M.~van~der Naald, J.~J. de~Pablo and H.~M. Jaeger,
  \emph{Physical Review Fluids}, 2022, \textbf{7}, 054302\relax
\mciteBstWouldAddEndPuncttrue
\mciteSetBstMidEndSepPunct{\mcitedefaultmidpunct}
{\mcitedefaultendpunct}{\mcitedefaultseppunct}\relax
\EndOfBibitem
\bibitem[Schneider \emph{et~al.}(2021)Schneider, Priestley, and
  Datta]{schneider2021using}
J.~Schneider, R.~D. Priestley and S.~S. Datta, \emph{Physical Review Fluids},
  2021, \textbf{6}, 014001\relax
\mciteBstWouldAddEndPuncttrue
\mciteSetBstMidEndSepPunct{\mcitedefaultmidpunct}
{\mcitedefaultendpunct}{\mcitedefaultseppunct}\relax
\EndOfBibitem
\bibitem[Cho \emph{et~al.}(2019)Cho, Lu, Howard, Adams, and
  Datta]{cho2019crack}
H.~J. Cho, N.~B. Lu, M.~P. Howard, R.~A. Adams and S.~S. Datta, \emph{Soft
  Matter}, 2019, \textbf{15}, 4689--4702\relax
\mciteBstWouldAddEndPuncttrue
\mciteSetBstMidEndSepPunct{\mcitedefaultmidpunct}
{\mcitedefaultendpunct}{\mcitedefaultseppunct}\relax
\EndOfBibitem
\bibitem[Bhattacharjee \emph{et~al.}(2022)Bhattacharjee, Amchin, Alert, Ott,
  and Datta]{bhattacharjee2022chemotactic}
T.~Bhattacharjee, D.~B. Amchin, R.~Alert, J.~A. Ott and S.~S. Datta,
  \emph{Elife}, 2022, \textbf{11}, e71226\relax
\mciteBstWouldAddEndPuncttrue
\mciteSetBstMidEndSepPunct{\mcitedefaultmidpunct}
{\mcitedefaultendpunct}{\mcitedefaultseppunct}\relax
\EndOfBibitem
\bibitem[Bhattacharjee \emph{et~al.}(2021)Bhattacharjee, Amchin, Ott, Kratz,
  and Datta]{bhattacharjee2021chemotactic}
T.~Bhattacharjee, D.~B. Amchin, J.~A. Ott, F.~Kratz and S.~S. Datta,
  \emph{Biophysical Journal}, 2021, \textbf{120}, 3483--3497\relax
\mciteBstWouldAddEndPuncttrue
\mciteSetBstMidEndSepPunct{\mcitedefaultmidpunct}
{\mcitedefaultendpunct}{\mcitedefaultseppunct}\relax
\EndOfBibitem
\bibitem[Anbari \emph{et~al.}(2018)Anbari, Chien, Datta, Deng, Weitz, and
  Fan]{anbari2018microfluidic}
A.~Anbari, H.-T. Chien, S.~S. Datta, W.~Deng, D.~A. Weitz and J.~Fan,
  \emph{Small}, 2018, \textbf{14}, 1703575\relax
\mciteBstWouldAddEndPuncttrue
\mciteSetBstMidEndSepPunct{\mcitedefaultmidpunct}
{\mcitedefaultendpunct}{\mcitedefaultseppunct}\relax
\EndOfBibitem
\bibitem[Larson \emph{et~al.}(1990)Larson, Shaqfeh, and Muller]{larson1990}
R.~G. Larson, E.~S. Shaqfeh and S.~J. Muller, \emph{Journal of Fluid
  Mechanics}, 1990, \textbf{218}, 573--600\relax
\mciteBstWouldAddEndPuncttrue
\mciteSetBstMidEndSepPunct{\mcitedefaultmidpunct}
{\mcitedefaultendpunct}{\mcitedefaultseppunct}\relax
\EndOfBibitem
\bibitem[Shaqfeh(1996)]{shaqfeh1996}
E.~S. Shaqfeh, \emph{Annual Review of Fluid Mechanics}, 1996, \textbf{28},
  129--185\relax
\mciteBstWouldAddEndPuncttrue
\mciteSetBstMidEndSepPunct{\mcitedefaultmidpunct}
{\mcitedefaultendpunct}{\mcitedefaultseppunct}\relax
\EndOfBibitem
\bibitem[McKinley \emph{et~al.}(1996)McKinley, Pakdel, and
  {\"O}ztekin]{mckinley1996}
G.~H. McKinley, P.~Pakdel and A.~{\"O}ztekin, \emph{Journal of Non-Newtonian
  Fluid Mechanics}, 1996, \textbf{67}, 19--47\relax
\mciteBstWouldAddEndPuncttrue
\mciteSetBstMidEndSepPunct{\mcitedefaultmidpunct}
{\mcitedefaultendpunct}{\mcitedefaultseppunct}\relax
\EndOfBibitem
\bibitem[Pakdel and McKinley(1996)]{pakdel1996}
P.~Pakdel and G.~H. McKinley, \emph{{Physical Review Letters}}, 1996,
  \textbf{77}, 2459\relax
\mciteBstWouldAddEndPuncttrue
\mciteSetBstMidEndSepPunct{\mcitedefaultmidpunct}
{\mcitedefaultendpunct}{\mcitedefaultseppunct}\relax
\EndOfBibitem
\bibitem[Burghelea \emph{et~al.}(2004)Burghelea, Segre, Bar-Joseph, Groisman,
  and Steinberg]{burghelea2004chaotic}
T.~Burghelea, E.~Segre, I.~Bar-Joseph, A.~Groisman and V.~Steinberg,
  \emph{Physical Review E}, 2004, \textbf{69}, 066305\relax
\mciteBstWouldAddEndPuncttrue
\mciteSetBstMidEndSepPunct{\mcitedefaultmidpunct}
{\mcitedefaultendpunct}{\mcitedefaultseppunct}\relax
\EndOfBibitem
\bibitem[Rodd \emph{et~al.}(2007)Rodd, Cooper-White, Boger, and
  McKinley]{rodd2007}
L.~Rodd, J.~Cooper-White, D.~Boger and G.~H. McKinley, \emph{Journal of
  Non-Newtonian Fluid Mechanics}, 2007, \textbf{143}, 170--191\relax
\mciteBstWouldAddEndPuncttrue
\mciteSetBstMidEndSepPunct{\mcitedefaultmidpunct}
{\mcitedefaultendpunct}{\mcitedefaultseppunct}\relax
\EndOfBibitem
\bibitem[Afonso \emph{et~al.}(2010)Afonso, Alves, and Pinho]{afonso2010purely}
A.~Afonso, M.~Alves and F.~Pinho, \emph{Journal of Non-Newtonian Fluid
  Mechanics}, 2010, \textbf{165}, 743--751\relax
\mciteBstWouldAddEndPuncttrue
\mciteSetBstMidEndSepPunct{\mcitedefaultmidpunct}
{\mcitedefaultendpunct}{\mcitedefaultseppunct}\relax
\EndOfBibitem
\bibitem[Zilz \emph{et~al.}(2012)Zilz, Poole, Alves, Bartolo, Levach{\'e}, and
  Lindner]{zilz2012}
J.~Zilz, R.~Poole, M.~Alves, D.~Bartolo, B.~Levach{\'e} and A.~Lindner,
  \emph{Journal of Fluid Mechanics}, 2012, \textbf{712}, 203--218\relax
\mciteBstWouldAddEndPuncttrue
\mciteSetBstMidEndSepPunct{\mcitedefaultmidpunct}
{\mcitedefaultendpunct}{\mcitedefaultseppunct}\relax
\EndOfBibitem
\bibitem[Galindo-Rosales \emph{et~al.}(2012)Galindo-Rosales, Campo-Dea{\~n}o,
  Pinho, Van~Bokhorst, Hamersma, Oliveira, and Alves]{galindo2012}
F.~J. Galindo-Rosales, L.~Campo-Dea{\~n}o, F.~Pinho, E.~Van~Bokhorst,
  P.~Hamersma, M.~S. Oliveira and M.~Alves, \emph{Microfluidics and
  Nanofluidics}, 2012, \textbf{12}, 485--498\relax
\mciteBstWouldAddEndPuncttrue
\mciteSetBstMidEndSepPunct{\mcitedefaultmidpunct}
{\mcitedefaultendpunct}{\mcitedefaultseppunct}\relax
\EndOfBibitem
\bibitem[Ribeiro \emph{et~al.}(2014)Ribeiro, Coelho, Pinho, and
  Alves]{ribeiro2014}
V.~Ribeiro, P.~Coelho, F.~Pinho and M.~Alves, \emph{Chemical Engineering
  Science}, 2014, \textbf{111}, 364--380\relax
\mciteBstWouldAddEndPuncttrue
\mciteSetBstMidEndSepPunct{\mcitedefaultmidpunct}
{\mcitedefaultendpunct}{\mcitedefaultseppunct}\relax
\EndOfBibitem
\bibitem[Clarke \emph{et~al.}(2016)Clarke, Howe, Mitchell, Staniland,
  Hawkes,\emph{et~al.}]{clarke2016}
A.~Clarke, A.~M. Howe, J.~Mitchell, J.~Staniland, L.~A. Hawkes \emph{et~al.},
  \emph{SPE Journal}, 2016, \textbf{21}, 675--687\relax
\mciteBstWouldAddEndPuncttrue
\mciteSetBstMidEndSepPunct{\mcitedefaultmidpunct}
{\mcitedefaultendpunct}{\mcitedefaultseppunct}\relax
\EndOfBibitem
\bibitem[Machado \emph{et~al.}(2016)Machado, Bodiguel, Beaumont, Clisson, and
  Colin]{machado2016extra}
A.~Machado, H.~Bodiguel, J.~Beaumont, G.~Clisson and A.~Colin,
  \emph{Biomicrofluidics}, 2016, \textbf{10}, 043507\relax
\mciteBstWouldAddEndPuncttrue
\mciteSetBstMidEndSepPunct{\mcitedefaultmidpunct}
{\mcitedefaultendpunct}{\mcitedefaultseppunct}\relax
\EndOfBibitem
\bibitem[Kawale \emph{et~al.}(2017)Kawale, Marques, Zitha, Kreutzer, Rossen,
  and Boukany]{kawale2017a}
D.~Kawale, E.~Marques, P.~L. Zitha, M.~T. Kreutzer, W.~R. Rossen and P.~E.
  Boukany, \emph{Soft Matter}, 2017, \textbf{13}, 765--775\relax
\mciteBstWouldAddEndPuncttrue
\mciteSetBstMidEndSepPunct{\mcitedefaultmidpunct}
{\mcitedefaultendpunct}{\mcitedefaultseppunct}\relax
\EndOfBibitem
\bibitem[Qin \emph{et~al.}(2019)Qin, Salipante, Hudson, and
  Arratia]{qin2019flow}
B.~Qin, P.~F. Salipante, S.~D. Hudson and P.~E. Arratia, \emph{{Physical Review
  Letters}}, 2019, \textbf{123}, 194501\relax
\mciteBstWouldAddEndPuncttrue
\mciteSetBstMidEndSepPunct{\mcitedefaultmidpunct}
{\mcitedefaultendpunct}{\mcitedefaultseppunct}\relax
\EndOfBibitem
\bibitem[Sousa \emph{et~al.}(2018)Sousa, Pinho, and Alves]{sousa2018purely}
P.~Sousa, F.~Pinho and M.~Alves, \emph{Soft Matter}, 2018, \textbf{14},
  1344--1354\relax
\mciteBstWouldAddEndPuncttrue
\mciteSetBstMidEndSepPunct{\mcitedefaultmidpunct}
{\mcitedefaultendpunct}{\mcitedefaultseppunct}\relax
\EndOfBibitem
\bibitem[Browne \emph{et~al.}(2020)Browne, Shih, and Datta]{browne2020pore}
C.~A. Browne, A.~Shih and S.~S. Datta, \emph{Small}, 2020, \textbf{16},
  1903944\relax
\mciteBstWouldAddEndPuncttrue
\mciteSetBstMidEndSepPunct{\mcitedefaultmidpunct}
{\mcitedefaultendpunct}{\mcitedefaultseppunct}\relax
\EndOfBibitem
\bibitem[Browne \emph{et~al.}(2020)Browne, Shih, and
  Datta]{browne2020bistability}
C.~A. Browne, A.~Shih and S.~S. Datta, \emph{Journal of Fluid Mechanics}, 2020,
  \textbf{890}, A2\relax
\mciteBstWouldAddEndPuncttrue
\mciteSetBstMidEndSepPunct{\mcitedefaultmidpunct}
{\mcitedefaultendpunct}{\mcitedefaultseppunct}\relax
\EndOfBibitem
\bibitem[Walkama \emph{et~al.}(2020)Walkama, Waisbord, and
  Guasto]{walkama2020disorder}
D.~M. Walkama, N.~Waisbord and J.~S. Guasto, \emph{{Physical Review Letters}},
  2020, \textbf{124}, 164501\relax
\mciteBstWouldAddEndPuncttrue
\mciteSetBstMidEndSepPunct{\mcitedefaultmidpunct}
{\mcitedefaultendpunct}{\mcitedefaultseppunct}\relax
\EndOfBibitem
\bibitem[Haward \emph{et~al.}(2021)Haward, Hopkins, and
  Shen]{haward2021stagnation}
S.~J. Haward, C.~C. Hopkins and A.~Q. Shen, \emph{arXiv preprint
  arXiv:2105.11063}, 2021\relax
\mciteBstWouldAddEndPuncttrue
\mciteSetBstMidEndSepPunct{\mcitedefaultmidpunct}
{\mcitedefaultendpunct}{\mcitedefaultseppunct}\relax
\EndOfBibitem
\bibitem[Datta \emph{et~al.}(2022)Datta, Ardekani, Arratia, Beris,
  Bischofberger, McKinley, Eggers, L{\'o}pez-Aguilar, Fielding,
  Frishman,\emph{et~al.}]{datta2022perspectives}
S.~S. Datta, A.~M. Ardekani, P.~E. Arratia, A.~N. Beris, I.~Bischofberger,
  G.~H. McKinley, J.~G. Eggers, J.~E. L{\'o}pez-Aguilar, S.~M. Fielding,
  A.~Frishman \emph{et~al.}, \emph{Physical Review Fluids}, 2022, \textbf{7},
  080701\relax
\mciteBstWouldAddEndPuncttrue
\mciteSetBstMidEndSepPunct{\mcitedefaultmidpunct}
{\mcitedefaultendpunct}{\mcitedefaultseppunct}\relax
\EndOfBibitem
\bibitem[Browne and Datta(2021)]{Browne2021}
C.~A. Browne and S.~S. Datta, \emph{Science Advances}, 2021, \textbf{7},
  eabj2619\relax
\mciteBstWouldAddEndPuncttrue
\mciteSetBstMidEndSepPunct{\mcitedefaultmidpunct}
{\mcitedefaultendpunct}{\mcitedefaultseppunct}\relax
\EndOfBibitem
\bibitem[Marshall and Metzner(1967)]{marshall1967flow}
R.~Marshall and A.~Metzner, \emph{Industrial \& Engineering Chemistry
  Fundamentals}, 1967, \textbf{6}, 393--400\relax
\mciteBstWouldAddEndPuncttrue
\mciteSetBstMidEndSepPunct{\mcitedefaultmidpunct}
{\mcitedefaultendpunct}{\mcitedefaultseppunct}\relax
\EndOfBibitem
\bibitem[James and McLaren(1975)]{james1975laminar}
D.~F. James and D.~McLaren, \emph{Journal of Fluid Mechanics}, 1975,
  \textbf{70}, 733--752\relax
\mciteBstWouldAddEndPuncttrue
\mciteSetBstMidEndSepPunct{\mcitedefaultmidpunct}
{\mcitedefaultendpunct}{\mcitedefaultseppunct}\relax
\EndOfBibitem
\bibitem[Durst \emph{et~al.}(1981)Durst, Haas, and Kaczmar]{durst1981}
F.~Durst, R.~Haas and B.~Kaczmar, \emph{Journal of Applied Polymer Science},
  1981, \textbf{26}, 3125--3149\relax
\mciteBstWouldAddEndPuncttrue
\mciteSetBstMidEndSepPunct{\mcitedefaultmidpunct}
{\mcitedefaultendpunct}{\mcitedefaultseppunct}\relax
\EndOfBibitem
\bibitem[Durst and Haas(1981)]{dursthaas1981}
F.~Durst and R.~Haas, \emph{Rheologica Acta}, 1981, \textbf{20}, 179--192\relax
\mciteBstWouldAddEndPuncttrue
\mciteSetBstMidEndSepPunct{\mcitedefaultmidpunct}
{\mcitedefaultendpunct}{\mcitedefaultseppunct}\relax
\EndOfBibitem
\bibitem[Chauveteau and Moan(1981)]{ChauveteauMoan}
G.~Chauveteau and M.~Moan, \emph{Journal de Physique}, 1981, \textbf{42},
  L--201\relax
\mciteBstWouldAddEndPuncttrue
\mciteSetBstMidEndSepPunct{\mcitedefaultmidpunct}
{\mcitedefaultendpunct}{\mcitedefaultseppunct}\relax
\EndOfBibitem
\bibitem[Kauser \emph{et~al.}(1999)Kauser, Dos~Santos, Delgado, Muller, and
  Saez]{kauser}
N.~Kauser, L.~Dos~Santos, M.~Delgado, A.~Muller and A.~Saez, \emph{Journal of
  Applied Polymer Science}, 1999, \textbf{72}, 783--795\relax
\mciteBstWouldAddEndPuncttrue
\mciteSetBstMidEndSepPunct{\mcitedefaultmidpunct}
{\mcitedefaultendpunct}{\mcitedefaultseppunct}\relax
\EndOfBibitem
\bibitem[Haward and Odell(2003)]{hawardodell}
S.~J. Haward and J.~A. Odell, \emph{Rheologica Acta}, 2003, \textbf{42},
  516\relax
\mciteBstWouldAddEndPuncttrue
\mciteSetBstMidEndSepPunct{\mcitedefaultmidpunct}
{\mcitedefaultendpunct}{\mcitedefaultseppunct}\relax
\EndOfBibitem
\bibitem[Odell and Haward(2006)]{odellhaward}
J.~A. Odell and S.~J. Haward, \emph{Rheologica Acta}, 2006, \textbf{45},
  853\relax
\mciteBstWouldAddEndPuncttrue
\mciteSetBstMidEndSepPunct{\mcitedefaultmidpunct}
{\mcitedefaultendpunct}{\mcitedefaultseppunct}\relax
\EndOfBibitem
\bibitem[Zamani \emph{et~al.}(2015)Zamani, Bondino, Kaufmann, and
  Skauge]{zamani2015effect}
N.~Zamani, I.~Bondino, R.~Kaufmann and A.~Skauge, \emph{Journal of Petroleum
  Science and Engineering}, 2015, \textbf{133}, 483--495\relax
\mciteBstWouldAddEndPuncttrue
\mciteSetBstMidEndSepPunct{\mcitedefaultmidpunct}
{\mcitedefaultendpunct}{\mcitedefaultseppunct}\relax
\EndOfBibitem
\bibitem[Skauge \emph{et~al.}(2018)Skauge, Zamani, Gausdal~Jacobsen,
  Shaker~Shiran, Al-Shakry, and Skauge]{skauge2018polymer}
A.~Skauge, N.~Zamani, J.~Gausdal~Jacobsen, B.~Shaker~Shiran, B.~Al-Shakry and
  T.~Skauge, \emph{Colloids and Interfaces}, 2018, \textbf{2}, 27\relax
\mciteBstWouldAddEndPuncttrue
\mciteSetBstMidEndSepPunct{\mcitedefaultmidpunct}
{\mcitedefaultendpunct}{\mcitedefaultseppunct}\relax
\EndOfBibitem
\bibitem[Ibezim \emph{et~al.}(2021)Ibezim, Poole, and
  Dennis]{ibezim2021viscoelastic}
V.~C. Ibezim, R.~J. Poole and D.~J. Dennis, \emph{Journal of Non-Newtonian
  Fluid Mechanics}, 2021, \textbf{296}, 104638\relax
\mciteBstWouldAddEndPuncttrue
\mciteSetBstMidEndSepPunct{\mcitedefaultmidpunct}
{\mcitedefaultendpunct}{\mcitedefaultseppunct}\relax
\EndOfBibitem
\bibitem[Browne \emph{et~al.}(2023)Browne, Huang, Zheng, and
  Datta]{browne2023homogenizing}
C.~A. Browne, R.~B. Huang, C.~W. Zheng and S.~S. Datta, \emph{Journal of Fluid
  Mechanics}, 2023, \textbf{963}, A30\relax
\mciteBstWouldAddEndPuncttrue
\mciteSetBstMidEndSepPunct{\mcitedefaultmidpunct}
{\mcitedefaultendpunct}{\mcitedefaultseppunct}\relax
\EndOfBibitem
\bibitem[Phenrat \emph{et~al.}(2009)Phenrat, Kim, Fagerlund, Illangasekare,
  Tilton, and Lowry]{phenrat2009particle}
T.~Phenrat, H.-J. Kim, F.~Fagerlund, T.~Illangasekare, R.~D. Tilton and G.~V.
  Lowry, \emph{Environmental Science \& Technology}, 2009, \textbf{43},
  5079--5085\relax
\mciteBstWouldAddEndPuncttrue
\mciteSetBstMidEndSepPunct{\mcitedefaultmidpunct}
{\mcitedefaultendpunct}{\mcitedefaultseppunct}\relax
\EndOfBibitem
\bibitem[Zhao \emph{et~al.}(2016)Zhao, Liu, Cai, Han, Qian, and
  Zhao]{zhao2016overview}
X.~Zhao, W.~Liu, Z.~Cai, B.~Han, T.~Qian and D.~Zhao, \emph{Water Research},
  2016, \textbf{100}, 245--266\relax
\mciteBstWouldAddEndPuncttrue
\mciteSetBstMidEndSepPunct{\mcitedefaultmidpunct}
{\mcitedefaultendpunct}{\mcitedefaultseppunct}\relax
\EndOfBibitem
\bibitem[Kanel \emph{et~al.}(2006)Kanel, Greneche, and Choi]{kanel2006arsenic}
S.~R. Kanel, J.-M. Greneche and H.~Choi, \emph{Environmental Science \&
  Technology}, 2006, \textbf{40}, 2045--2050\relax
\mciteBstWouldAddEndPuncttrue
\mciteSetBstMidEndSepPunct{\mcitedefaultmidpunct}
{\mcitedefaultendpunct}{\mcitedefaultseppunct}\relax
\EndOfBibitem
\bibitem[Schijven \emph{et~al.}(2003)Schijven, De~Bruin, Hassanizadeh, and
  de~Roda~Husman]{schijven2003bacteriophages}
J.~Schijven, H.~De~Bruin, S.~Hassanizadeh and A.~de~Roda~Husman, \emph{Water
  Research}, 2003, \textbf{37}, 2186--2194\relax
\mciteBstWouldAddEndPuncttrue
\mciteSetBstMidEndSepPunct{\mcitedefaultmidpunct}
{\mcitedefaultendpunct}{\mcitedefaultseppunct}\relax
\EndOfBibitem
\bibitem[Zhong \emph{et~al.}(2017)Zhong, Liu, Jiang, Yang, Liu, Yang, Liu, and
  Zeng]{zhong2017transport}
H.~Zhong, G.~Liu, Y.~Jiang, J.~Yang, Y.~Liu, X.~Yang, Z.~Liu and G.~Zeng,
  \emph{Biotechnology Advances}, 2017, \textbf{35}, 490--504\relax
\mciteBstWouldAddEndPuncttrue
\mciteSetBstMidEndSepPunct{\mcitedefaultmidpunct}
{\mcitedefaultendpunct}{\mcitedefaultseppunct}\relax
\EndOfBibitem
\bibitem[Harvey and Garabedian(1991)]{harvey1991use}
R.~W. Harvey and S.~P. Garabedian, \emph{Environmental Science \& Technology},
  1991, \textbf{25}, 178--185\relax
\mciteBstWouldAddEndPuncttrue
\mciteSetBstMidEndSepPunct{\mcitedefaultmidpunct}
{\mcitedefaultendpunct}{\mcitedefaultseppunct}\relax
\EndOfBibitem
\bibitem[Bizmark \emph{et~al.}(2019)Bizmark, Schneider, de~Jong, and
  Datta]{bizmark2019transport}
N.~Bizmark, J.~Schneider, E.~de~Jong and S.~S. Datta, \emph{Polymer Colloids:
  Formation, Characterization and Applications}, The Royal Society of
  Chemistry, 2019, pp. 289--321\relax
\mciteBstWouldAddEndPuncttrue
\mciteSetBstMidEndSepPunct{\mcitedefaultmidpunct}
{\mcitedefaultendpunct}{\mcitedefaultseppunct}\relax
\EndOfBibitem
\bibitem[Dressaire and Sauret(2016)]{DressaireEmilie2016Coms}
E.~Dressaire and A.~Sauret, \emph{Soft Matter}, 2016, \textbf{13}, 37--48\relax
\mciteBstWouldAddEndPuncttrue
\mciteSetBstMidEndSepPunct{\mcitedefaultmidpunct}
{\mcitedefaultendpunct}{\mcitedefaultseppunct}\relax
\EndOfBibitem
\bibitem[Sahimi and Imdakm(1991)]{sahimi1991hydrodynamics}
M.~Sahimi and A.~Imdakm, \emph{Physical Review Letters}, 1991, \textbf{66},
  1169\relax
\mciteBstWouldAddEndPuncttrue
\mciteSetBstMidEndSepPunct{\mcitedefaultmidpunct}
{\mcitedefaultendpunct}{\mcitedefaultseppunct}\relax
\EndOfBibitem
\bibitem[Zeman and Zydney(2017)]{zeman2017microfiltration}
L.~J. Zeman and A.~L. Zydney, \emph{Microfiltration and ultrafiltration:
  principles and applications}, Routledge, 2017\relax
\mciteBstWouldAddEndPuncttrue
\mciteSetBstMidEndSepPunct{\mcitedefaultmidpunct}
{\mcitedefaultendpunct}{\mcitedefaultseppunct}\relax
\EndOfBibitem
\bibitem[Linkhorst \emph{et~al.}(2016)Linkhorst, Beckmann, Go, Kuehne, and
  Wessling]{linkhorst2016microfluidic}
J.~Linkhorst, T.~Beckmann, D.~Go, A.~J. Kuehne and M.~Wessling,
  \emph{Scientific Reports}, 2016, \textbf{6}, 22376\relax
\mciteBstWouldAddEndPuncttrue
\mciteSetBstMidEndSepPunct{\mcitedefaultmidpunct}
{\mcitedefaultendpunct}{\mcitedefaultseppunct}\relax
\EndOfBibitem
\bibitem[Molnar \emph{et~al.}(2015)Molnar, Johnson, Gerhard, Willson, and
  O'carroll]{molnar2015predicting}
I.~L. Molnar, W.~P. Johnson, J.~I. Gerhard, C.~S. Willson and D.~M. O'carroll,
  \emph{Water Resources Research}, 2015, \textbf{51}, 6804--6845\relax
\mciteBstWouldAddEndPuncttrue
\mciteSetBstMidEndSepPunct{\mcitedefaultmidpunct}
{\mcitedefaultendpunct}{\mcitedefaultseppunct}\relax
\EndOfBibitem
\bibitem[Gerber \emph{et~al.}(2019)Gerber, Bensouda, Weitz, and
  Coussot]{gerber2019self}
G.~Gerber, M.~Bensouda, D.~A. Weitz and P.~Coussot, \emph{Physical Review
  Letters}, 2019, \textbf{123}, 158005\relax
\mciteBstWouldAddEndPuncttrue
\mciteSetBstMidEndSepPunct{\mcitedefaultmidpunct}
{\mcitedefaultendpunct}{\mcitedefaultseppunct}\relax
\EndOfBibitem
\bibitem[Kusaka \emph{et~al.}(2010)Kusaka, Duval, and
  Adachi]{kusaka2010morphology}
Y.~Kusaka, J.~F. Duval and Y.~Adachi, \emph{Environmental Science \&
  Technology}, 2010, \textbf{44}, 9413--9418\relax
\mciteBstWouldAddEndPuncttrue
\mciteSetBstMidEndSepPunct{\mcitedefaultmidpunct}
{\mcitedefaultendpunct}{\mcitedefaultseppunct}\relax
\EndOfBibitem
\bibitem[Lin \emph{et~al.}(2016)Lin, He, Tavakkoli, Mathew, Fatt, Chai,
  Goharzadeh, Vargas, and Biswal]{lin2016examining}
Y.-J. Lin, P.~He, M.~Tavakkoli, N.~T. Mathew, Y.~Y. Fatt, J.~C. Chai,
  A.~Goharzadeh, F.~M. Vargas and S.~L. Biswal, \emph{Langmuir}, 2016,
  \textbf{32}, 8729--8734\relax
\mciteBstWouldAddEndPuncttrue
\mciteSetBstMidEndSepPunct{\mcitedefaultmidpunct}
{\mcitedefaultendpunct}{\mcitedefaultseppunct}\relax
\EndOfBibitem
\bibitem[Auset and Keller(2006)]{auset2006pore}
M.~Auset and A.~A. Keller, \emph{Water Resources Research}, 2006, \textbf{42},
  W12S02\relax
\mciteBstWouldAddEndPuncttrue
\mciteSetBstMidEndSepPunct{\mcitedefaultmidpunct}
{\mcitedefaultendpunct}{\mcitedefaultseppunct}\relax
\EndOfBibitem
\bibitem[Wyss \emph{et~al.}(2006)Wyss, Blair, Morris, Stone, and
  Weitz]{wyss2006mechanism}
H.~M. Wyss, D.~L. Blair, J.~F. Morris, H.~A. Stone and D.~A. Weitz,
  \emph{Physical Review E}, 2006, \textbf{74}, 061402\relax
\mciteBstWouldAddEndPuncttrue
\mciteSetBstMidEndSepPunct{\mcitedefaultmidpunct}
{\mcitedefaultendpunct}{\mcitedefaultseppunct}\relax
\EndOfBibitem
\bibitem[de~Saint~Vincent \emph{et~al.}(2016)de~Saint~Vincent, Abkarian, and
  Tabuteau]{de2016dynamics}
M.~R. de~Saint~Vincent, M.~Abkarian and H.~Tabuteau, \emph{Soft Matter}, 2016,
  \textbf{12}, 1041--1050\relax
\mciteBstWouldAddEndPuncttrue
\mciteSetBstMidEndSepPunct{\mcitedefaultmidpunct}
{\mcitedefaultendpunct}{\mcitedefaultseppunct}\relax
\EndOfBibitem
\bibitem[Lin \emph{et~al.}(2017)Lin, He, Tavakkoli, Mathew, Fatt, Chai,
  Goharzadeh, Vargas, and Biswal]{lin2017characterizing}
Y.-J. Lin, P.~He, M.~Tavakkoli, N.~T. Mathew, Y.~Y. Fatt, J.~C. Chai,
  A.~Goharzadeh, F.~M. Vargas and S.~L. Biswal, \emph{Energy \& Fuels}, 2017,
  \textbf{31}, 11660--11668\relax
\mciteBstWouldAddEndPuncttrue
\mciteSetBstMidEndSepPunct{\mcitedefaultmidpunct}
{\mcitedefaultendpunct}{\mcitedefaultseppunct}\relax
\EndOfBibitem
\bibitem[Mays \emph{et~al.}(2011)Mays, Cannon, Kanold, Harris, Lei, and
  Gilbert]{mays2011static}
D.~C. Mays, O.~T. Cannon, A.~W. Kanold, K.~J. Harris, T.~C. Lei and B.~Gilbert,
  \emph{Journal of Colloid and Interface Science}, 2011, \textbf{363},
  418--424\relax
\mciteBstWouldAddEndPuncttrue
\mciteSetBstMidEndSepPunct{\mcitedefaultmidpunct}
{\mcitedefaultendpunct}{\mcitedefaultseppunct}\relax
\EndOfBibitem
\bibitem[Roth \emph{et~al.}(2015)Roth, Gilbert, and Mays]{roth2015colloid}
E.~J. Roth, B.~Gilbert and D.~C. Mays, \emph{Environmental Science \&
  Technology}, 2015, \textbf{49}, 12263--12270\relax
\mciteBstWouldAddEndPuncttrue
\mciteSetBstMidEndSepPunct{\mcitedefaultmidpunct}
{\mcitedefaultendpunct}{\mcitedefaultseppunct}\relax
\EndOfBibitem
\bibitem[Li \emph{et~al.}(2006)Li, Lin, Miller, and Johnson]{li2006role}
X.~Li, C.-L. Lin, J.~D. Miller and W.~P. Johnson, \emph{Environmental Science
  \& Technology}, 2006, \textbf{40}, 3769--3774\relax
\mciteBstWouldAddEndPuncttrue
\mciteSetBstMidEndSepPunct{\mcitedefaultmidpunct}
{\mcitedefaultendpunct}{\mcitedefaultseppunct}\relax
\EndOfBibitem
\bibitem[Gerber \emph{et~al.}(2018)Gerber, Rodts, Aimedieu, Faure, and
  Coussot]{gerber2018prl}
G.~Gerber, S.~Rodts, P.~Aimedieu, P.~Faure and P.~Coussot, \emph{Physical
  Review Letters}, 2018, \textbf{120}, 148001\relax
\mciteBstWouldAddEndPuncttrue
\mciteSetBstMidEndSepPunct{\mcitedefaultmidpunct}
{\mcitedefaultendpunct}{\mcitedefaultseppunct}\relax
\EndOfBibitem
\bibitem[Bizmark \emph{et~al.}(2020)Bizmark, Schneider, Priestley, and
  Datta]{bizmark2020multiscale}
N.~Bizmark, J.~Schneider, R.~D. Priestley and S.~S. Datta, \emph{Science
  Advances}, 2020, \textbf{6}, eabc2530\relax
\mciteBstWouldAddEndPuncttrue
\mciteSetBstMidEndSepPunct{\mcitedefaultmidpunct}
{\mcitedefaultendpunct}{\mcitedefaultseppunct}\relax
\EndOfBibitem
\bibitem[Braun \emph{et~al.}(2021)Braun, Coquery, Kieffer, Blondel, Favero,
  Besset, Mesnager, Voelker, Delorme, and Matioszek]{braun2021spotlight}
O.~Braun, C.~Coquery, J.~Kieffer, F.~Blondel, C.~Favero, C.~Besset,
  J.~Mesnager, F.~Voelker, C.~Delorme and D.~Matioszek, \emph{Molecules}, 2021,
  \textbf{27}, 42\relax
\mciteBstWouldAddEndPuncttrue
\mciteSetBstMidEndSepPunct{\mcitedefaultmidpunct}
{\mcitedefaultendpunct}{\mcitedefaultseppunct}\relax
\EndOfBibitem
\bibitem[Said \emph{et~al.}(2018)Said, Atassi, Tally, and
  Khatib]{said2018environmentally}
M.~Said, Y.~Atassi, M.~Tally and H.~Khatib, \emph{Journal of Polymers and the
  Environment}, 2018, \textbf{26}, 3937--3948\relax
\mciteBstWouldAddEndPuncttrue
\mciteSetBstMidEndSepPunct{\mcitedefaultmidpunct}
{\mcitedefaultendpunct}{\mcitedefaultseppunct}\relax
\EndOfBibitem
\bibitem[Wei \emph{et~al.}(2016)Wei, Yang, Cao, and Tan]{wei2016using}
J.~Wei, H.~Yang, H.~Cao and T.~Tan, \emph{Saudi Journal of Biological
  Sciences}, 2016, \textbf{23}, 654--659\relax
\mciteBstWouldAddEndPuncttrue
\mciteSetBstMidEndSepPunct{\mcitedefaultmidpunct}
{\mcitedefaultendpunct}{\mcitedefaultseppunct}\relax
\EndOfBibitem
\bibitem[Abedi-Koupai \emph{et~al.}(2008)Abedi-Koupai, Sohrab, and
  Swarbrick]{abedi2008evaluation}
J.~Abedi-Koupai, F.~Sohrab and G.~Swarbrick, \emph{Journal of Plant Nutrition},
  2008, \textbf{31}, 317--331\relax
\mciteBstWouldAddEndPuncttrue
\mciteSetBstMidEndSepPunct{\mcitedefaultmidpunct}
{\mcitedefaultendpunct}{\mcitedefaultseppunct}\relax
\EndOfBibitem
\bibitem[Hemvichian \emph{et~al.}(2014)Hemvichian, Chanthawong, and
  Suwanmala]{hemvichian2014synthesis}
K.~Hemvichian, A.~Chanthawong and P.~Suwanmala, \emph{Radiation Physics and
  Chemistry}, 2014, \textbf{103}, 167--171\relax
\mciteBstWouldAddEndPuncttrue
\mciteSetBstMidEndSepPunct{\mcitedefaultmidpunct}
{\mcitedefaultendpunct}{\mcitedefaultseppunct}\relax
\EndOfBibitem
\bibitem[Nascimento \emph{et~al.}(2021)Nascimento, Mota, Nascimento,
  da~Silva~Albuquerque, Simmons, dos Santos~Dias, and
  Costa]{nascimento2021temperature}
C.~D.~V. Nascimento, J.~C.~A. Mota, {\'I}.~V. Nascimento, G.~H.
  da~Silva~Albuquerque, R.~W. Simmons, C.~T. dos Santos~Dias and M.~C.~G.
  Costa, \emph{Geoderma Regional}, 2021, \textbf{26}, e00407\relax
\mciteBstWouldAddEndPuncttrue
\mciteSetBstMidEndSepPunct{\mcitedefaultmidpunct}
{\mcitedefaultendpunct}{\mcitedefaultseppunct}\relax
\EndOfBibitem
\bibitem[Guilherme \emph{et~al.}(2015)Guilherme, Aouada, Fajardo, Martins,
  Paulino, Davi, Rubira, and Muniz]{guilherme2015superabsorbent}
M.~R. Guilherme, F.~A. Aouada, A.~R. Fajardo, A.~F. Martins, A.~T. Paulino,
  M.~F. Davi, A.~F. Rubira and E.~C. Muniz, \emph{European Polymer Journal},
  2015, \textbf{72}, 365--385\relax
\mciteBstWouldAddEndPuncttrue
\mciteSetBstMidEndSepPunct{\mcitedefaultmidpunct}
{\mcitedefaultendpunct}{\mcitedefaultseppunct}\relax
\EndOfBibitem
\bibitem[Bandak \emph{et~al.}(2021)Bandak, Naeini, Zeinali, and
  Bandak]{bandak2021effects}
S.~Bandak, S.~A. R.~M. Naeini, E.~Zeinali and I.~Bandak, \emph{Arabian Journal
  of Geosciences}, 2021, \textbf{14}, 1--10\relax
\mciteBstWouldAddEndPuncttrue
\mciteSetBstMidEndSepPunct{\mcitedefaultmidpunct}
{\mcitedefaultendpunct}{\mcitedefaultseppunct}\relax
\EndOfBibitem
\bibitem[Souza \emph{et~al.}(2016)Souza, Guimar{\~a}es, Dominghetti, Scalco,
  and Rezende]{souza2016water}
A.~J.~J. Souza, R.~J. Guimar{\~a}es, A.~W. Dominghetti, M.~S. Scalco and T.~T.
  Rezende, \emph{Revista Ci{\^e}ncia Agron{\^o}mica}, 2016, \textbf{47},
  334--343\relax
\mciteBstWouldAddEndPuncttrue
\mciteSetBstMidEndSepPunct{\mcitedefaultmidpunct}
{\mcitedefaultendpunct}{\mcitedefaultseppunct}\relax
\EndOfBibitem
\bibitem[Banedjschafie and Durner(2015)]{banedjschafie2015water}
S.~Banedjschafie and W.~Durner, \emph{Journal of Plant Nutrition and Soil
  Science}, 2015, \textbf{178}, 798--806\relax
\mciteBstWouldAddEndPuncttrue
\mciteSetBstMidEndSepPunct{\mcitedefaultmidpunct}
{\mcitedefaultendpunct}{\mcitedefaultseppunct}\relax
\EndOfBibitem
\bibitem[Garbowski \emph{et~al.}(2020)Garbowski, Brown, and
  Johnston]{garbowski2020soil}
M.~Garbowski, C.~S. Brown and D.~B. Johnston, \emph{Restoration Ecology}, 2020,
  \textbf{28}, A13--A23\relax
\mciteBstWouldAddEndPuncttrue
\mciteSetBstMidEndSepPunct{\mcitedefaultmidpunct}
{\mcitedefaultendpunct}{\mcitedefaultseppunct}\relax
\EndOfBibitem
\bibitem[Sojka \emph{et~al.}(1998)Sojka, Lentz, and Westermann]{sojka1998water}
R.~Sojka, R.~Lentz and D.~Westermann, \emph{Soil Science Society of America
  Journal}, 1998, \textbf{62}, 1672--1680\relax
\mciteBstWouldAddEndPuncttrue
\mciteSetBstMidEndSepPunct{\mcitedefaultmidpunct}
{\mcitedefaultendpunct}{\mcitedefaultseppunct}\relax
\EndOfBibitem
\bibitem[Wei and Durian(2013)]{wei2013effect}
Y.~Wei and D.~J. Durian, \emph{Physical Review E}, 2013, \textbf{87},
  053013\relax
\mciteBstWouldAddEndPuncttrue
\mciteSetBstMidEndSepPunct{\mcitedefaultmidpunct}
{\mcitedefaultendpunct}{\mcitedefaultseppunct}\relax
\EndOfBibitem
\bibitem[Cejas \emph{et~al.}(2014)Cejas, Wei, Barrois, Fr{\'e}tigny, Durian,
  and Dreyfus]{cejas2014kinetics}
C.~M. Cejas, Y.~Wei, R.~Barrois, C.~Fr{\'e}tigny, D.~J. Durian and R.~Dreyfus,
  \emph{Physical Review E}, 2014, \textbf{90}, 042205\relax
\mciteBstWouldAddEndPuncttrue
\mciteSetBstMidEndSepPunct{\mcitedefaultmidpunct}
{\mcitedefaultendpunct}{\mcitedefaultseppunct}\relax
\EndOfBibitem
\bibitem[Wei and Durian(2014)]{wei2014rain}
Y.~Wei and D.~Durian, \emph{The European Physical Journal E}, 2014,
  \textbf{37}, 1--11\relax
\mciteBstWouldAddEndPuncttrue
\mciteSetBstMidEndSepPunct{\mcitedefaultmidpunct}
{\mcitedefaultendpunct}{\mcitedefaultseppunct}\relax
\EndOfBibitem
\bibitem[Wei \emph{et~al.}(2014)Wei, Cejas, Barrois, Dreyfus, and
  Durian]{wei2014morphology}
Y.~Wei, C.~M. Cejas, R.~Barrois, R.~Dreyfus and D.~J. Durian, \emph{Physical
  Review Applied}, 2014, \textbf{2}, 044004\relax
\mciteBstWouldAddEndPuncttrue
\mciteSetBstMidEndSepPunct{\mcitedefaultmidpunct}
{\mcitedefaultendpunct}{\mcitedefaultseppunct}\relax
\EndOfBibitem
\bibitem[Woodhouse and Johnson(1991)]{woodhouse1991effect}
J.~Woodhouse and M.~Johnson, \emph{Agricultural Water Management}, 1991,
  \textbf{20}, 63--70\relax
\mciteBstWouldAddEndPuncttrue
\mciteSetBstMidEndSepPunct{\mcitedefaultmidpunct}
{\mcitedefaultendpunct}{\mcitedefaultseppunct}\relax
\EndOfBibitem
\bibitem[Bai \emph{et~al.}(2010)Bai, Zhang, Liu, Wu, and Song]{bai2010effects}
W.~Bai, H.~Zhang, B.~Liu, Y.~Wu and J.~Song, \emph{Soil Use and Management},
  2010, \textbf{26}, 253--260\relax
\mciteBstWouldAddEndPuncttrue
\mciteSetBstMidEndSepPunct{\mcitedefaultmidpunct}
{\mcitedefaultendpunct}{\mcitedefaultseppunct}\relax
\EndOfBibitem
\bibitem[Frantz \emph{et~al.}(2005)Frantz, Locke, Pitchay, and
  Krause]{frantz2005actual}
J.~M. Frantz, J.~C. Locke, D.~S. Pitchay and C.~R. Krause, \emph{HortScience},
  2005, \textbf{40}, 2040--2046\relax
\mciteBstWouldAddEndPuncttrue
\mciteSetBstMidEndSepPunct{\mcitedefaultmidpunct}
{\mcitedefaultendpunct}{\mcitedefaultseppunct}\relax
\EndOfBibitem
\bibitem[Hejduk \emph{et~al.}(2012)Hejduk, Baker, and
  Spring]{hejduk2012evaluation}
S.~Hejduk, S.~W. Baker and C.~A. Spring, \emph{Acta Agriculturae Scandinavica,
  Section B-Soil \& Plant Science}, 2012, \textbf{62}, 155--164\relax
\mciteBstWouldAddEndPuncttrue
\mciteSetBstMidEndSepPunct{\mcitedefaultmidpunct}
{\mcitedefaultendpunct}{\mcitedefaultseppunct}\relax
\EndOfBibitem
\bibitem[Louf \emph{et~al.}(2021)Louf, Lu, O’Connell, Cho, and
  Datta]{louf2021under}
J.-F. Louf, N.~B. Lu, M.~G. O’Connell, H.~J. Cho and S.~S. Datta,
  \emph{Science Advances}, 2021, \textbf{7}, eabd2711\relax
\mciteBstWouldAddEndPuncttrue
\mciteSetBstMidEndSepPunct{\mcitedefaultmidpunct}
{\mcitedefaultendpunct}{\mcitedefaultseppunct}\relax
\EndOfBibitem
\bibitem[Misiewicz \emph{et~al.}(2022)Misiewicz, Datta, Lejcu{\'s}, and
  Marczak]{misiewicz2022characteristics}
J.~Misiewicz, S.~S. Datta, K.~Lejcu{\'s} and D.~Marczak, \emph{Materials},
  2022, \textbf{15}, 4465\relax
\mciteBstWouldAddEndPuncttrue
\mciteSetBstMidEndSepPunct{\mcitedefaultmidpunct}
{\mcitedefaultendpunct}{\mcitedefaultseppunct}\relax
\EndOfBibitem
\bibitem[Goehring \emph{et~al.}(2015)Goehring, Nakahara, Dutta, Kitsunezaki,
  and Tarafdar]{goehring2015desiccation}
L.~Goehring, A.~Nakahara, T.~Dutta, S.~Kitsunezaki and S.~Tarafdar,
  \emph{Desiccation cracks and their patterns : formation and modeling in
  science and nature}, John Wiley and Sons, Inc, Weinheim, Germany, 2015\relax
\mciteBstWouldAddEndPuncttrue
\mciteSetBstMidEndSepPunct{\mcitedefaultmidpunct}
{\mcitedefaultendpunct}{\mcitedefaultseppunct}\relax
\EndOfBibitem
\bibitem[Gates \emph{et~al.}(2009)Gates, Bouazza, and
  Churchman]{doi:10.2113/gselements.5.2.105}
W.~P. Gates, A.~Bouazza and G.~J. Churchman, \emph{Elements}, 2009, \textbf{5},
  105\relax
\mciteBstWouldAddEndPuncttrue
\mciteSetBstMidEndSepPunct{\mcitedefaultmidpunct}
{\mcitedefaultendpunct}{\mcitedefaultseppunct}\relax
\EndOfBibitem
\bibitem[Espinoza and Santamarina(2012)]{Espinoza:2012iq}
D.~N. Espinoza and J.~C. Santamarina, \emph{International Journal of Greenhouse
  Gas Control}, 2012, \textbf{10}, 351--362\relax
\mciteBstWouldAddEndPuncttrue
\mciteSetBstMidEndSepPunct{\mcitedefaultmidpunct}
{\mcitedefaultendpunct}{\mcitedefaultseppunct}\relax
\EndOfBibitem
\bibitem[Cho and Datta(2019)]{cho2019scaling}
H.~J. Cho and S.~S. Datta, \emph{Physical Review Letters}, 2019, \textbf{123},
  158004\relax
\mciteBstWouldAddEndPuncttrue
\mciteSetBstMidEndSepPunct{\mcitedefaultmidpunct}
{\mcitedefaultendpunct}{\mcitedefaultseppunct}\relax
\EndOfBibitem
\bibitem[Morrow and Mason(2001)]{morrow2001recovery}
N.~R. Morrow and G.~Mason, \emph{Current Opinion in Colloid \& Interface
  Science}, 2001, \textbf{6}, 321--337\relax
\mciteBstWouldAddEndPuncttrue
\mciteSetBstMidEndSepPunct{\mcitedefaultmidpunct}
{\mcitedefaultendpunct}{\mcitedefaultseppunct}\relax
\EndOfBibitem
\bibitem[Mattax \emph{et~al.}(1962)Mattax,
  Kyte,\emph{et~al.}]{mattax1962imbibition}
C.~C. Mattax, J.~Kyte \emph{et~al.}, \emph{Society of Petroleum Engineers
  Journal}, 1962, \textbf{2}, 177--184\relax
\mciteBstWouldAddEndPuncttrue
\mciteSetBstMidEndSepPunct{\mcitedefaultmidpunct}
{\mcitedefaultendpunct}{\mcitedefaultseppunct}\relax
\EndOfBibitem
\bibitem[Li \emph{et~al.}(2000)Li, Horne,\emph{et~al.}]{li2000characterization}
K.~Li, R.~N. Horne \emph{et~al.}, SPE/AAPG Western Regional Meeting, 2000\relax
\mciteBstWouldAddEndPuncttrue
\mciteSetBstMidEndSepPunct{\mcitedefaultmidpunct}
{\mcitedefaultendpunct}{\mcitedefaultseppunct}\relax
\EndOfBibitem
\bibitem[Nicolaides \emph{et~al.}(2015)Nicolaides, Jha, Cueto-Felgueroso, and
  Juanes]{nicolaides2015impact}
C.~Nicolaides, B.~Jha, L.~Cueto-Felgueroso and R.~Juanes, \emph{Water Resources
  Research}, 2015, \textbf{51}, 2634--2647\relax
\mciteBstWouldAddEndPuncttrue
\mciteSetBstMidEndSepPunct{\mcitedefaultmidpunct}
{\mcitedefaultendpunct}{\mcitedefaultseppunct}\relax
\EndOfBibitem
\bibitem[Bennion \emph{et~al.}(2006)Bennion,
  Bachu,\emph{et~al.}]{bennion2006supercritical}
D.~B. Bennion, S.~Bachu \emph{et~al.}, SPE Europec/EAGE Annual Conference and
  Exhibition, 2006\relax
\mciteBstWouldAddEndPuncttrue
\mciteSetBstMidEndSepPunct{\mcitedefaultmidpunct}
{\mcitedefaultendpunct}{\mcitedefaultseppunct}\relax
\EndOfBibitem
\bibitem[Bennion \emph{et~al.}(2010)Bennion,
  Bachu,\emph{et~al.}]{bennion2010drainage}
D.~B. Bennion, S.~Bachu \emph{et~al.}, SPE Annual Technical Conference and
  Exhibition, 2010\relax
\mciteBstWouldAddEndPuncttrue
\mciteSetBstMidEndSepPunct{\mcitedefaultmidpunct}
{\mcitedefaultendpunct}{\mcitedefaultseppunct}\relax
\EndOfBibitem
\bibitem[Hatiboglu and Babadagli(2008)]{hatiboglu2008pore}
C.~U. Hatiboglu and T.~Babadagli, \emph{Physical Review E}, 2008, \textbf{77},
  066311\relax
\mciteBstWouldAddEndPuncttrue
\mciteSetBstMidEndSepPunct{\mcitedefaultmidpunct}
{\mcitedefaultendpunct}{\mcitedefaultseppunct}\relax
\EndOfBibitem
\bibitem[Celia \emph{et~al.}(2015)Celia, Bachu, Nordbotten, and
  Bandilla]{celia2015status}
M.~A. Celia, S.~Bachu, J.~Nordbotten and K.~Bandilla, \emph{Water Resources
  Research}, 2015, \textbf{51}, 6846--6892\relax
\mciteBstWouldAddEndPuncttrue
\mciteSetBstMidEndSepPunct{\mcitedefaultmidpunct}
{\mcitedefaultendpunct}{\mcitedefaultseppunct}\relax
\EndOfBibitem
\bibitem[Schaefer \emph{et~al.}(2000)Schaefer, DiCarlo, and
  Blunt]{schaefer2000experimental}
C.~Schaefer, D.~DiCarlo and M.~Blunt, \emph{Water Resources Research}, 2000,
  \textbf{36}, 885--890\relax
\mciteBstWouldAddEndPuncttrue
\mciteSetBstMidEndSepPunct{\mcitedefaultmidpunct}
{\mcitedefaultendpunct}{\mcitedefaultseppunct}\relax
\EndOfBibitem
\bibitem[Esposito and Benson(2011)]{esposito2011remediation}
A.~Esposito and S.~M. Benson, \emph{Energy Procedia}, 2011, \textbf{4},
  3216--3223\relax
\mciteBstWouldAddEndPuncttrue
\mciteSetBstMidEndSepPunct{\mcitedefaultmidpunct}
{\mcitedefaultendpunct}{\mcitedefaultseppunct}\relax
\EndOfBibitem
\bibitem[Penvern \emph{et~al.}(2020)Penvern, Zhou, Maillet, Courtier-Murias,
  Scheel, Perrin, Weitkamp, Bardet, Car\'e, and Coussot]{penvern2020}
H.~Penvern, M.~Zhou, B.~Maillet, D.~Courtier-Murias, M.~Scheel, J.~Perrin,
  T.~Weitkamp, S.~Bardet, S.~Car\'e and P.~Coussot, \emph{Physical Review
  Applied}, 2020, \textbf{14}, 054051\relax
\mciteBstWouldAddEndPuncttrue
\mciteSetBstMidEndSepPunct{\mcitedefaultmidpunct}
{\mcitedefaultendpunct}{\mcitedefaultseppunct}\relax
\EndOfBibitem
\bibitem[Zhou \emph{et~al.}(2019)Zhou, Car\'e, King, Courtier-Murias, Rodts,
  Gerber, Aimedieu, Bonnet, Bornert, and Coussot]{zhou2019}
M.~Zhou, S.~Car\'e, A.~King, D.~Courtier-Murias, S.~Rodts, G.~Gerber,
  P.~Aimedieu, M.~Bonnet, M.~Bornert and P.~Coussot, \emph{Physical Review
  Research}, 2019, \textbf{1}, 033190\relax
\mciteBstWouldAddEndPuncttrue
\mciteSetBstMidEndSepPunct{\mcitedefaultmidpunct}
{\mcitedefaultendpunct}{\mcitedefaultseppunct}\relax
\EndOfBibitem
\bibitem[Weisbrod \emph{et~al.}(2002)Weisbrod, Niemet, and
  Selker]{weisbrod2002imbibition}
N.~Weisbrod, M.~R. Niemet and J.~S. Selker, \emph{Advances in Water Resources},
  2002, \textbf{25}, 841--855\relax
\mciteBstWouldAddEndPuncttrue
\mciteSetBstMidEndSepPunct{\mcitedefaultmidpunct}
{\mcitedefaultendpunct}{\mcitedefaultseppunct}\relax
\EndOfBibitem
\bibitem[Tesoro \emph{et~al.}(2007)Tesoro, Choong, and
  Kimbler]{tesoro2007relative}
F.~Tesoro, E.~Choong and O.~Kimbler, \emph{Wood and Fiber Science}, 2007,
  \textbf{6}, 226--236\relax
\mciteBstWouldAddEndPuncttrue
\mciteSetBstMidEndSepPunct{\mcitedefaultmidpunct}
{\mcitedefaultendpunct}{\mcitedefaultseppunct}\relax
\EndOfBibitem
\bibitem[Hassanein \emph{et~al.}(2006)Hassanein, Meyer, Carminati, Estermann,
  Lehmann, and Vontobel]{hassanein2006investigation}
R.~Hassanein, H.~Meyer, A.~Carminati, M.~Estermann, E.~Lehmann and P.~Vontobel,
  \emph{Journal of Physics D: Applied Physics}, 2006, \textbf{39}, 4284\relax
\mciteBstWouldAddEndPuncttrue
\mciteSetBstMidEndSepPunct{\mcitedefaultmidpunct}
{\mcitedefaultendpunct}{\mcitedefaultseppunct}\relax
\EndOfBibitem
\bibitem[Lenormand \emph{et~al.}(1984)Lenormand,
  Zarcone,\emph{et~al.}]{lenormand1984role}
R.~Lenormand, C.~Zarcone \emph{et~al.}, SPE Annual Technical Conference and
  Exhibition, 1984\relax
\mciteBstWouldAddEndPuncttrue
\mciteSetBstMidEndSepPunct{\mcitedefaultmidpunct}
{\mcitedefaultendpunct}{\mcitedefaultseppunct}\relax
\EndOfBibitem
\bibitem[Chang \emph{et~al.}(2009)Chang, Tsai, Shan, and
  Chen]{chang2009experimental}
L.-C. Chang, J.-P. Tsai, H.-Y. Shan and H.-H. Chen, \emph{Environmental Earth
  Sciences}, 2009, \textbf{59}, 901\relax
\mciteBstWouldAddEndPuncttrue
\mciteSetBstMidEndSepPunct{\mcitedefaultmidpunct}
{\mcitedefaultendpunct}{\mcitedefaultseppunct}\relax
\EndOfBibitem
\bibitem[Sun \emph{et~al.}(2016)Sun, Kharaghani, and Tsotsas]{sun2016micro}
Y.~Sun, A.~Kharaghani and E.~Tsotsas, \emph{Chemical Engineering Science},
  2016, \textbf{150}, 41--53\relax
\mciteBstWouldAddEndPuncttrue
\mciteSetBstMidEndSepPunct{\mcitedefaultmidpunct}
{\mcitedefaultendpunct}{\mcitedefaultseppunct}\relax
\EndOfBibitem
\bibitem[Hughes and Blunt(2000)]{hughes2000pore}
R.~G. Hughes and M.~J. Blunt, \emph{Transport in Porous Media}, 2000,
  \textbf{40}, 295--322\relax
\mciteBstWouldAddEndPuncttrue
\mciteSetBstMidEndSepPunct{\mcitedefaultmidpunct}
{\mcitedefaultendpunct}{\mcitedefaultseppunct}\relax
\EndOfBibitem
\bibitem[Lenormand \emph{et~al.}(1988)Lenormand, Touboul, and
  Zarcone]{lenormand1988numerical}
R.~Lenormand, E.~Touboul and C.~Zarcone, \emph{Journal of Fluid Mechanics},
  1988, \textbf{189}, 165--187\relax
\mciteBstWouldAddEndPuncttrue
\mciteSetBstMidEndSepPunct{\mcitedefaultmidpunct}
{\mcitedefaultendpunct}{\mcitedefaultseppunct}\relax
\EndOfBibitem
\bibitem[Lenormand(1990)]{lenormand1990liquids}
R.~Lenormand, \emph{Journal of Physics: Condensed Matter}, 1990, \textbf{2},
  SA79\relax
\mciteBstWouldAddEndPuncttrue
\mciteSetBstMidEndSepPunct{\mcitedefaultmidpunct}
{\mcitedefaultendpunct}{\mcitedefaultseppunct}\relax
\EndOfBibitem
\bibitem[Sahimi(1993)]{sahimi1993flow}
M.~Sahimi, \emph{Reviews of Modern Physics}, 1993, \textbf{65}, 1393\relax
\mciteBstWouldAddEndPuncttrue
\mciteSetBstMidEndSepPunct{\mcitedefaultmidpunct}
{\mcitedefaultendpunct}{\mcitedefaultseppunct}\relax
\EndOfBibitem
\bibitem[Alava \emph{et~al.}(2004)Alava, Dub{\'e}, and
  Rost]{alava2004imbibition}
M.~Alava, M.~Dub{\'e} and M.~Rost, \emph{Advances in Physics}, 2004,
  \textbf{53}, 83--175\relax
\mciteBstWouldAddEndPuncttrue
\mciteSetBstMidEndSepPunct{\mcitedefaultmidpunct}
{\mcitedefaultendpunct}{\mcitedefaultseppunct}\relax
\EndOfBibitem
\bibitem[Stokes \emph{et~al.}(1986)Stokes, Weitz, Gollub, Dougherty, Robbins,
  Chaikin, and Lindsay]{stokes1986interfacial}
J.~Stokes, D.~Weitz, J.~P. Gollub, A.~Dougherty, M.~Robbins, P.~Chaikin and
  H.~Lindsay, \emph{Physical Review Letters}, 1986, \textbf{57}, 1718\relax
\mciteBstWouldAddEndPuncttrue
\mciteSetBstMidEndSepPunct{\mcitedefaultmidpunct}
{\mcitedefaultendpunct}{\mcitedefaultseppunct}\relax
\EndOfBibitem
\bibitem[Weitz \emph{et~al.}(1987)Weitz, Stokes, Ball, and
  Kushnick]{weitz1987dynamic}
D.~Weitz, J.~Stokes, R.~Ball and A.~Kushnick, \emph{Physical Review Letters},
  1987, \textbf{59}, 2967\relax
\mciteBstWouldAddEndPuncttrue
\mciteSetBstMidEndSepPunct{\mcitedefaultmidpunct}
{\mcitedefaultendpunct}{\mcitedefaultseppunct}\relax
\EndOfBibitem
\bibitem[Hultmark \emph{et~al.}(2011)Hultmark, Aristoff, and
  Stone]{hultmark2011influence}
M.~Hultmark, J.~M. Aristoff and H.~A. Stone, \emph{Journal of Fluid Mechanics},
  2011, \textbf{678}, 600--606\relax
\mciteBstWouldAddEndPuncttrue
\mciteSetBstMidEndSepPunct{\mcitedefaultmidpunct}
{\mcitedefaultendpunct}{\mcitedefaultseppunct}\relax
\EndOfBibitem
\bibitem[Zhao \emph{et~al.}(2016)Zhao, MacMinn, and
  Juanes]{zhao2016wettability}
B.~Zhao, C.~W. MacMinn and R.~Juanes, \emph{Proceedings of the National Academy
  of Sciences}, 2016, \textbf{113}, 10251--10256\relax
\mciteBstWouldAddEndPuncttrue
\mciteSetBstMidEndSepPunct{\mcitedefaultmidpunct}
{\mcitedefaultendpunct}{\mcitedefaultseppunct}\relax
\EndOfBibitem
\bibitem[Tanino \emph{et~al.}(2018)Tanino, Zacarias-Hernandez, and
  Christensen]{tanino2018oil}
Y.~Tanino, X.~Zacarias-Hernandez and M.~Christensen, \emph{Experiments in
  Fluids}, 2018, \textbf{59}, 35\relax
\mciteBstWouldAddEndPuncttrue
\mciteSetBstMidEndSepPunct{\mcitedefaultmidpunct}
{\mcitedefaultendpunct}{\mcitedefaultseppunct}\relax
\EndOfBibitem
\bibitem[Odier \emph{et~al.}(2017)Odier, Levach{\'e}, Santanach-Carreras, and
  Bartolo]{odier2017forced}
C.~Odier, B.~Levach{\'e}, E.~Santanach-Carreras and D.~Bartolo, \emph{Physical
  Review Letters}, 2017, \textbf{119}, 208005\relax
\mciteBstWouldAddEndPuncttrue
\mciteSetBstMidEndSepPunct{\mcitedefaultmidpunct}
{\mcitedefaultendpunct}{\mcitedefaultseppunct}\relax
\EndOfBibitem
\bibitem[Bear(2013)]{bear2013dynamics}
J.~Bear, \emph{Dynamics of Fluids in Porous Media}, Courier Corporation,
  2013\relax
\mciteBstWouldAddEndPuncttrue
\mciteSetBstMidEndSepPunct{\mcitedefaultmidpunct}
{\mcitedefaultendpunct}{\mcitedefaultseppunct}\relax
\EndOfBibitem
\bibitem[Galloway and Hobday(2012)]{galloway2012terrigenous}
W.~E. Galloway and D.~K. Hobday, \emph{Terrigenous Clastic Depositional
  Systems: {A}pplications to Fossil Fuel and Groundwater Resources}, Springer
  Science \& Business Media, 2012\relax
\mciteBstWouldAddEndPuncttrue
\mciteSetBstMidEndSepPunct{\mcitedefaultmidpunct}
{\mcitedefaultendpunct}{\mcitedefaultseppunct}\relax
\EndOfBibitem
\bibitem[Gasda and Celia(2005)]{gasda2005upscaling}
S.~Gasda and M.~A. Celia, \emph{Advances in Water Resources}, 2005,
  \textbf{28}, 493--506\relax
\mciteBstWouldAddEndPuncttrue
\mciteSetBstMidEndSepPunct{\mcitedefaultmidpunct}
{\mcitedefaultendpunct}{\mcitedefaultseppunct}\relax
\EndOfBibitem
\bibitem[King \emph{et~al.}(2018)King, Sansone, Kortunov, Xu, Callen, Chhatre,
  Sahoo, Buono,\emph{et~al.}]{king2018microstructural}
H.~King, M.~Sansone, P.~Kortunov, Y.~Xu, N.~Callen, S.~Chhatre, H.~Sahoo,
  A.~Buono \emph{et~al.}, \emph{Petrophysics}, 2018, \textbf{59}, 35--43\relax
\mciteBstWouldAddEndPuncttrue
\mciteSetBstMidEndSepPunct{\mcitedefaultmidpunct}
{\mcitedefaultendpunct}{\mcitedefaultseppunct}\relax
\EndOfBibitem
\bibitem[Lu \emph{et~al.}(2019)Lu, Browne, Amchin, Nunes, and
  Datta]{lu2019controlling}
N.~B. Lu, C.~A. Browne, D.~B. Amchin, J.~K. Nunes and S.~S. Datta,
  \emph{Physical Review Fluids}, 2019, \textbf{4}, 084303\relax
\mciteBstWouldAddEndPuncttrue
\mciteSetBstMidEndSepPunct{\mcitedefaultmidpunct}
{\mcitedefaultendpunct}{\mcitedefaultseppunct}\relax
\EndOfBibitem
\bibitem[Rabbani \emph{et~al.}(2018)Rabbani, Or, Liu, Lai, Lu, Datta, Stone,
  and Shokri]{rabbani2018suppressing}
H.~S. Rabbani, D.~Or, Y.~Liu, C.-Y. Lai, N.~B. Lu, S.~S. Datta, H.~A. Stone and
  N.~Shokri, \emph{Proceedings of the National Academy of Sciences}, 2018,
  \textbf{115}, 4833--4838\relax
\mciteBstWouldAddEndPuncttrue
\mciteSetBstMidEndSepPunct{\mcitedefaultmidpunct}
{\mcitedefaultendpunct}{\mcitedefaultseppunct}\relax
\EndOfBibitem
\bibitem[Lu \emph{et~al.}(2020)Lu, Pahlavan, Browne, Amchin, Stone, and
  Datta]{lu2020forced}
N.~B. Lu, A.~A. Pahlavan, C.~A. Browne, D.~B. Amchin, H.~A. Stone and S.~S.
  Datta, \emph{Physical Review Applied}, 2020, \textbf{14}, 054009\relax
\mciteBstWouldAddEndPuncttrue
\mciteSetBstMidEndSepPunct{\mcitedefaultmidpunct}
{\mcitedefaultendpunct}{\mcitedefaultseppunct}\relax
\EndOfBibitem
\bibitem[Lu \emph{et~al.}(2021)Lu, Amchin, and Datta]{lu2021forced}
N.~B. Lu, D.~B. Amchin and S.~S. Datta, \emph{Physical Review Fluids}, 2021,
  \textbf{6}, 114007\relax
\mciteBstWouldAddEndPuncttrue
\mciteSetBstMidEndSepPunct{\mcitedefaultmidpunct}
{\mcitedefaultendpunct}{\mcitedefaultseppunct}\relax
\EndOfBibitem
\bibitem[Dechesne \emph{et~al.}(2010)Dechesne, Wang, Gaolez, Or, and
  Smets]{dechesne}
A.~Dechesne, G.~Wang, G.~Gaolez, D.~Or and B.~F. Smets, \emph{Proceedings of
  the National Academy of Sciences}, 2010, \textbf{107}, 14369--14372\relax
\mciteBstWouldAddEndPuncttrue
\mciteSetBstMidEndSepPunct{\mcitedefaultmidpunct}
{\mcitedefaultendpunct}{\mcitedefaultseppunct}\relax
\EndOfBibitem
\bibitem[Souza \emph{et~al.}(2015)Souza, Ambrosini, and Passaglia]{souza}
R.~d. Souza, A.~Ambrosini and L.~M. Passaglia, \emph{Genetics and Molecular
  Biology}, 2015, \textbf{38}, 401--419\relax
\mciteBstWouldAddEndPuncttrue
\mciteSetBstMidEndSepPunct{\mcitedefaultmidpunct}
{\mcitedefaultendpunct}{\mcitedefaultseppunct}\relax
\EndOfBibitem
\bibitem[Turnbull \emph{et~al.}(2001)Turnbull, Morgan, Whipps, and
  Saunders]{turnbull}
G.~A. Turnbull, J.~A.~W. Morgan, J.~M. Whipps and J.~R. Saunders, \emph{FEMS
  Microbiology Ecology}, 2001, \textbf{36}, 21--31\relax
\mciteBstWouldAddEndPuncttrue
\mciteSetBstMidEndSepPunct{\mcitedefaultmidpunct}
{\mcitedefaultendpunct}{\mcitedefaultseppunct}\relax
\EndOfBibitem
\bibitem[Watt \emph{et~al.}(2006)Watt, Kirkegaard, and Passioura]{watt}
M.~Watt, J.~Kirkegaard and J.~Passioura, \emph{Soil Research}, 2006,
  \textbf{44}, 299--317\relax
\mciteBstWouldAddEndPuncttrue
\mciteSetBstMidEndSepPunct{\mcitedefaultmidpunct}
{\mcitedefaultendpunct}{\mcitedefaultseppunct}\relax
\EndOfBibitem
\bibitem[Babalola(2010)]{babalola}
O.~O. Babalola, \emph{Biotechnol. Lett}, 2010, \textbf{32}, 1559--1570\relax
\mciteBstWouldAddEndPuncttrue
\mciteSetBstMidEndSepPunct{\mcitedefaultmidpunct}
{\mcitedefaultendpunct}{\mcitedefaultseppunct}\relax
\EndOfBibitem
\bibitem[Adadevoh \emph{et~al.}(2018)Adadevoh, Ramsburg, and Ford]{roseanne18}
J.~S. Adadevoh, C.~A. Ramsburg and R.~M. Ford, \emph{Environmental Science \&
  Technology}, 2018, \textbf{52}, 7289--7295\relax
\mciteBstWouldAddEndPuncttrue
\mciteSetBstMidEndSepPunct{\mcitedefaultmidpunct}
{\mcitedefaultendpunct}{\mcitedefaultseppunct}\relax
\EndOfBibitem
\bibitem[Adadevoh \emph{et~al.}(2016)Adadevoh, Triolo, Ramsburg, and
  Ford]{Adadevoh:2016}
J.~S.~T. Adadevoh, S.~Triolo, C.~A. Ramsburg and R.~M. Ford,
  \emph{Environmental Science \& Technology}, 2016, \textbf{50}, 181--187\relax
\mciteBstWouldAddEndPuncttrue
\mciteSetBstMidEndSepPunct{\mcitedefaultmidpunct}
{\mcitedefaultendpunct}{\mcitedefaultseppunct}\relax
\EndOfBibitem
\bibitem[Ford and Harvey(2007)]{ford07}
R.~M. Ford and R.~W. Harvey, \emph{Advances in Water Resources}, 2007,
  \textbf{30}, 1608--1617\relax
\mciteBstWouldAddEndPuncttrue
\mciteSetBstMidEndSepPunct{\mcitedefaultmidpunct}
{\mcitedefaultendpunct}{\mcitedefaultseppunct}\relax
\EndOfBibitem
\bibitem[Wang \emph{et~al.}(2008)Wang, Ford, and Harvey]{wang08}
M.~Wang, R.~M. Ford and R.~W. Harvey, \emph{Environmental Science \&
  Technology}, 2008, \textbf{42}, 3556--3562\relax
\mciteBstWouldAddEndPuncttrue
\mciteSetBstMidEndSepPunct{\mcitedefaultmidpunct}
{\mcitedefaultendpunct}{\mcitedefaultseppunct}\relax
\EndOfBibitem
\bibitem[Raina \emph{et~al.}(2019)Raina, Fernandez, Lambert, Stocker, and
  Seymour]{raina2019role}
J.-B. Raina, V.~Fernandez, B.~Lambert, R.~Stocker and J.~R. Seymour,
  \emph{Nature Reviews Microbiology}, 2019, \textbf{17}, 284--294\relax
\mciteBstWouldAddEndPuncttrue
\mciteSetBstMidEndSepPunct{\mcitedefaultmidpunct}
{\mcitedefaultendpunct}{\mcitedefaultseppunct}\relax
\EndOfBibitem
\bibitem[Bhattacharjee and Datta(2019)]{bhattacharjee2019confinement}
T.~Bhattacharjee and S.~S. Datta, \emph{Soft Matter}, 2019, \textbf{15},
  9920--9930\relax
\mciteBstWouldAddEndPuncttrue
\mciteSetBstMidEndSepPunct{\mcitedefaultmidpunct}
{\mcitedefaultendpunct}{\mcitedefaultseppunct}\relax
\EndOfBibitem
\bibitem[Bhattacharjee and Datta(2019)]{bhattacharjee2019bacterial}
T.~Bhattacharjee and S.~S. Datta, \emph{Nature Communications}, 2019,
  \textbf{10}, 2075\relax
\mciteBstWouldAddEndPuncttrue
\mciteSetBstMidEndSepPunct{\mcitedefaultmidpunct}
{\mcitedefaultendpunct}{\mcitedefaultseppunct}\relax
\EndOfBibitem
\bibitem[Amchin \emph{et~al.}(2022)Amchin, Ott, Bhattacharjee, and
  Datta]{amchin2022influence}
D.~B. Amchin, J.~A. Ott, T.~Bhattacharjee and S.~S. Datta, \emph{PLoS
  Computational Biology}, 2022, \textbf{18}, e1010063\relax
\mciteBstWouldAddEndPuncttrue
\mciteSetBstMidEndSepPunct{\mcitedefaultmidpunct}
{\mcitedefaultendpunct}{\mcitedefaultseppunct}\relax
\EndOfBibitem
\bibitem[Li \emph{et~al.}(2021)Li, Meng, Primkulov, and Juanes]{li2021}
W.~Li, Y.~Meng, B.~K. Primkulov and R.~Juanes, \emph{Physical Review Applied},
  2021, \textbf{16}, 024043\relax
\mciteBstWouldAddEndPuncttrue
\mciteSetBstMidEndSepPunct{\mcitedefaultmidpunct}
{\mcitedefaultendpunct}{\mcitedefaultseppunct}\relax
\EndOfBibitem
\bibitem[Frocht(1941)]{frocht1941}
M.~M. Frocht, \emph{Photoelasticity}, John Wiley \& Sons, 1941\relax
\mciteBstWouldAddEndPuncttrue
\mciteSetBstMidEndSepPunct{\mcitedefaultmidpunct}
{\mcitedefaultendpunct}{\mcitedefaultseppunct}\relax
\EndOfBibitem
\bibitem[{Abed Zadeh} \emph{et~al.}(2019){Abed Zadeh}, Bar{\'{e}}s, Brzinski,
  Daniels, Dijksman, Docquier, Everitt, Kollmer, Lantsoght, Wang, Workamp,
  Zhao, and Zheng]{AbedZadeh2019}
A.~{Abed Zadeh}, J.~Bar{\'{e}}s, T.~A. Brzinski, K.~E. Daniels, J.~Dijksman,
  N.~Docquier, H.~O. Everitt, J.~E. Kollmer, O.~Lantsoght, D.~Wang, M.~Workamp,
  Y.~Zhao and H.~Zheng, \emph{Granular Matter}, 2019, \textbf{21}, 83\relax
\mciteBstWouldAddEndPuncttrue
\mciteSetBstMidEndSepPunct{\mcitedefaultmidpunct}
{\mcitedefaultendpunct}{\mcitedefaultseppunct}\relax
\EndOfBibitem
\bibitem[Daniels \emph{et~al.}(2017)Daniels, Kollmer, and Puckett]{Daniels2017}
K.~E. Daniels, J.~E. Kollmer and J.~G. Puckett, \emph{Review of Scientific
  Instruments}, 2017, \textbf{88}, 051808\relax
\mciteBstWouldAddEndPuncttrue
\mciteSetBstMidEndSepPunct{\mcitedefaultmidpunct}
{\mcitedefaultendpunct}{\mcitedefaultseppunct}\relax
\EndOfBibitem
\bibitem[Majmudar and Behringer(2005)]{majmudar2005}
T.~S. Majmudar and R.~P. Behringer, \emph{Nature}, 2005, \textbf{435},
  1079--1082\relax
\mciteBstWouldAddEndPuncttrue
\mciteSetBstMidEndSepPunct{\mcitedefaultmidpunct}
{\mcitedefaultendpunct}{\mcitedefaultseppunct}\relax
\EndOfBibitem
\bibitem[Biot(1941)]{Biot1941}
M.~A. Biot, \emph{Journal of Applied Physics}, 1941, \textbf{12},
  155--164\relax
\mciteBstWouldAddEndPuncttrue
\mciteSetBstMidEndSepPunct{\mcitedefaultmidpunct}
{\mcitedefaultendpunct}{\mcitedefaultseppunct}\relax
\EndOfBibitem
\bibitem[Juanes \emph{et~al.}(2020)Juanes, Meng, and Primkulov]{juanes2020}
R.~Juanes, Y.~Meng and B.~K. Primkulov, \emph{Physical Review Fluids}, 2020,
  \textbf{5}, 110516\relax
\mciteBstWouldAddEndPuncttrue
\mciteSetBstMidEndSepPunct{\mcitedefaultmidpunct}
{\mcitedefaultendpunct}{\mcitedefaultseppunct}\relax
\EndOfBibitem
\bibitem[Terzaghi(1925)]{terzaghi1925}
K.~Terzaghi, \emph{Erdbaumechanik auf Bodenphysikalischer Grundlage}, F.
  Deuticke, 1925\relax
\mciteBstWouldAddEndPuncttrue
\mciteSetBstMidEndSepPunct{\mcitedefaultmidpunct}
{\mcitedefaultendpunct}{\mcitedefaultseppunct}\relax
\EndOfBibitem
\bibitem[Terzaghi(1943)]{terzaghi1943}
K.~Terzaghi, \emph{Theoretical Soil Mechanics}, John Wiley \& Sons, Ltd, 1943,
  pp. 265--296\relax
\mciteBstWouldAddEndPuncttrue
\mciteSetBstMidEndSepPunct{\mcitedefaultmidpunct}
{\mcitedefaultendpunct}{\mcitedefaultseppunct}\relax
\EndOfBibitem
\bibitem[Palmer(2017)]{palmer2017}
J.~Palmer, \emph{Nature}, 2017, \textbf{548}, 384--386\relax
\mciteBstWouldAddEndPuncttrue
\mciteSetBstMidEndSepPunct{\mcitedefaultmidpunct}
{\mcitedefaultendpunct}{\mcitedefaultseppunct}\relax
\EndOfBibitem
\bibitem[Skarke \emph{et~al.}(2014)Skarke, Ruppel, Kodis, Brothers, and
  Lobecker]{skarke2014}
A.~Skarke, C.~Ruppel, M.~Kodis, D.~Brothers and E.~Lobecker, \emph{Nature
  Geoscience}, 2014, \textbf{7}, 657\relax
\mciteBstWouldAddEndPuncttrue
\mciteSetBstMidEndSepPunct{\mcitedefaultmidpunct}
{\mcitedefaultendpunct}{\mcitedefaultseppunct}\relax
\EndOfBibitem
\bibitem[Guglielmi \emph{et~al.}(2015)Guglielmi, Cappa, Avouac, Henry, and
  Elsworth]{guglielmi2015}
Y.~Guglielmi, F.~Cappa, J.-P. Avouac, P.~Henry and D.~Elsworth, \emph{Science},
  2015, \textbf{348}, 1224--1226\relax
\mciteBstWouldAddEndPuncttrue
\mciteSetBstMidEndSepPunct{\mcitedefaultmidpunct}
{\mcitedefaultendpunct}{\mcitedefaultseppunct}\relax
\EndOfBibitem
\bibitem[Frocht and Guernsey(1952)]{frocht1952}
M.~M. Frocht and R.~Guernsey, \emph{{A special investigation to develop a
  general method for three-dimensional photoelastic stress analysis}}, National
  {A}dvisory {C}ommittee for {A}eronautics technical report, 1952\relax
\mciteBstWouldAddEndPuncttrue
\mciteSetBstMidEndSepPunct{\mcitedefaultmidpunct}
{\mcitedefaultendpunct}{\mcitedefaultseppunct}\relax
\EndOfBibitem
\bibitem[Li and Juanes(2023)]{li2023}
W.~Li and R.~Juanes, 2023\relax
\mciteBstWouldAddEndPuncttrue
\mciteSetBstMidEndSepPunct{\mcitedefaultmidpunct}
{\mcitedefaultendpunct}{\mcitedefaultseppunct}\relax
\EndOfBibitem
\bibitem[Ferdowsi and Rubin(2020)]{FerdowsiRubin2020}
B.~Ferdowsi and A.~M. Rubin, \emph{Journal of Geophysical Research: Solid
  Earth}, 2020, \textbf{125}, e2019JB019016\relax
\mciteBstWouldAddEndPuncttrue
\mciteSetBstMidEndSepPunct{\mcitedefaultmidpunct}
{\mcitedefaultendpunct}{\mcitedefaultseppunct}\relax
\EndOfBibitem
\bibitem[Bhattacharya \emph{et~al.}(2015)Bhattacharya, Rubin, Bayart, Savage,
  and Marone]{Bhattacharya2015}
P.~Bhattacharya, A.~M. Rubin, E.~Bayart, H.~M. Savage and C.~Marone,
  \emph{Journal of Geophysical Research: Solid Earth}, 2015, \textbf{120},
  6365--6385\relax
\mciteBstWouldAddEndPuncttrue
\mciteSetBstMidEndSepPunct{\mcitedefaultmidpunct}
{\mcitedefaultendpunct}{\mcitedefaultseppunct}\relax
\EndOfBibitem
\bibitem[Sammis and Biegel(1989)]{sammis1989fractals}
C.~G. Sammis and R.~L. Biegel, \emph{PAGEOPH}, 1989, \textbf{131},
  255--271\relax
\mciteBstWouldAddEndPuncttrue
\mciteSetBstMidEndSepPunct{\mcitedefaultmidpunct}
{\mcitedefaultendpunct}{\mcitedefaultseppunct}\relax
\EndOfBibitem
\bibitem[Marone(1998)]{marone1998laboratory}
C.~Marone, \emph{Annual Review of Earth and Planetary Sciences}, 1998,
  \textbf{26}, 643--696\relax
\mciteBstWouldAddEndPuncttrue
\mciteSetBstMidEndSepPunct{\mcitedefaultmidpunct}
{\mcitedefaultendpunct}{\mcitedefaultseppunct}\relax
\EndOfBibitem
\bibitem[Bhattacharya \emph{et~al.}(2022)Bhattacharya, Rubin, Tullis, Beeler,
  and Okazaki]{bhattacharya2022evolution}
P.~Bhattacharya, A.~M. Rubin, T.~E. Tullis, N.~M. Beeler and K.~Okazaki,
  \emph{Proceedings of the National Academy of Sciences}, 2022, \textbf{119},
  e2119462119\relax
\mciteBstWouldAddEndPuncttrue
\mciteSetBstMidEndSepPunct{\mcitedefaultmidpunct}
{\mcitedefaultendpunct}{\mcitedefaultseppunct}\relax
\EndOfBibitem
\bibitem[Daub and Carlson(2010)]{daub2010friction}
E.~G. Daub and J.~M. Carlson, \emph{Annual Review of Condensed Matter Physics},
  2010, \textbf{1}, 397--418\relax
\mciteBstWouldAddEndPuncttrue
\mciteSetBstMidEndSepPunct{\mcitedefaultmidpunct}
{\mcitedefaultendpunct}{\mcitedefaultseppunct}\relax
\EndOfBibitem
\bibitem[Rice(2006)]{rice2006heating}
J.~R. Rice, \emph{Journal of Geophysical Research: Solid Earth}, 2006,
  \textbf{111}, B05311\relax
\mciteBstWouldAddEndPuncttrue
\mciteSetBstMidEndSepPunct{\mcitedefaultmidpunct}
{\mcitedefaultendpunct}{\mcitedefaultseppunct}\relax
\EndOfBibitem
\bibitem[Rempel(2006)]{Rempel2006}
A.~W. Rempel, \emph{Earthquakes: Radiated Energy and the Physics of Faulting},
  American Geophysical Union, 2006, pp. 263--270\relax
\mciteBstWouldAddEndPuncttrue
\mciteSetBstMidEndSepPunct{\mcitedefaultmidpunct}
{\mcitedefaultendpunct}{\mcitedefaultseppunct}\relax
\EndOfBibitem
\bibitem[Kato and Tullis(2001)]{kato2001composite}
N.~Kato and T.~E. Tullis, \emph{Geophysical Research Letters}, 2001,
  \textbf{28}, 1103--1106\relax
\mciteBstWouldAddEndPuncttrue
\mciteSetBstMidEndSepPunct{\mcitedefaultmidpunct}
{\mcitedefaultendpunct}{\mcitedefaultseppunct}\relax
\EndOfBibitem
\bibitem[Ampuero and Rubin(2008)]{AmpueroRubin2008}
J.-P. Ampuero and A.~M. Rubin, \emph{Journal of Geophysical Research: Solid
  Earth}, 2008, \textbf{113}, \relax
\mciteBstWouldAddEndPuncttrue
\mciteSetBstMidEndSepPunct{\mcitedefaultmidpunct}
{\mcitedefaultendpunct}{\mcitedefaultseppunct}\relax
\EndOfBibitem
\bibitem[Ruina(1983)]{ruina1983slip}
A.~Ruina, \emph{Journal of Geophysical Research: Solid Earth}, 1983,
  \textbf{88}, 10359--10370\relax
\mciteBstWouldAddEndPuncttrue
\mciteSetBstMidEndSepPunct{\mcitedefaultmidpunct}
{\mcitedefaultendpunct}{\mcitedefaultseppunct}\relax
\EndOfBibitem
\bibitem[Rice \emph{et~al.}(2001)Rice, Lapusta, and Ranjith]{rice2001rate}
J.~R. Rice, N.~Lapusta and K.~Ranjith, \emph{Journal of the Mechanics and
  Physics of Solids}, 2001, \textbf{49}, 1865--1898\relax
\mciteBstWouldAddEndPuncttrue
\mciteSetBstMidEndSepPunct{\mcitedefaultmidpunct}
{\mcitedefaultendpunct}{\mcitedefaultseppunct}\relax
\EndOfBibitem
\bibitem[Kim and Kamrin(2020)]{kim2020power}
S.~Kim and K.~Kamrin, \emph{Physical Review Letters}, 2020, \textbf{125},
  088002\relax
\mciteBstWouldAddEndPuncttrue
\mciteSetBstMidEndSepPunct{\mcitedefaultmidpunct}
{\mcitedefaultendpunct}{\mcitedefaultseppunct}\relax
\EndOfBibitem
\bibitem[Jop(2015)]{jop2015rheological}
P.~Jop, \emph{Comptes Rendus Physique}, 2015, \textbf{16}, 62--72\relax
\mciteBstWouldAddEndPuncttrue
\mciteSetBstMidEndSepPunct{\mcitedefaultmidpunct}
{\mcitedefaultendpunct}{\mcitedefaultseppunct}\relax
\EndOfBibitem
\bibitem[Henann and Kamrin(2013)]{henann2013predictive}
D.~L. Henann and K.~Kamrin, \emph{Proceedings of the National Academy of
  Sciences}, 2013, \textbf{110}, 6730--6735\relax
\mciteBstWouldAddEndPuncttrue
\mciteSetBstMidEndSepPunct{\mcitedefaultmidpunct}
{\mcitedefaultendpunct}{\mcitedefaultseppunct}\relax
\EndOfBibitem
\bibitem[Forterre and Pouliquen(2008)]{forterre2008flows}
Y.~Forterre and O.~Pouliquen, \emph{Annual Review of Fluid Mechanics}, 2008,
  \textbf{40}, 1--24\relax
\mciteBstWouldAddEndPuncttrue
\mciteSetBstMidEndSepPunct{\mcitedefaultmidpunct}
{\mcitedefaultendpunct}{\mcitedefaultseppunct}\relax
\EndOfBibitem
\bibitem[Lois \emph{et~al.}(2005)Lois, Lema{\^\i}tre, and
  Carlson]{lois2005numerical}
G.~Lois, A.~Lema{\^\i}tre and J.~M. Carlson, \emph{Physical Review E}, 2005,
  \textbf{72}, 051303\relax
\mciteBstWouldAddEndPuncttrue
\mciteSetBstMidEndSepPunct{\mcitedefaultmidpunct}
{\mcitedefaultendpunct}{\mcitedefaultseppunct}\relax
\EndOfBibitem
\bibitem[Silbert \emph{et~al.}(2001)Silbert, Erta{\c{s}}, Grest, Halsey,
  Levine, and Plimpton]{silbert2001granular}
L.~E. Silbert, D.~Erta{\c{s}}, G.~S. Grest, T.~C. Halsey, D.~Levine and S.~J.
  Plimpton, \emph{Physical Review E}, 2001, \textbf{64}, 051302\relax
\mciteBstWouldAddEndPuncttrue
\mciteSetBstMidEndSepPunct{\mcitedefaultmidpunct}
{\mcitedefaultendpunct}{\mcitedefaultseppunct}\relax
\EndOfBibitem
\bibitem[Thompson and Grest(1991)]{thompson1991granular}
P.~A. Thompson and G.~S. Grest, \emph{Physical Review Letters}, 1991,
  \textbf{67}, 1751\relax
\mciteBstWouldAddEndPuncttrue
\mciteSetBstMidEndSepPunct{\mcitedefaultmidpunct}
{\mcitedefaultendpunct}{\mcitedefaultseppunct}\relax
\EndOfBibitem
\bibitem[Ghezzehei and Or(2001)]{Ghezzehei2001}
T.~A. Ghezzehei and D.~Or, \emph{Soil Science Society of America Journal},
  2001, \textbf{65}, 624--637\relax
\mciteBstWouldAddEndPuncttrue
\mciteSetBstMidEndSepPunct{\mcitedefaultmidpunct}
{\mcitedefaultendpunct}{\mcitedefaultseppunct}\relax
\EndOfBibitem
\bibitem[Grabowski \emph{et~al.}(2011)Grabowski, Droppo, and
  Wharton]{Grabowski2011}
R.~C. Grabowski, I.~G. Droppo and G.~Wharton, \emph{Earth-Science Reviews},
  2011, \textbf{105}, 101--120\relax
\mciteBstWouldAddEndPuncttrue
\mciteSetBstMidEndSepPunct{\mcitedefaultmidpunct}
{\mcitedefaultendpunct}{\mcitedefaultseppunct}\relax
\EndOfBibitem
\bibitem[Carrillo and Bourg(2019)]{Carrillo2019}
F.~J. Carrillo and I.~C. Bourg, \emph{Water Resources Research}, 2019,
  \textbf{55}, 8096--8121\relax
\mciteBstWouldAddEndPuncttrue
\mciteSetBstMidEndSepPunct{\mcitedefaultmidpunct}
{\mcitedefaultendpunct}{\mcitedefaultseppunct}\relax
\EndOfBibitem
\bibitem[Kleber \emph{et~al.}(2021)Kleber, Bourg, Coward, Hansel, Myneni, and
  Nunan]{Kleber2021}
M.~Kleber, I.~C. Bourg, E.~K. Coward, C.~M. Hansel, S.~C.~B. Myneni and
  N.~Nunan, \emph{Nature Reviews Earth and Environment}, 2021, \textbf{2},
  402--421\relax
\mciteBstWouldAddEndPuncttrue
\mciteSetBstMidEndSepPunct{\mcitedefaultmidpunct}
{\mcitedefaultendpunct}{\mcitedefaultseppunct}\relax
\EndOfBibitem
\bibitem[Bourg and Ajo-Franklin(2017)]{Bourg2017}
I.~C. Bourg and J.~B. Ajo-Franklin, \emph{Accounts of Chemical Research}, 2017,
  \textbf{50}, 2067--2074\relax
\mciteBstWouldAddEndPuncttrue
\mciteSetBstMidEndSepPunct{\mcitedefaultmidpunct}
{\mcitedefaultendpunct}{\mcitedefaultseppunct}\relax
\EndOfBibitem
\bibitem[Neuzil(2019)]{Neuzil2019}
C.~E. Neuzil, \emph{Annual Review of Earth and Planetary Sciences}, 2019,
  \textbf{47}, 247--273\relax
\mciteBstWouldAddEndPuncttrue
\mciteSetBstMidEndSepPunct{\mcitedefaultmidpunct}
{\mcitedefaultendpunct}{\mcitedefaultseppunct}\relax
\EndOfBibitem
\bibitem[Iverson(1997)]{Iverson1997}
R.~M. Iverson, \emph{Reviews of Geophysics}, 1997, \textbf{35}, 245--296\relax
\mciteBstWouldAddEndPuncttrue
\mciteSetBstMidEndSepPunct{\mcitedefaultmidpunct}
{\mcitedefaultendpunct}{\mcitedefaultseppunct}\relax
\EndOfBibitem
\bibitem[Kaitna \emph{et~al.}(2016)Kaitna, Palucis, Yohannes, Hill, and
  Dietrich]{Kaitna2016}
R.~Kaitna, M.~C. Palucis, B.~Yohannes, K.~M. Hill and W.~E. Dietrich,
  \emph{Journal of Geophysical Research: Earth Surface}, 2016, \textbf{121},
  415--441\relax
\mciteBstWouldAddEndPuncttrue
\mciteSetBstMidEndSepPunct{\mcitedefaultmidpunct}
{\mcitedefaultendpunct}{\mcitedefaultseppunct}\relax
\EndOfBibitem
\bibitem[Koos and Willenbacher(2011)]{Koos2011}
E.~Koos and N.~Willenbacher, \emph{Science}, 2011, \textbf{331}, 897--900\relax
\mciteBstWouldAddEndPuncttrue
\mciteSetBstMidEndSepPunct{\mcitedefaultmidpunct}
{\mcitedefaultendpunct}{\mcitedefaultseppunct}\relax
\EndOfBibitem
\bibitem[Underwood and Bourg(2020)]{Underwood2020}
T.~R. Underwood and I.~C. Bourg, \emph{Journal of Physical Chemistry C}, 2020,
  \textbf{124}, 3702--3714\relax
\mciteBstWouldAddEndPuncttrue
\mciteSetBstMidEndSepPunct{\mcitedefaultmidpunct}
{\mcitedefaultendpunct}{\mcitedefaultseppunct}\relax
\EndOfBibitem
\bibitem[Shen and Bourg(2021)]{Shen2021}
X.~Shen and I.~C. Bourg, \emph{Journal of Colloid and Interface Science}, 2021,
  \textbf{584}, 610--621\relax
\mciteBstWouldAddEndPuncttrue
\mciteSetBstMidEndSepPunct{\mcitedefaultmidpunct}
{\mcitedefaultendpunct}{\mcitedefaultseppunct}\relax
\EndOfBibitem
\bibitem[Ancey(2007)]{Ancey2007}
C.~Ancey, \emph{Journal of Non-Newtonian Fluid Mechanics}, 2007, \textbf{142},
  4--35\relax
\mciteBstWouldAddEndPuncttrue
\mciteSetBstMidEndSepPunct{\mcitedefaultmidpunct}
{\mcitedefaultendpunct}{\mcitedefaultseppunct}\relax
\EndOfBibitem
\bibitem[Seiphoori \emph{et~al.}(2020)Seiphoori, Ma, Arratia, and
  Jerolmack]{Seiphoori2020}
A.~Seiphoori, X.-g. Ma, P.~E. Arratia and D.~J. Jerolmack, \emph{Proceedings of
  the National Academy of Sciences}, 2020, \textbf{117}, 2275--2281\relax
\mciteBstWouldAddEndPuncttrue
\mciteSetBstMidEndSepPunct{\mcitedefaultmidpunct}
{\mcitedefaultendpunct}{\mcitedefaultseppunct}\relax
\EndOfBibitem
\bibitem[Dash \emph{et~al.}(2006)Dash, Rempel, and Wettlaufer]{Dash2006}
J.~G. Dash, A.~W. Rempel and J.~S. Wettlaufer, \emph{Reviews of Modern
  Physics}, 2006, \textbf{78}, 695--741\relax
\mciteBstWouldAddEndPuncttrue
\mciteSetBstMidEndSepPunct{\mcitedefaultmidpunct}
{\mcitedefaultendpunct}{\mcitedefaultseppunct}\relax
\EndOfBibitem
\bibitem[Jin \emph{et~al.}(2020)Jin, Yang, Gao, Zhao, Li, and Jiang]{Jin2020}
X.~Jin, W.~Yang, X.~Gao, J.-Q. Zhao, Z.~Li and J.~Jiang, \emph{Water Resources
  Research}, 2020, \textbf{56}, e2020WR027482\relax
\mciteBstWouldAddEndPuncttrue
\mciteSetBstMidEndSepPunct{\mcitedefaultmidpunct}
{\mcitedefaultendpunct}{\mcitedefaultseppunct}\relax
\EndOfBibitem
\bibitem[Gratier \emph{et~al.}(2013)Gratier, Dysthe, and Renard]{Gratier2013}
J.-P. Gratier, D.~K. Dysthe and F.~Renard, \emph{Advances in Geophysics}, 2013,
  \textbf{54}, 47--179\relax
\mciteBstWouldAddEndPuncttrue
\mciteSetBstMidEndSepPunct{\mcitedefaultmidpunct}
{\mcitedefaultendpunct}{\mcitedefaultseppunct}\relax
\EndOfBibitem
\bibitem[Bourg(2015)]{Bourg2015}
I.~C. Bourg, \emph{Environmental Science \& Technology Letters}, 2015,
  \textbf{20}, 255--259\relax
\mciteBstWouldAddEndPuncttrue
\mciteSetBstMidEndSepPunct{\mcitedefaultmidpunct}
{\mcitedefaultendpunct}{\mcitedefaultseppunct}\relax
\EndOfBibitem
\bibitem[Vachier and Wettlaufer(2022)]{Vachier_Wettlaufer_2022}
J.~Vachier and J.~S. Wettlaufer, \emph{Physical Review E}, 2022, \textbf{105},
  024601\relax
\mciteBstWouldAddEndPuncttrue
\mciteSetBstMidEndSepPunct{\mcitedefaultmidpunct}
{\mcitedefaultendpunct}{\mcitedefaultseppunct}\relax
\EndOfBibitem
\bibitem[Wettlaufer(2019)]{Wettlaufer2019}
J.~S. Wettlaufer, \emph{Philosophical Transactions: Mathematical, Physical and
  Engineering Sciences}, 2019, \textbf{377}, pp. 1--17\relax
\mciteBstWouldAddEndPuncttrue
\mciteSetBstMidEndSepPunct{\mcitedefaultmidpunct}
{\mcitedefaultendpunct}{\mcitedefaultseppunct}\relax
\EndOfBibitem
\bibitem[Marath and Wettlaufer(2020)]{Navaneeth_Wettlaufer_2020}
N.~K. Marath and J.~S. Wettlaufer, \emph{Soft Matter}, 2020, \textbf{16},
  5886--5891\relax
\mciteBstWouldAddEndPuncttrue
\mciteSetBstMidEndSepPunct{\mcitedefaultmidpunct}
{\mcitedefaultendpunct}{\mcitedefaultseppunct}\relax
\EndOfBibitem
\bibitem[Glaser(2012)]{glaserBiophysicsIntroduction2012}
R.~Glaser, \emph{Biophysics: {{An Introduction}}}, {Springer Science \&
  Business Media}, 2012\relax
\mciteBstWouldAddEndPuncttrue
\mciteSetBstMidEndSepPunct{\mcitedefaultmidpunct}
{\mcitedefaultendpunct}{\mcitedefaultseppunct}\relax
\EndOfBibitem
\bibitem[Gupta(2011)]{guptaEncyclopediaSolidEarth2011}
H.~Gupta, \emph{Encyclopedia of {{Solid Earth Geophysics}}}, {Springer Science
  \& Business Media}, 2011\relax
\mciteBstWouldAddEndPuncttrue
\mciteSetBstMidEndSepPunct{\mcitedefaultmidpunct}
{\mcitedefaultendpunct}{\mcitedefaultseppunct}\relax
\EndOfBibitem
\bibitem[Pedlosky(2013)]{pedloskyGeophysicalFluidDynamics2013}
J.~Pedlosky, \emph{Geophysical {{Fluid Dynamics}}}, {Springer Science \&
  Business Media}, 2013\relax
\mciteBstWouldAddEndPuncttrue
\mciteSetBstMidEndSepPunct{\mcitedefaultmidpunct}
{\mcitedefaultendpunct}{\mcitedefaultseppunct}\relax
\EndOfBibitem
\bibitem[Jerolmack and Daniels(2019)]{jerolmackViewingEarthSurface2019}
D.~J. Jerolmack and K.~E. Daniels, \emph{Nature Reviews Physics}, 2019,
  \textbf{1}, 716--730\relax
\mciteBstWouldAddEndPuncttrue
\mciteSetBstMidEndSepPunct{\mcitedefaultmidpunct}
{\mcitedefaultendpunct}{\mcitedefaultseppunct}\relax
\EndOfBibitem
\bibitem[Regmi \emph{et~al.}(2015)Regmi, Giardino, McDonald, and
  Vitek]{regmiChapter11Review2015}
N.~R. Regmi, J.~R. Giardino, E.~V. McDonald and J.~D. Vitek, \emph{Developments
  in {{Earth Surface Processes}}}, {Elsevier}, 2015, vol.~19, pp.
  319--362\relax
\mciteBstWouldAddEndPuncttrue
\mciteSetBstMidEndSepPunct{\mcitedefaultmidpunct}
{\mcitedefaultendpunct}{\mcitedefaultseppunct}\relax
\EndOfBibitem
\bibitem[Hubbert \emph{et~al.}(2012)Hubbert, Wohlgemuth, Beyers, Narog, and
  Gerrard]{hubbertPostFireSoilWater2012}
K.~R. Hubbert, P.~M. Wohlgemuth, J.~L. Beyers, M.~G. Narog and R.~Gerrard,
  \emph{Fire Ecology}, 2012, \textbf{8}, 143--162\relax
\mciteBstWouldAddEndPuncttrue
\mciteSetBstMidEndSepPunct{\mcitedefaultmidpunct}
{\mcitedefaultendpunct}{\mcitedefaultseppunct}\relax
\EndOfBibitem
\bibitem[Gabet(2003)]{gabetPostfireThinDebris2003}
E.~J. Gabet, \emph{Earth Surface Processes and Landforms}, 2003, \textbf{28},
  1341--1348\relax
\mciteBstWouldAddEndPuncttrue
\mciteSetBstMidEndSepPunct{\mcitedefaultmidpunct}
{\mcitedefaultendpunct}{\mcitedefaultseppunct}\relax
\EndOfBibitem
\bibitem[Alessio \emph{et~al.}(2021)Alessio, Dunne, and
  Morell]{alessioPostWildfireGenerationDebrisFlow2021}
P.~Alessio, T.~Dunne and K.~Morell, \emph{Journal of Geophysical Research:
  Earth Surface}, 2021, \textbf{126}, e2021JF006108\relax
\mciteBstWouldAddEndPuncttrue
\mciteSetBstMidEndSepPunct{\mcitedefaultmidpunct}
{\mcitedefaultendpunct}{\mcitedefaultseppunct}\relax
\EndOfBibitem
\bibitem[Shen \emph{et~al.}(2018)Shen, Zhang, Chen, and
  Fan]{shenEDDAIntegratedSimulation2018}
P.~Shen, L.~Zhang, H.~Chen and R.~Fan, \emph{Geoscientific Model Development},
  2018, \textbf{11}, 2841--2856\relax
\mciteBstWouldAddEndPuncttrue
\mciteSetBstMidEndSepPunct{\mcitedefaultmidpunct}
{\mcitedefaultendpunct}{\mcitedefaultseppunct}\relax
\EndOfBibitem
\bibitem[Cannon \emph{et~al.}(2001)Cannon, Bigio, and
  Mine]{cannonProcessFirerelatedDebris2001}
S.~H. Cannon, E.~R. Bigio and E.~Mine, \emph{Hydrological Processes}, 2001,
  \textbf{15}, 3011--3023\relax
\mciteBstWouldAddEndPuncttrue
\mciteSetBstMidEndSepPunct{\mcitedefaultmidpunct}
{\mcitedefaultendpunct}{\mcitedefaultseppunct}\relax
\EndOfBibitem
\bibitem[Schippa(2018)]{schippaEffectsSedimentSize2018}
L.~Schippa, \emph{Granularity in {{Materials Science}}}, {IntechOpen},
  2018\relax
\mciteBstWouldAddEndPuncttrue
\mciteSetBstMidEndSepPunct{\mcitedefaultmidpunct}
{\mcitedefaultendpunct}{\mcitedefaultseppunct}\relax
\EndOfBibitem
\bibitem[Sosio and Crosta(2009)]{sosioRheologyConcentratedGranular2009}
R.~Sosio and G.~B. Crosta, \emph{Water Resources Research}, 2009, \textbf{45},
  W03412\relax
\mciteBstWouldAddEndPuncttrue
\mciteSetBstMidEndSepPunct{\mcitedefaultmidpunct}
{\mcitedefaultendpunct}{\mcitedefaultseppunct}\relax
\EndOfBibitem
\bibitem[Parsons \emph{et~al.}(2001)Parsons, Whipple, and
  Simoni]{parsonsExperimentalStudyGrain2001}
J.~D. Parsons, K.~X. Whipple and A.~Simoni, \emph{The Journal of Geology},
  2001, \textbf{109}, 427--447\relax
\mciteBstWouldAddEndPuncttrue
\mciteSetBstMidEndSepPunct{\mcitedefaultmidpunct}
{\mcitedefaultendpunct}{\mcitedefaultseppunct}\relax
\EndOfBibitem
\bibitem[Kostynick \emph{et~al.}(2022)Kostynick, Matinpour, Pradeep, Haber,
  Sauret, Meiburg, Dunne, Arratia, and
  Jerolmack]{kostynickRheologyDebrisFlow2022}
R.~Kostynick, H.~Matinpour, S.~Pradeep, S.~Haber, A.~Sauret, E.~Meiburg,
  T.~Dunne, P.~Arratia and D.~Jerolmack, \emph{Proceedings of the National
  Academy of Sciences}, 2022, \textbf{119}, e2209109119\relax
\mciteBstWouldAddEndPuncttrue
\mciteSetBstMidEndSepPunct{\mcitedefaultmidpunct}
{\mcitedefaultendpunct}{\mcitedefaultseppunct}\relax
\EndOfBibitem
\bibitem[Baker \emph{et~al.}(2016)Baker, Gray, and
  Kokelaar]{bakerParticleSizeSegregationSpontaneous2016}
J.~Baker, N.~Gray and P.~Kokelaar, \emph{International Journal of Erosion
  Control Engineering}, 2016, \textbf{9}, 174--178\relax
\mciteBstWouldAddEndPuncttrue
\mciteSetBstMidEndSepPunct{\mcitedefaultmidpunct}
{\mcitedefaultendpunct}{\mcitedefaultseppunct}\relax
\EndOfBibitem
\bibitem[Kean \emph{et~al.}(2019)Kean, Staley, Lancaster, Rengers, Swanson,
  Coe, Hernandez, Sigman, Allstadt, and
  Lindsay]{keanInundationFlowDynamics2019}
J.~Kean, D.~Staley, J.~Lancaster, F.~Rengers, B.~Swanson, J.~Coe, J.~Hernandez,
  A.~Sigman, K.~Allstadt and D.~Lindsay, \emph{Geosphere}, 2019, \textbf{15},
  1140--1163\relax
\mciteBstWouldAddEndPuncttrue
\mciteSetBstMidEndSepPunct{\mcitedefaultmidpunct}
{\mcitedefaultendpunct}{\mcitedefaultseppunct}\relax
\EndOfBibitem
\bibitem[Eyles and Ho(1970)]{eylesSoilCreepHumid1970}
R.~J. Eyles and R.~Ho, \emph{Journal of Tropical Geography}, 1970, \textbf{31},
  40--42\relax
\mciteBstWouldAddEndPuncttrue
\mciteSetBstMidEndSepPunct{\mcitedefaultmidpunct}
{\mcitedefaultendpunct}{\mcitedefaultseppunct}\relax
\EndOfBibitem
\bibitem[Fleming and Johnson(1975)]{flemingRatesSeasonalCreep1975}
R.~W. Fleming and A.~M. Johnson, \emph{Quarterly Journal of Engineering
  Geology}, 1975, \textbf{8}, 1--29\relax
\mciteBstWouldAddEndPuncttrue
\mciteSetBstMidEndSepPunct{\mcitedefaultmidpunct}
{\mcitedefaultendpunct}{\mcitedefaultseppunct}\relax
\EndOfBibitem
\bibitem[Auzet and Ambroise(1996)]{auzetSoilCreepDynamics1996}
A.-V. Auzet and B.~Ambroise, \emph{Earth Surface Processes and Landforms},
  1996, \textbf{21}, 531--542\relax
\mciteBstWouldAddEndPuncttrue
\mciteSetBstMidEndSepPunct{\mcitedefaultmidpunct}
{\mcitedefaultendpunct}{\mcitedefaultseppunct}\relax
\EndOfBibitem
\bibitem[Matsuoka(1998)]{matsuokaRelationshipFrostHeave1998}
N.~Matsuoka, \emph{Permafrost and Periglacial Processes}, 1998, \textbf{9},
  121--133\relax
\mciteBstWouldAddEndPuncttrue
\mciteSetBstMidEndSepPunct{\mcitedefaultmidpunct}
{\mcitedefaultendpunct}{\mcitedefaultseppunct}\relax
\EndOfBibitem
\bibitem[Houssais \emph{et~al.}(2015)Houssais, Ortiz, Durian, and
  Jerolmack]{houssaisOnsetSedimentTransport2015}
M.~Houssais, C.~P. Ortiz, D.~J. Durian and D.~J. Jerolmack, \emph{Nature
  Communications}, 2015, \textbf{6}, 1--8\relax
\mciteBstWouldAddEndPuncttrue
\mciteSetBstMidEndSepPunct{\mcitedefaultmidpunct}
{\mcitedefaultendpunct}{\mcitedefaultseppunct}\relax
\EndOfBibitem
\bibitem[Houssais \emph{et~al.}(2021)Houssais, Maldarelli, and
  Morris]{Houssais2021}
M.~Houssais, C.~Maldarelli and J.~F. Morris, \emph{Physical Review Fluids},
  2021, \textbf{6}, L012301\relax
\mciteBstWouldAddEndPuncttrue
\mciteSetBstMidEndSepPunct{\mcitedefaultmidpunct}
{\mcitedefaultendpunct}{\mcitedefaultseppunct}\relax
\EndOfBibitem
\bibitem[Cúñez \emph{et~al.}(2022)Cúñez, Franklin, Houssais, Arratia, and
  Jerolmack]{cunez_strain_2022}
F.~D. Cúñez, E.~M. Franklin, M.~Houssais, P.~Arratia and D.~J. Jerolmack,
  \emph{Physical Review Research}, 2022, \textbf{4}, L022055\relax
\mciteBstWouldAddEndPuncttrue
\mciteSetBstMidEndSepPunct{\mcitedefaultmidpunct}
{\mcitedefaultendpunct}{\mcitedefaultseppunct}\relax
\EndOfBibitem
\bibitem[Ferdowsi \emph{et~al.}(2018)Ferdowsi, Ortiz, and
  Jerolmack]{Ferdowsi-PNAS2018}
B.~Ferdowsi, C.~P. Ortiz and D.~J. Jerolmack, \emph{Proceedings of the National
  Academy of Sciences}, 2018, \textbf{115}, 4827--4832\relax
\mciteBstWouldAddEndPuncttrue
\mciteSetBstMidEndSepPunct{\mcitedefaultmidpunct}
{\mcitedefaultendpunct}{\mcitedefaultseppunct}\relax
\EndOfBibitem
\bibitem[Deshpande \emph{et~al.}(2021)Deshpande, Furbish, Arratia, and
  Jerolmack]{Deshpande2021}
N.~S. Deshpande, D.~J. Furbish, P.~E. Arratia and D.~J. Jerolmack, \emph{Nature
  Communications}, 2021, \textbf{12}, 3909\relax
\mciteBstWouldAddEndPuncttrue
\mciteSetBstMidEndSepPunct{\mcitedefaultmidpunct}
{\mcitedefaultendpunct}{\mcitedefaultseppunct}\relax
\EndOfBibitem
\bibitem[Jónsson \emph{et~al.}(2003)Jónsson, Segall, Pedersen, and
  Björnsson]{jonssonPostearthquakeGroundMovements2003}
S.~Jónsson, P.~Segall, R.~Pedersen and G.~Björnsson, \emph{Nature}, 2003,
  \textbf{424}, 179--183\relax
\mciteBstWouldAddEndPuncttrue
\mciteSetBstMidEndSepPunct{\mcitedefaultmidpunct}
{\mcitedefaultendpunct}{\mcitedefaultseppunct}\relax
\EndOfBibitem
\bibitem[Murphy \emph{et~al.}(2022)Murphy, Finnegan, and
  Oberle]{murphyVadoseZoneThickness2022}
C.~R. Murphy, N.~J. Finnegan and F.~K.~J. Oberle, \emph{Journal of Geophysical
  Research: Earth Surface}, 2022, \textbf{127}, e2021JF006415\relax
\mciteBstWouldAddEndPuncttrue
\mciteSetBstMidEndSepPunct{\mcitedefaultmidpunct}
{\mcitedefaultendpunct}{\mcitedefaultseppunct}\relax
\EndOfBibitem
\bibitem[Cascini \emph{et~al.}(2022)Cascini, Scoppettuolo, and
  Babilio]{Cascini2022}
L.~Cascini, M.~R. Scoppettuolo and E.~Babilio, \emph{Landslides}, 2022,
  2839--2851\relax
\mciteBstWouldAddEndPuncttrue
\mciteSetBstMidEndSepPunct{\mcitedefaultmidpunct}
{\mcitedefaultendpunct}{\mcitedefaultseppunct}\relax
\EndOfBibitem
\bibitem[Wyart(2005)]{Wyart2005}
M.~Wyart, \emph{Annales De Physique}, 2005, \textbf{30}, 1\relax
\mciteBstWouldAddEndPuncttrue
\mciteSetBstMidEndSepPunct{\mcitedefaultmidpunct}
{\mcitedefaultendpunct}{\mcitedefaultseppunct}\relax
\EndOfBibitem
\bibitem[Mao and Lubensky(2018)]{Mao2018}
X.~Mao and T.~C. Lubensky, \emph{Annual Review of Condensed Matter Physics},
  2018, \textbf{9}, 413--433\relax
\mciteBstWouldAddEndPuncttrue
\mciteSetBstMidEndSepPunct{\mcitedefaultmidpunct}
{\mcitedefaultendpunct}{\mcitedefaultseppunct}\relax
\EndOfBibitem
\bibitem[Liu \emph{et~al.}(2021)Liu, Kollmer, Daniels, Schwarz, and
  Henkes]{Liu2021}
K.~Liu, J.~E. Kollmer, K.~E. Daniels, J.~M. Schwarz and S.~Henkes,
  \emph{Physical Review Letters}, 2021, \textbf{126}, 088002\relax
\mciteBstWouldAddEndPuncttrue
\mciteSetBstMidEndSepPunct{\mcitedefaultmidpunct}
{\mcitedefaultendpunct}{\mcitedefaultseppunct}\relax
\EndOfBibitem
\bibitem[Brzinski and Daniels(2018)]{Brzinski2018}
T.~A. Brzinski and K.~E. Daniels, \emph{Physical Review Letters}, 2018,
  \textbf{120}, 218003\relax
\mciteBstWouldAddEndPuncttrue
\mciteSetBstMidEndSepPunct{\mcitedefaultmidpunct}
{\mcitedefaultendpunct}{\mcitedefaultseppunct}\relax
\EndOfBibitem
\bibitem[O'Hern \emph{et~al.}(2003)O'Hern, Silbert, Liu, and Nagel]{Ohern2003}
C.~S. O'Hern, L.~E. Silbert, A.~J. Liu and S.~R. Nagel, \emph{Physical Review
  E}, 2003, \textbf{68}, 011306\relax
\mciteBstWouldAddEndPuncttrue
\mciteSetBstMidEndSepPunct{\mcitedefaultmidpunct}
{\mcitedefaultendpunct}{\mcitedefaultseppunct}\relax
\EndOfBibitem
\bibitem[Berthier \emph{et~al.}(2019)Berthier, Porter, and
  Daniels]{Berthier2019}
E.~Berthier, M.~A. Porter and K.~E. Daniels, \emph{Proceedings of the National
  Academy of Sciences}, 2019, \textbf{116}, 16742--16749\relax
\mciteBstWouldAddEndPuncttrue
\mciteSetBstMidEndSepPunct{\mcitedefaultmidpunct}
{\mcitedefaultendpunct}{\mcitedefaultseppunct}\relax
\EndOfBibitem
\bibitem[Owens and Daniels(2013)]{Owens2013}
E.~T. Owens and K.~E. Daniels, \emph{Soft Matter}, 2013, \textbf{9},
  1214--1219\relax
\mciteBstWouldAddEndPuncttrue
\mciteSetBstMidEndSepPunct{\mcitedefaultmidpunct}
{\mcitedefaultendpunct}{\mcitedefaultseppunct}\relax
\EndOfBibitem
\bibitem[Tordesillas \emph{et~al.}(2018)Tordesillas, Zhou, and
  Batterham]{Tordesillas2018}
A.~Tordesillas, Z.~Zhou and R.~Batterham, \emph{Mechanics Research
  Communications}, 2018, \textbf{92}, 137--141\relax
\mciteBstWouldAddEndPuncttrue
\mciteSetBstMidEndSepPunct{\mcitedefaultmidpunct}
{\mcitedefaultendpunct}{\mcitedefaultseppunct}\relax
\EndOfBibitem
\bibitem[Cassotto \emph{et~al.}(2021)Cassotto, Burton, Amundson, Fahnestock,
  and Truffer]{cassotto2021granular}
R.~K. Cassotto, J.~C. Burton, J.~M. Amundson, M.~A. Fahnestock and M.~Truffer,
  \emph{Nature Geoscience}, 2021, \textbf{14}, 417--422\relax
\mciteBstWouldAddEndPuncttrue
\mciteSetBstMidEndSepPunct{\mcitedefaultmidpunct}
{\mcitedefaultendpunct}{\mcitedefaultseppunct}\relax
\EndOfBibitem
\bibitem[Deshpande and Crosby(2019)]{deshpande2019logjams}
N.~S. Deshpande and B.~T. Crosby, \emph{arXiv preprint arXiv:1911.01518},
  2019\relax
\mciteBstWouldAddEndPuncttrue
\mciteSetBstMidEndSepPunct{\mcitedefaultmidpunct}
{\mcitedefaultendpunct}{\mcitedefaultseppunct}\relax
\EndOfBibitem
\bibitem[Ettema(1990)]{ettema1990jam}
R.~Ettema, \emph{Journal of Hydraulic Research}, 1990, \textbf{28},
  673--684\relax
\mciteBstWouldAddEndPuncttrue
\mciteSetBstMidEndSepPunct{\mcitedefaultmidpunct}
{\mcitedefaultendpunct}{\mcitedefaultseppunct}\relax
\EndOfBibitem
\bibitem[Herman(2013)]{herman2013numerical}
A.~Herman, \emph{Annals of Glaciology}, 2013, \textbf{54}, 114--120\relax
\mciteBstWouldAddEndPuncttrue
\mciteSetBstMidEndSepPunct{\mcitedefaultmidpunct}
{\mcitedefaultendpunct}{\mcitedefaultseppunct}\relax
\EndOfBibitem
\bibitem[Jutzeler \emph{et~al.}(2014)Jutzeler, Marsh, Carey, White, Talling,
  and Karlstrom]{jutzeler2014fate}
M.~Jutzeler, R.~Marsh, R.~J. Carey, J.~D. White, P.~J. Talling and
  L.~Karlstrom, \emph{Nature Communications}, 2014, \textbf{5}, 1--10\relax
\mciteBstWouldAddEndPuncttrue
\mciteSetBstMidEndSepPunct{\mcitedefaultmidpunct}
{\mcitedefaultendpunct}{\mcitedefaultseppunct}\relax
\EndOfBibitem
\bibitem[Mlot \emph{et~al.}(2011)Mlot, Tovey, and Hu]{mlot2011fire}
N.~J. Mlot, C.~A. Tovey and D.~L. Hu, \emph{Proceedings of the National Academy
  of Sciences}, 2011, \textbf{108}, 7669--7673\relax
\mciteBstWouldAddEndPuncttrue
\mciteSetBstMidEndSepPunct{\mcitedefaultmidpunct}
{\mcitedefaultendpunct}{\mcitedefaultseppunct}\relax
\EndOfBibitem
\bibitem[Burton \emph{et~al.}(2018)Burton, Amundson, Cassotto, Kuo, and
  Dennin]{burton2018quantifying}
J.~C. Burton, J.~M. Amundson, R.~Cassotto, C.-C. Kuo and M.~Dennin,
  \emph{Proceedings of the National Academy of Sciences}, 2018, \textbf{115},
  5105--5110\relax
\mciteBstWouldAddEndPuncttrue
\mciteSetBstMidEndSepPunct{\mcitedefaultmidpunct}
{\mcitedefaultendpunct}{\mcitedefaultseppunct}\relax
\EndOfBibitem
\bibitem[Robel(2017)]{robel2017thinning}
A.~A. Robel, \emph{Nature Communications}, 2017, \textbf{8}, 1--7\relax
\mciteBstWouldAddEndPuncttrue
\mciteSetBstMidEndSepPunct{\mcitedefaultmidpunct}
{\mcitedefaultendpunct}{\mcitedefaultseppunct}\relax
\EndOfBibitem
\bibitem[Amundson and Burton(2018)]{amundson2018quasi}
J.~M. Amundson and J.~Burton, \emph{Journal of Geophysical Research: Earth
  Surface}, 2018, \textbf{123}, 2243--2257\relax
\mciteBstWouldAddEndPuncttrue
\mciteSetBstMidEndSepPunct{\mcitedefaultmidpunct}
{\mcitedefaultendpunct}{\mcitedefaultseppunct}\relax
\EndOfBibitem
\bibitem[Engelund and Fredsoe(1976)]{engelund_sediment_1976}
F.~Engelund and J.~Fredsoe, \emph{Hydrology Research}, 1976, \textbf{7},
  293–--306\relax
\mciteBstWouldAddEndPuncttrue
\mciteSetBstMidEndSepPunct{\mcitedefaultmidpunct}
{\mcitedefaultendpunct}{\mcitedefaultseppunct}\relax
\EndOfBibitem
\bibitem[Luque and Beek(1976)]{luque_erosion_1976}
R.~F. Luque and R.~V. Beek, \emph{Journal of Hydraulic Research}, 1976,
  \textbf{14}, 127--144\relax
\mciteBstWouldAddEndPuncttrue
\mciteSetBstMidEndSepPunct{\mcitedefaultmidpunct}
{\mcitedefaultendpunct}{\mcitedefaultseppunct}\relax
\EndOfBibitem
\bibitem[Meyer-Peter and M{\"u}ller(1948)]{Meyer-Peter1948}
E.~Meyer-Peter and R.~M{\"u}ller, IAHSR 2nd meeting, Stockholm, appendix 2,
  1948\relax
\mciteBstWouldAddEndPuncttrue
\mciteSetBstMidEndSepPunct{\mcitedefaultmidpunct}
{\mcitedefaultendpunct}{\mcitedefaultseppunct}\relax
\EndOfBibitem
\bibitem[Wong and Parker(2006)]{wong_reanalysis_2006}
M.~Wong and G.~Parker, \emph{Journal of Hydraulic Engineering}, 2006,
  \textbf{132}, 1159--1168\relax
\mciteBstWouldAddEndPuncttrue
\mciteSetBstMidEndSepPunct{\mcitedefaultmidpunct}
{\mcitedefaultendpunct}{\mcitedefaultseppunct}\relax
\EndOfBibitem
\bibitem[Shields(1936)]{shields_application_1936}
A.~Shields, \emph{Application of similarity principles and turbulence research
  to bed-load movement}, 1936,
  \url{https://resolver.caltech.edu/CaltechKHR:HydroLabpub167}, Num Pages: 47
  Place: Pasadena, CA Publisher: California Institute of Technology\relax
\mciteBstWouldAddEndPuncttrue
\mciteSetBstMidEndSepPunct{\mcitedefaultmidpunct}
{\mcitedefaultendpunct}{\mcitedefaultseppunct}\relax
\EndOfBibitem
\bibitem[Wiberg and Smith(1987)]{wiberg_calculations_1987}
P.~L. Wiberg and J.~D. Smith, \emph{Water Resources Research}, 1987,
  \textbf{23}, 1471--1480\relax
\mciteBstWouldAddEndPuncttrue
\mciteSetBstMidEndSepPunct{\mcitedefaultmidpunct}
{\mcitedefaultendpunct}{\mcitedefaultseppunct}\relax
\EndOfBibitem
\bibitem[Parker(1978)]{parker_self-formed_1978}
G.~Parker, \emph{Journal of Fluid Mechanics}, 1978, \textbf{89}, 127--146\relax
\mciteBstWouldAddEndPuncttrue
\mciteSetBstMidEndSepPunct{\mcitedefaultmidpunct}
{\mcitedefaultendpunct}{\mcitedefaultseppunct}\relax
\EndOfBibitem
\bibitem[Wilcock and Crowe(2003)]{wilcock_surface-based_2003}
P.~R. Wilcock and J.~C. Crowe, \emph{Journal of Hydraulic Engineering}, 2003,
  \textbf{129}, 120--128\relax
\mciteBstWouldAddEndPuncttrue
\mciteSetBstMidEndSepPunct{\mcitedefaultmidpunct}
{\mcitedefaultendpunct}{\mcitedefaultseppunct}\relax
\EndOfBibitem
\bibitem[Phillips and Jerolmack(2016)]{phillips_self-organization_2016}
C.~B. Phillips and D.~J. Jerolmack, \emph{Science}, 2016, \textbf{352},
  694--697\relax
\mciteBstWouldAddEndPuncttrue
\mciteSetBstMidEndSepPunct{\mcitedefaultmidpunct}
{\mcitedefaultendpunct}{\mcitedefaultseppunct}\relax
\EndOfBibitem
\bibitem[Phillips \emph{et~al.}(2022)Phillips, Masteller, Slater, Dunne,
  Francalanci, Lanzoni, Merritts, Lajeunesse, and Jerolmack]{Phillips2022}
C.~B. Phillips, C.~C. Masteller, L.~J. Slater, K.~B.~J. Dunne, S.~Francalanci,
  S.~Lanzoni, D.~J. Merritts, E.~Lajeunesse and D.~J. Jerolmack, \emph{Nature
  Reviews Earth \& Environment}, 2022, \textbf{3}, 406--419\relax
\mciteBstWouldAddEndPuncttrue
\mciteSetBstMidEndSepPunct{\mcitedefaultmidpunct}
{\mcitedefaultendpunct}{\mcitedefaultseppunct}\relax
\EndOfBibitem
\bibitem[Métivier \emph{et~al.}(2017)Métivier, Lajeunesse, and
  Devauchelle]{metivier_laboratory_2017}
F.~Métivier, E.~Lajeunesse and O.~Devauchelle, \emph{Earth Surface Dynamics},
  2017, \textbf{5}, 187--198\relax
\mciteBstWouldAddEndPuncttrue
\mciteSetBstMidEndSepPunct{\mcitedefaultmidpunct}
{\mcitedefaultendpunct}{\mcitedefaultseppunct}\relax
\EndOfBibitem
\bibitem[Mueller \emph{et~al.}(2005)Mueller, Pitlick, and
  Nelson]{mueller_variation_2005}
E.~R. Mueller, J.~Pitlick and J.~M. Nelson, \emph{Water Resources Research},
  2005, \textbf{41}, W04006\relax
\mciteBstWouldAddEndPuncttrue
\mciteSetBstMidEndSepPunct{\mcitedefaultmidpunct}
{\mcitedefaultendpunct}{\mcitedefaultseppunct}\relax
\EndOfBibitem
\bibitem[Buffington and Montgomery(1997)]{buffington_systematic_1997}
J.~M. Buffington and D.~R. Montgomery, \emph{Water Resources Research}, 1997,
  \textbf{33}, 1993--2029\relax
\mciteBstWouldAddEndPuncttrue
\mciteSetBstMidEndSepPunct{\mcitedefaultmidpunct}
{\mcitedefaultendpunct}{\mcitedefaultseppunct}\relax
\EndOfBibitem
\bibitem[Reid \emph{et~al.}(1985)Reid, Frostick, and
  Layman]{reid_incidence_1985}
I.~Reid, L.~E. Frostick and J.~T. Layman, \emph{Earth Surface Processes and
  Landforms}, 1985, \textbf{10}, 33--44\relax
\mciteBstWouldAddEndPuncttrue
\mciteSetBstMidEndSepPunct{\mcitedefaultmidpunct}
{\mcitedefaultendpunct}{\mcitedefaultseppunct}\relax
\EndOfBibitem
\bibitem[Masteller and Finnegan(2017)]{masteller_interplay_2017}
C.~C. Masteller and N.~J. Finnegan, \emph{Journal of Geophysical Research:
  Earth Surface}, 2017, \textbf{122}, 274--289\relax
\mciteBstWouldAddEndPuncttrue
\mciteSetBstMidEndSepPunct{\mcitedefaultmidpunct}
{\mcitedefaultendpunct}{\mcitedefaultseppunct}\relax
\EndOfBibitem
\bibitem[Masteller \emph{et~al.}(2019)Masteller, Finnegan, Turowski, Yager, and
  Rickenmann]{Masteller2019}
C.~C. Masteller, N.~J. Finnegan, J.~M. Turowski, E.~M. Yager and D.~Rickenmann,
  \emph{Geophysical Research Letters}, 2019, \textbf{46}, 2583--2591\relax
\mciteBstWouldAddEndPuncttrue
\mciteSetBstMidEndSepPunct{\mcitedefaultmidpunct}
{\mcitedefaultendpunct}{\mcitedefaultseppunct}\relax
\EndOfBibitem
\bibitem[{Masteller} and {Johnson}(2020)]{2020AGUFMEP008..04M}
C.~C. {Masteller} and J.~P. {Johnson}, AGU Fall Meeting Abstracts, 2020, pp.
  EP008--04\relax
\mciteBstWouldAddEndPuncttrue
\mciteSetBstMidEndSepPunct{\mcitedefaultmidpunct}
{\mcitedefaultendpunct}{\mcitedefaultseppunct}\relax
\EndOfBibitem
\bibitem[Deal \emph{et~al.}(2023)Deal, Venditti, Benavides, Bradley, Zhang,
  Kamrin, and Perron]{deal_grain_2023}
E.~Deal, J.~G. Venditti, S.~J. Benavides, R.~Bradley, Q.~Zhang, K.~Kamrin and
  J.~T. Perron, \emph{Nature}, 2023, \textbf{613}, 298--302\relax
\mciteBstWouldAddEndPuncttrue
\mciteSetBstMidEndSepPunct{\mcitedefaultmidpunct}
{\mcitedefaultendpunct}{\mcitedefaultseppunct}\relax
\EndOfBibitem
\bibitem[Burtin \emph{et~al.}(2013)Burtin, Hovius, Milodowski, Chen, Wu, Lin,
  Chen, Emberson, and Leu]{burtin_continuous_2013}
A.~Burtin, N.~Hovius, D.~T. Milodowski, Y.-G. Chen, Y.-M. Wu, C.-W. Lin,
  H.~Chen, R.~Emberson and P.-L. Leu, \emph{Journal of Geophysical Research:
  Earth Surface}, 2013, \textbf{118}, 1956--1974\relax
\mciteBstWouldAddEndPuncttrue
\mciteSetBstMidEndSepPunct{\mcitedefaultmidpunct}
{\mcitedefaultendpunct}{\mcitedefaultseppunct}\relax
\EndOfBibitem
\bibitem[Cook and Dietze(2022)]{Cook2022}
K.~L. Cook and M.~Dietze, \emph{Annual Review of Earth and Planetary Sciences},
  2022, \textbf{50}, 183--204\relax
\mciteBstWouldAddEndPuncttrue
\mciteSetBstMidEndSepPunct{\mcitedefaultmidpunct}
{\mcitedefaultendpunct}{\mcitedefaultseppunct}\relax
\EndOfBibitem
\bibitem[Larose \emph{et~al.}(2015)Larose, Carri{\`{e}}re, Voisin, Bottelin,
  Baillet, Gu{\'{e}}guen, Walter, Jongmans, Guillier, Garambois, Gimbert, and
  Massey]{Larose2015}
E.~Larose, S.~Carri{\`{e}}re, C.~Voisin, P.~Bottelin, L.~Baillet,
  P.~Gu{\'{e}}guen, F.~Walter, D.~Jongmans, B.~Guillier, S.~Garambois,
  F.~Gimbert and C.~Massey, \emph{Journal of Applied Geophysics}, 2015,
  \textbf{116}, 62--74\relax
\mciteBstWouldAddEndPuncttrue
\mciteSetBstMidEndSepPunct{\mcitedefaultmidpunct}
{\mcitedefaultendpunct}{\mcitedefaultseppunct}\relax
\EndOfBibitem
\bibitem[Schmandt \emph{et~al.}(2017)Schmandt, Gaeuman, Stewart, Hansen, Tsai,
  and Smith]{Schmandt2017}
B.~Schmandt, D.~Gaeuman, R.~Stewart, S.~M. Hansen, V.~C. Tsai and J.~Smith,
  \emph{Geology}, 2017, \textbf{45}, 299--302\relax
\mciteBstWouldAddEndPuncttrue
\mciteSetBstMidEndSepPunct{\mcitedefaultmidpunct}
{\mcitedefaultendpunct}{\mcitedefaultseppunct}\relax
\EndOfBibitem
\bibitem[{Roth} \emph{et~al.}(2022){Roth}, {Jin}, {Bezada}, {Titov},
  {Masteller}, {Tate}, and {Siegfried}]{2022AGUFMEP33A..01R}
D.~L. {Roth}, G.~{Jin}, M.~{Bezada}, A.~{Titov}, C.~C. {Masteller}, B.~{Tate}
  and M.~{Siegfried}, AGU Fall Meeting Abstracts, 2022, pp. EP33A--01\relax
\mciteBstWouldAddEndPuncttrue
\mciteSetBstMidEndSepPunct{\mcitedefaultmidpunct}
{\mcitedefaultendpunct}{\mcitedefaultseppunct}\relax
\EndOfBibitem
\bibitem[Schwindinger and Anderson(1989)]{schwindinger1989synneusis}
K.~R. Schwindinger and A.~T. Anderson, \emph{Contributions to Mineralogy and
  Petrology}, 1989, \textbf{103}, 187--198\relax
\mciteBstWouldAddEndPuncttrue
\mciteSetBstMidEndSepPunct{\mcitedefaultmidpunct}
{\mcitedefaultendpunct}{\mcitedefaultseppunct}\relax
\EndOfBibitem
\bibitem[Albert \emph{et~al.}(2016)Albert, Costa, and
  Mart{\'\i}]{albert2016years}
H.~Albert, F.~Costa and J.~Mart{\'\i}, \emph{Geology}, 2016, \textbf{44},
  211--214\relax
\mciteBstWouldAddEndPuncttrue
\mciteSetBstMidEndSepPunct{\mcitedefaultmidpunct}
{\mcitedefaultendpunct}{\mcitedefaultseppunct}\relax
\EndOfBibitem
\bibitem[Passarelli and Brodsky(2012)]{passarelli2012correlation}
L.~Passarelli and E.~E. Brodsky, \emph{Geophysical Journal International},
  2012, \textbf{188}, 1025--1045\relax
\mciteBstWouldAddEndPuncttrue
\mciteSetBstMidEndSepPunct{\mcitedefaultmidpunct}
{\mcitedefaultendpunct}{\mcitedefaultseppunct}\relax
\EndOfBibitem
\bibitem[Ripepe \emph{et~al.}(2017)Ripepe, Pistolesi, Coppola, Delle~Donne,
  Genco, Lacanna, Laiolo, Marchetti, Ulivieri, and
  Valade]{ripepe2017forecasting}
M.~Ripepe, M.~Pistolesi, D.~Coppola, D.~Delle~Donne, R.~Genco, G.~Lacanna,
  M.~Laiolo, E.~Marchetti, G.~Ulivieri and S.~Valade, \emph{Scientific
  Reports}, 2017, \textbf{7}, 1--9\relax
\mciteBstWouldAddEndPuncttrue
\mciteSetBstMidEndSepPunct{\mcitedefaultmidpunct}
{\mcitedefaultendpunct}{\mcitedefaultseppunct}\relax
\EndOfBibitem
\bibitem[Richter \emph{et~al.}(1970)Richter, Eaton, Murata, Ault, and
  Krivoy]{richter1970chronological}
D.~H. Richter, J.~Eaton, K.~Murata, W.~Ault and H.~Krivoy, \emph{US Geological
  Survey Professional Paper}, 1970, \textbf{537-E}, E1--E73\relax
\mciteBstWouldAddEndPuncttrue
\mciteSetBstMidEndSepPunct{\mcitedefaultmidpunct}
{\mcitedefaultendpunct}{\mcitedefaultseppunct}\relax
\EndOfBibitem
\bibitem[Swanson \emph{et~al.}(1976)Swanson, Jackson, Koyanagi, and
  Wright]{swanson1976february}
D.~A. Swanson, D.~B. Jackson, R.~Y. Koyanagi and T.~L. Wright, \emph{The
  February 1969 East Rift Eruption of Kilauea Volcano, Hawaii}, {US}
  {G}eological {S}urvey technical report, 1976\relax
\mciteBstWouldAddEndPuncttrue
\mciteSetBstMidEndSepPunct{\mcitedefaultmidpunct}
{\mcitedefaultendpunct}{\mcitedefaultseppunct}\relax
\EndOfBibitem
\bibitem[Wolfe \emph{et~al.}(1987)Wolfe, Garcia, Jackson, Koyanagi, Neal, and
  Okamura]{wolfe1987puu}
E.~W. Wolfe, M.~O. Garcia, D.~B. Jackson, R.~Y. Koyanagi, C.~A. Neal and
  A.~Okamura, \emph{US Geol Surv Prof Pap}, 1987, \textbf{1350}, 471--508\relax
\mciteBstWouldAddEndPuncttrue
\mciteSetBstMidEndSepPunct{\mcitedefaultmidpunct}
{\mcitedefaultendpunct}{\mcitedefaultseppunct}\relax
\EndOfBibitem
\bibitem[Heliker \emph{et~al.}(2003)Heliker, Mattox, Swanson, and
  Takahashi]{heliker2003first}
C.~Heliker, T.~N. Mattox, D.~Swanson and T.~Takahashi, \emph{The Pu’u
  ‘{\=O}’o-K{\=u}paianaha Eruption of Kilauea Volcano, Hawai’i: The First
  20 Years}, US Geological Survey, 2003, pp. 1--28\relax
\mciteBstWouldAddEndPuncttrue
\mciteSetBstMidEndSepPunct{\mcitedefaultmidpunct}
{\mcitedefaultendpunct}{\mcitedefaultseppunct}\relax
\EndOfBibitem
\bibitem[National Academies~of Sciences and Medicine(2017)]{NAS2017}
E.~National Academies~of Sciences and Medicine, \emph{{Volcanic Eruptions and
  Their Repose, Unrest, Precursors, and Timing}}, {National Academies Press,
  Washington, D.C.} technical report, 2017\relax
\mciteBstWouldAddEndPuncttrue
\mciteSetBstMidEndSepPunct{\mcitedefaultmidpunct}
{\mcitedefaultendpunct}{\mcitedefaultseppunct}\relax
\EndOfBibitem
\bibitem[Stoiber and Williams(1986)]{StoiberandWilliams1986}
R.~Stoiber and S.~Williams, \emph{Journal of Geophysical Research}, 1986,
  \textbf{91}, 12,215--12,231\relax
\mciteBstWouldAddEndPuncttrue
\mciteSetBstMidEndSepPunct{\mcitedefaultmidpunct}
{\mcitedefaultendpunct}{\mcitedefaultseppunct}\relax
\EndOfBibitem
\bibitem[Allard \emph{et~al.}(1994)Allard, Carbonnelle, M\'{e}trich, Loyer, and
  Zettwoog]{ALLARD1994}
P.~Allard, J.~Carbonnelle, N.~M\'{e}trich, H.~Loyer and P.~Zettwoog,
  \emph{Nature}, 1994, \textbf{368}, 326--330\relax
\mciteBstWouldAddEndPuncttrue
\mciteSetBstMidEndSepPunct{\mcitedefaultmidpunct}
{\mcitedefaultendpunct}{\mcitedefaultseppunct}\relax
\EndOfBibitem
\bibitem[Kazahaya \emph{et~al.}(1994)Kazahaya, Shinohara, and
  Saito]{KAZAHAYA1994}
K.~Kazahaya, H.~Shinohara and G.~Saito, \emph{Bulletin of Volcanology}, 1994,
  \textbf{56}, 207--216\relax
\mciteBstWouldAddEndPuncttrue
\mciteSetBstMidEndSepPunct{\mcitedefaultmidpunct}
{\mcitedefaultendpunct}{\mcitedefaultseppunct}\relax
\EndOfBibitem
\bibitem[Palma \emph{et~al.}(2008)Palma, Calder, Basualto, Blake, and
  Rothery]{Palmaetal2008}
J.~L. Palma, E.~S. Calder, D.~Basualto, S.~Blake and D.~A. Rothery,
  \emph{Journal of Geophysical Research: Solid Earth}, 2008, \textbf{113},
  201\relax
\mciteBstWouldAddEndPuncttrue
\mciteSetBstMidEndSepPunct{\mcitedefaultmidpunct}
{\mcitedefaultendpunct}{\mcitedefaultseppunct}\relax
\EndOfBibitem
\bibitem[Oppenheimer \emph{et~al.}(2009)Oppenheimer, Lomakina, Kyle, Kingsbury,
  and Boichu]{OPPENHEIMER2009}
C.~Oppenheimer, A.~S. Lomakina, P.~R. Kyle, N.~G. Kingsbury and M.~Boichu,
  \emph{Earth and Planetary Science Letters}, 2009, \textbf{284}, 392 --
  398\relax
\mciteBstWouldAddEndPuncttrue
\mciteSetBstMidEndSepPunct{\mcitedefaultmidpunct}
{\mcitedefaultendpunct}{\mcitedefaultseppunct}\relax
\EndOfBibitem
\bibitem[Woitischek \emph{et~al.}(2020)Woitischek, Woods, Edmonds, Oppenheimer,
  Aiuppa, Pering,\emph{et~al.}]{woitischek2020strombolian}
J.~Woitischek, A.~W. Woods, M.~Edmonds, C.~Oppenheimer, A.~Aiuppa, T.~D. Pering
  \emph{et~al.}, \emph{Journal of Volcanology and Geothermal Research}, 2020,
  \textbf{398}, 106869\relax
\mciteBstWouldAddEndPuncttrue
\mciteSetBstMidEndSepPunct{\mcitedefaultmidpunct}
{\mcitedefaultendpunct}{\mcitedefaultseppunct}\relax
\EndOfBibitem
\bibitem[M{\'e}trich \emph{et~al.}(2001)M{\'e}trich, Bertagnini, Landi, and
  Rosi]{metrich2001crystallization}
N.~M{\'e}trich, A.~Bertagnini, P.~Landi and M.~Rosi, \emph{Journal of
  Petrology}, 2001, \textbf{42}, 1471--1490\relax
\mciteBstWouldAddEndPuncttrue
\mciteSetBstMidEndSepPunct{\mcitedefaultmidpunct}
{\mcitedefaultendpunct}{\mcitedefaultseppunct}\relax
\EndOfBibitem
\bibitem[Francalanci \emph{et~al.}(2004)Francalanci, Tommasini, and
  Conticelli]{francalanci2004volcanic}
L.~Francalanci, S.~Tommasini and S.~Conticelli, \emph{Journal of Volcanology
  and Geothermal Research}, 2004, \textbf{131}, 179--211\relax
\mciteBstWouldAddEndPuncttrue
\mciteSetBstMidEndSepPunct{\mcitedefaultmidpunct}
{\mcitedefaultendpunct}{\mcitedefaultseppunct}\relax
\EndOfBibitem
\bibitem[Francalanci \emph{et~al.}(2005)Francalanci, Davies, Lustenhouwer,
  Tommasini, Mason, and Conticelli]{francalanci2005intra}
L.~Francalanci, G.~R. Davies, W.~Lustenhouwer, S.~Tommasini, P.~R. Mason and
  S.~Conticelli, \emph{Journal of Petrology}, 2005, \textbf{46},
  1997--2021\relax
\mciteBstWouldAddEndPuncttrue
\mciteSetBstMidEndSepPunct{\mcitedefaultmidpunct}
{\mcitedefaultendpunct}{\mcitedefaultseppunct}\relax
\EndOfBibitem
\bibitem[Suckale \emph{et~al.}(2016)Suckale, Keller, Cashman, and
  Persson]{suckale2016flow}
J.~Suckale, T.~Keller, K.~V. Cashman and P.-O. Persson, \emph{Geophysical
  Research Letters}, 2016, \textbf{43}, 12--071\relax
\mciteBstWouldAddEndPuncttrue
\mciteSetBstMidEndSepPunct{\mcitedefaultmidpunct}
{\mcitedefaultendpunct}{\mcitedefaultseppunct}\relax
\EndOfBibitem
\bibitem[Qin and Suckale(2020)]{qin2020flow}
Z.~Qin and J.~Suckale, \emph{Journal of Geophysical Research: Solid Earth},
  2020, \textbf{125}, e2019JB018549\relax
\mciteBstWouldAddEndPuncttrue
\mciteSetBstMidEndSepPunct{\mcitedefaultmidpunct}
{\mcitedefaultendpunct}{\mcitedefaultseppunct}\relax
\EndOfBibitem
\bibitem[Suckale \emph{et~al.}(2012)Suckale, Sethian, Yu, and
  Elkins-Tanton]{suckale2012crystals}
J.~Suckale, J.~A. Sethian, J.-d. Yu and L.~T. Elkins-Tanton, \emph{Journal of
  Geophysical Research: Planets}, 2012, \textbf{117}, E08004\relax
\mciteBstWouldAddEndPuncttrue
\mciteSetBstMidEndSepPunct{\mcitedefaultmidpunct}
{\mcitedefaultendpunct}{\mcitedefaultseppunct}\relax
\EndOfBibitem
\bibitem[Qin \emph{et~al.}(2020)Qin, Allison, and Suckale]{qin2020direct}
Z.~Qin, K.~Allison and J.~Suckale, \emph{Journal of Computational Physics},
  2020, \textbf{401}, 109021\relax
\mciteBstWouldAddEndPuncttrue
\mciteSetBstMidEndSepPunct{\mcitedefaultmidpunct}
{\mcitedefaultendpunct}{\mcitedefaultseppunct}\relax
\EndOfBibitem
\bibitem[DiBenedetto \emph{et~al.}(2020)DiBenedetto, Qin, and
  Suckale]{dibenedetto2020crystal}
M.~DiBenedetto, Z.~Qin and J.~Suckale, \emph{Science Advances}, 2020,
  \textbf{6}, eabd4850\relax
\mciteBstWouldAddEndPuncttrue
\mciteSetBstMidEndSepPunct{\mcitedefaultmidpunct}
{\mcitedefaultendpunct}{\mcitedefaultseppunct}\relax
\EndOfBibitem
\bibitem[Wieser \emph{et~al.}(2019)Wieser, Vukmanovic, Kilian, Ringe, Holness,
  Maclennan, and Edmonds]{wieser2019sink}
P.~E. Wieser, Z.~Vukmanovic, R.~Kilian, E.~Ringe, M.~B. Holness, J.~Maclennan
  and M.~Edmonds, \emph{Geology}, 2019, \textbf{47}, 948--952\relax
\mciteBstWouldAddEndPuncttrue
\mciteSetBstMidEndSepPunct{\mcitedefaultmidpunct}
{\mcitedefaultendpunct}{\mcitedefaultseppunct}\relax
\EndOfBibitem
\bibitem[Jeffery(1922)]{jeffery1922}
G.~B. Jeffery, \emph{Proceedings of the Royal Society B}, 1922, \textbf{102},
  161--179\relax
\mciteBstWouldAddEndPuncttrue
\mciteSetBstMidEndSepPunct{\mcitedefaultmidpunct}
{\mcitedefaultendpunct}{\mcitedefaultseppunct}\relax
\EndOfBibitem
\bibitem[DiBenedetto and Ouellette(2018)]{DiBenedetto2018b}
M.~H. DiBenedetto and N.~T. Ouellette, \emph{Journal of Fluid Mechanics}, 2018,
  \textbf{856}, 850--869\relax
\mciteBstWouldAddEndPuncttrue
\mciteSetBstMidEndSepPunct{\mcitedefaultmidpunct}
{\mcitedefaultendpunct}{\mcitedefaultseppunct}\relax
\EndOfBibitem
\bibitem[DiBenedetto \emph{et~al.}(2019)DiBenedetto, Koseff, and
  Ouellette]{DiBenedetto2019}
M.~H. DiBenedetto, J.~R. Koseff and N.~T. Ouellette, \emph{Physical Review
  Fluids}, 2019, \textbf{4}, 034301\relax
\mciteBstWouldAddEndPuncttrue
\mciteSetBstMidEndSepPunct{\mcitedefaultmidpunct}
{\mcitedefaultendpunct}{\mcitedefaultseppunct}\relax
\EndOfBibitem
\bibitem[Richter and Murata(1966)]{richter1966petrography}
D.~H. Richter and K.~J. Murata, \emph{US Geological Survey Professional Paper},
  1966, \textbf{537-D}, D1--D12\relax
\mciteBstWouldAddEndPuncttrue
\mciteSetBstMidEndSepPunct{\mcitedefaultmidpunct}
{\mcitedefaultendpunct}{\mcitedefaultseppunct}\relax
\EndOfBibitem
\bibitem[Lai \emph{et~al.}(2021)Lai, Stevens, Chase, Creyts, Behn, Das, and
  Stone]{lai2021hydraulic}
C.-Y. Lai, L.~A. Stevens, D.~L. Chase, T.~T. Creyts, M.~D. Behn, S.~B. Das and
  H.~A. Stone, \emph{Nature Communications}, 2021, \textbf{12}, 1--10\relax
\mciteBstWouldAddEndPuncttrue
\mciteSetBstMidEndSepPunct{\mcitedefaultmidpunct}
{\mcitedefaultendpunct}{\mcitedefaultseppunct}\relax
\EndOfBibitem
\bibitem[Chase \emph{et~al.}(2021)Chase, Lai, and Stone]{chase2021relaxation}
D.~L. Chase, C.-Y. Lai and H.~A. Stone, \emph{Physical Review Fluids}, 2021,
  \textbf{6}, 084101\relax
\mciteBstWouldAddEndPuncttrue
\mciteSetBstMidEndSepPunct{\mcitedefaultmidpunct}
{\mcitedefaultendpunct}{\mcitedefaultseppunct}\relax
\EndOfBibitem
\bibitem[Fowler(1986)]{fowler1986sliding}
A.~Fowler, \emph{Proceedings of the Royal Society of London. A. Mathematical
  and Physical Sciences}, 1986, \textbf{407}, 147--170\relax
\mciteBstWouldAddEndPuncttrue
\mciteSetBstMidEndSepPunct{\mcitedefaultmidpunct}
{\mcitedefaultendpunct}{\mcitedefaultseppunct}\relax
\EndOfBibitem
\bibitem[Schoof(2010)]{schoof2010ice}
C.~Schoof, \emph{Nature}, 2010, \textbf{468}, 803--806\relax
\mciteBstWouldAddEndPuncttrue
\mciteSetBstMidEndSepPunct{\mcitedefaultmidpunct}
{\mcitedefaultendpunct}{\mcitedefaultseppunct}\relax
\EndOfBibitem
\bibitem[Schoof and Hewitt(2013)]{schoof2013ice}
C.~Schoof and I.~Hewitt, \emph{Annual Review of Fluid Mechanics}, 2013,
  \textbf{45}, 217--239\relax
\mciteBstWouldAddEndPuncttrue
\mciteSetBstMidEndSepPunct{\mcitedefaultmidpunct}
{\mcitedefaultendpunct}{\mcitedefaultseppunct}\relax
\EndOfBibitem
\bibitem[Hewitt \emph{et~al.}(2012)Hewitt, Schoof, and
  Werder]{hewitt2012flotation}
I.~J. Hewitt, C.~Schoof and M.~A. Werder, \emph{Journal of Fluid Mechanics},
  2012, \textbf{702}, 157--187\relax
\mciteBstWouldAddEndPuncttrue
\mciteSetBstMidEndSepPunct{\mcitedefaultmidpunct}
{\mcitedefaultendpunct}{\mcitedefaultseppunct}\relax
\EndOfBibitem
\bibitem[Schoof(2007)]{schoof2007marine}
C.~Schoof, \emph{Journal of Fluid Mechanics}, 2007, \textbf{573}, 27--55\relax
\mciteBstWouldAddEndPuncttrue
\mciteSetBstMidEndSepPunct{\mcitedefaultmidpunct}
{\mcitedefaultendpunct}{\mcitedefaultseppunct}\relax
\EndOfBibitem
\bibitem[Pegler(2018)]{pegler2018suppression}
S.~S. Pegler, \emph{Journal of Fluid Mechanics}, 2018, \textbf{857},
  648--680\relax
\mciteBstWouldAddEndPuncttrue
\mciteSetBstMidEndSepPunct{\mcitedefaultmidpunct}
{\mcitedefaultendpunct}{\mcitedefaultseppunct}\relax
\EndOfBibitem
\bibitem[Sayag and Worster(2019)]{sayag2019instability}
R.~Sayag and M.~G. Worster, \emph{Journal of Fluid Mechanics}, 2019,
  \textbf{881}, 722--738\relax
\mciteBstWouldAddEndPuncttrue
\mciteSetBstMidEndSepPunct{\mcitedefaultmidpunct}
{\mcitedefaultendpunct}{\mcitedefaultseppunct}\relax
\EndOfBibitem
\bibitem[Rallabandi \emph{et~al.}(2017)Rallabandi, Zheng, Winton, and
  Stone]{rallabandi2017wind}
B.~Rallabandi, Z.~Zheng, M.~Winton and H.~A. Stone, \emph{Physical Review
  Letters}, 2017, \textbf{118}, 128701\relax
\mciteBstWouldAddEndPuncttrue
\mciteSetBstMidEndSepPunct{\mcitedefaultmidpunct}
{\mcitedefaultendpunct}{\mcitedefaultseppunct}\relax
\EndOfBibitem
\bibitem[Wettlaufer and Worster(2006)]{wettlaufer2006premelting}
J.~Wettlaufer and M.~G. Worster, \emph{Annual Review of Fluid Mechanics}, 2006,
  \textbf{38}, 427--452\relax
\mciteBstWouldAddEndPuncttrue
\mciteSetBstMidEndSepPunct{\mcitedefaultmidpunct}
{\mcitedefaultendpunct}{\mcitedefaultseppunct}\relax
\EndOfBibitem
\bibitem[Vella and Wettlaufer(2007)]{vella2007finger}
D.~Vella and J.~Wettlaufer, \emph{Physical Review Letters}, 2007, \textbf{98},
  088303\relax
\mciteBstWouldAddEndPuncttrue
\mciteSetBstMidEndSepPunct{\mcitedefaultmidpunct}
{\mcitedefaultendpunct}{\mcitedefaultseppunct}\relax
\EndOfBibitem
\bibitem[Vella and Wettlaufer(2008)]{vella2008explaining}
D.~Vella and J.~Wettlaufer, \emph{Journal of Geophysical Research: Oceans},
  2008, \textbf{113}, C11011\relax
\mciteBstWouldAddEndPuncttrue
\mciteSetBstMidEndSepPunct{\mcitedefaultmidpunct}
{\mcitedefaultendpunct}{\mcitedefaultseppunct}\relax
\EndOfBibitem
\bibitem[Sayag and Worster(2013)]{sayag2013elastic}
R.~Sayag and M.~G. Worster, \emph{Geophysical Research Letters}, 2013,
  \textbf{40}, 5877--5881\relax
\mciteBstWouldAddEndPuncttrue
\mciteSetBstMidEndSepPunct{\mcitedefaultmidpunct}
{\mcitedefaultendpunct}{\mcitedefaultseppunct}\relax
\EndOfBibitem
\bibitem[Wagner \emph{et~al.}(2016)Wagner, James, Murray, and
  Vella]{wagner2016role}
T.~J. Wagner, T.~D. James, T.~Murray and D.~Vella, \emph{Geophysical Research
  Letters}, 2016, \textbf{43}, 232--240A\relax
\mciteBstWouldAddEndPuncttrue
\mciteSetBstMidEndSepPunct{\mcitedefaultmidpunct}
{\mcitedefaultendpunct}{\mcitedefaultseppunct}\relax
\EndOfBibitem
\bibitem[Wagner \emph{et~al.}(2014)Wagner, Wadhams, Bates, Elosegui, Stern,
  Vella, Abrahamsen, Crawford, and Nicholls]{wagner2014footloose}
T.~J. Wagner, P.~Wadhams, R.~Bates, P.~Elosegui, A.~Stern, D.~Vella, E.~P.
  Abrahamsen, A.~Crawford and K.~W. Nicholls, \emph{Geophysical Research
  Letters}, 2014, \textbf{41}, 5522--5529\relax
\mciteBstWouldAddEndPuncttrue
\mciteSetBstMidEndSepPunct{\mcitedefaultmidpunct}
{\mcitedefaultendpunct}{\mcitedefaultseppunct}\relax
\EndOfBibitem
\bibitem[Fannon \emph{et~al.}(2017)Fannon, Fowler, and
  Moyles]{fannon2017numerical}
J.~Fannon, A.~Fowler and I.~Moyles, \emph{Proceedings of the Royal Society A:
  Mathematical, Physical and Engineering Sciences}, 2017, \textbf{473},
  20170220\relax
\mciteBstWouldAddEndPuncttrue
\mciteSetBstMidEndSepPunct{\mcitedefaultmidpunct}
{\mcitedefaultendpunct}{\mcitedefaultseppunct}\relax
\EndOfBibitem
\bibitem[Zoet and Iverson(2020)]{zoet2020slip}
L.~K. Zoet and N.~R. Iverson, \emph{Science}, 2020, \textbf{368}, 76--78\relax
\mciteBstWouldAddEndPuncttrue
\mciteSetBstMidEndSepPunct{\mcitedefaultmidpunct}
{\mcitedefaultendpunct}{\mcitedefaultseppunct}\relax
\EndOfBibitem
\bibitem[Warburton \emph{et~al.}(2023)Warburton, Hewitt, and
  Neufeld]{warburton2023shear}
K.~Warburton, D.~Hewitt and J.~Neufeld, \emph{Proceedings of the Royal Society
  A}, 2023, \textbf{479}, 20220536\relax
\mciteBstWouldAddEndPuncttrue
\mciteSetBstMidEndSepPunct{\mcitedefaultmidpunct}
{\mcitedefaultendpunct}{\mcitedefaultseppunct}\relax
\EndOfBibitem
\bibitem[Rempel \emph{et~al.}(2004)Rempel, Wettlaufer, and
  Worster]{rempel2004premelting}
A.~W. Rempel, J.~Wettlaufer and M.~G. Worster, \emph{Journal of fluid
  mechanics}, 2004, \textbf{498}, 227--244\relax
\mciteBstWouldAddEndPuncttrue
\mciteSetBstMidEndSepPunct{\mcitedefaultmidpunct}
{\mcitedefaultendpunct}{\mcitedefaultseppunct}\relax
\EndOfBibitem
\bibitem[Meyer and Hewitt(2017)]{meyer2017continuum}
C.~R. Meyer and I.~J. Hewitt, \emph{The Cryosphere}, 2017, \textbf{11},
  2799--2813\relax
\mciteBstWouldAddEndPuncttrue
\mciteSetBstMidEndSepPunct{\mcitedefaultmidpunct}
{\mcitedefaultendpunct}{\mcitedefaultseppunct}\relax
\EndOfBibitem
\bibitem[Moure \emph{et~al.}(2023)Moure, Jones, Pawlak, Meyer, and
  Fu]{moure2023thermodynamic}
A.~Moure, N.~Jones, J.~Pawlak, C.~Meyer and X.~Fu, \emph{Water Resources
  Research}, 2023,  e2022WR034035\relax
\mciteBstWouldAddEndPuncttrue
\mciteSetBstMidEndSepPunct{\mcitedefaultmidpunct}
{\mcitedefaultendpunct}{\mcitedefaultseppunct}\relax
\EndOfBibitem
\bibitem[Benn \emph{et~al.}(2007)Benn, Warren, and Mottram]{benn2007calving}
D.~I. Benn, C.~R. Warren and R.~H. Mottram, \emph{Earth-Science Reviews}, 2007,
  \textbf{82}, 143--179\relax
\mciteBstWouldAddEndPuncttrue
\mciteSetBstMidEndSepPunct{\mcitedefaultmidpunct}
{\mcitedefaultendpunct}{\mcitedefaultseppunct}\relax
\EndOfBibitem
\bibitem[Bassis and Jacobs(2013)]{bassis2013diverse}
J.~N. Bassis and S.~Jacobs, \emph{Nature Geoscience}, 2013, \textbf{6},
  833--836\relax
\mciteBstWouldAddEndPuncttrue
\mciteSetBstMidEndSepPunct{\mcitedefaultmidpunct}
{\mcitedefaultendpunct}{\mcitedefaultseppunct}\relax
\EndOfBibitem
\bibitem[Scambos \emph{et~al.}(2000)Scambos, Hulbe, Fahnestock, and
  Bohlander]{scambos2000link}
T.~A. Scambos, C.~Hulbe, M.~Fahnestock and J.~Bohlander, \emph{Journal of
  Glaciology}, 2000, \textbf{46}, 516--530\relax
\mciteBstWouldAddEndPuncttrue
\mciteSetBstMidEndSepPunct{\mcitedefaultmidpunct}
{\mcitedefaultendpunct}{\mcitedefaultseppunct}\relax
\EndOfBibitem
\bibitem[Weertman(1973)]{weertman1973can}
J.~Weertman, 1973\relax
\mciteBstWouldAddEndPuncttrue
\mciteSetBstMidEndSepPunct{\mcitedefaultmidpunct}
{\mcitedefaultendpunct}{\mcitedefaultseppunct}\relax
\EndOfBibitem
\bibitem[Van~der Veen(1998)]{van1998fracture}
C.~Van~der Veen, \emph{Cold Regions Science and Technology}, 1998, \textbf{27},
  31--47\relax
\mciteBstWouldAddEndPuncttrue
\mciteSetBstMidEndSepPunct{\mcitedefaultmidpunct}
{\mcitedefaultendpunct}{\mcitedefaultseppunct}\relax
\EndOfBibitem
\bibitem[Banwell \emph{et~al.}(2013)Banwell, MacAyeal, and
  Sergienko]{banwell2013breakup}
A.~F. Banwell, D.~R. MacAyeal and O.~V. Sergienko, \emph{Geophysical Research
  Letters}, 2013, \textbf{40}, 5872--5876\relax
\mciteBstWouldAddEndPuncttrue
\mciteSetBstMidEndSepPunct{\mcitedefaultmidpunct}
{\mcitedefaultendpunct}{\mcitedefaultseppunct}\relax
\EndOfBibitem
\bibitem[Robel and Banwell(2019)]{robel2019speed}
A.~A. Robel and A.~F. Banwell, \emph{Geophysical Research Letters}, 2019,
  \textbf{46}, 12092--12100\relax
\mciteBstWouldAddEndPuncttrue
\mciteSetBstMidEndSepPunct{\mcitedefaultmidpunct}
{\mcitedefaultendpunct}{\mcitedefaultseppunct}\relax
\EndOfBibitem
\bibitem[DeConto and Pollard(2016)]{deconto2016contribution}
R.~M. DeConto and D.~Pollard, \emph{Nature}, 2016, \textbf{531}, 591--597\relax
\mciteBstWouldAddEndPuncttrue
\mciteSetBstMidEndSepPunct{\mcitedefaultmidpunct}
{\mcitedefaultendpunct}{\mcitedefaultseppunct}\relax
\EndOfBibitem
\bibitem[Tsai and Rice(2010)]{tsai2010model}
V.~C. Tsai and J.~R. Rice, \emph{Journal of Geophysical Research: Earth
  Surface}, 2010, \textbf{115}, F03007\relax
\mciteBstWouldAddEndPuncttrue
\mciteSetBstMidEndSepPunct{\mcitedefaultmidpunct}
{\mcitedefaultendpunct}{\mcitedefaultseppunct}\relax
\EndOfBibitem
\bibitem[Luckman \emph{et~al.}(2012)Luckman, Jansen, Kulessa, King, Sammonds,
  and Benn]{luckman2012basal}
A.~Luckman, D.~Jansen, B.~Kulessa, E.~King, P.~Sammonds and D.~Benn, \emph{The
  Cryosphere}, 2012, \textbf{6}, 113--123\relax
\mciteBstWouldAddEndPuncttrue
\mciteSetBstMidEndSepPunct{\mcitedefaultmidpunct}
{\mcitedefaultendpunct}{\mcitedefaultseppunct}\relax
\EndOfBibitem
\bibitem[McGrath \emph{et~al.}(2012)McGrath, Steffen, Scambos, Rajaram,
  Casassa, and Lagos]{mcgrath2012basal}
D.~McGrath, K.~Steffen, T.~Scambos, H.~Rajaram, G.~Casassa and J.~L.~R. Lagos,
  \emph{Annals of Glaciology}, 2012, \textbf{53}, 10--18\relax
\mciteBstWouldAddEndPuncttrue
\mciteSetBstMidEndSepPunct{\mcitedefaultmidpunct}
{\mcitedefaultendpunct}{\mcitedefaultseppunct}\relax
\EndOfBibitem
\bibitem[Buck and Lai(2021)]{buck2021flexural}
W.~R. Buck and C.-Y. Lai, \emph{Geophysical Research Letters}, 2021,
  \textbf{48}, e2021GL093110\relax
\mciteBstWouldAddEndPuncttrue
\mciteSetBstMidEndSepPunct{\mcitedefaultmidpunct}
{\mcitedefaultendpunct}{\mcitedefaultseppunct}\relax
\EndOfBibitem
\bibitem[Coffey \emph{et~al.}(2022)Coffey, MacAyeal, Copland, Mueller,
  Sergienko, Banwell, and Lai]{coffey2022enigmatic}
N.~B. Coffey, D.~R. MacAyeal, L.~Copland, D.~R. Mueller, O.~V. Sergienko, A.~F.
  Banwell and C.-Y. Lai, \emph{Journal of Glaciology}, 2022, \textbf{68},
  867--878\relax
\mciteBstWouldAddEndPuncttrue
\mciteSetBstMidEndSepPunct{\mcitedefaultmidpunct}
{\mcitedefaultendpunct}{\mcitedefaultseppunct}\relax
\EndOfBibitem
\bibitem[Jules \emph{et~al.}(2021)Jules, Lajeunesse, Devauchelle, Gu{\'e}rin,
  Jaupart, and Lagr{\'e}e]{jules2021flow}
V.~Jules, E.~Lajeunesse, O.~Devauchelle, A.~Gu{\'e}rin, C.~Jaupart and P.-Y.
  Lagr{\'e}e, \emph{Journal of Fluid Mechanics}, 2021, \textbf{917}, A13\relax
\mciteBstWouldAddEndPuncttrue
\mciteSetBstMidEndSepPunct{\mcitedefaultmidpunct}
{\mcitedefaultendpunct}{\mcitedefaultseppunct}\relax
\EndOfBibitem
\bibitem[Maher(2011)]{maher2011role}
K.~Maher, \emph{Earth and Planetary Science Letters}, 2011, \textbf{312},
  48--58\relax
\mciteBstWouldAddEndPuncttrue
\mciteSetBstMidEndSepPunct{\mcitedefaultmidpunct}
{\mcitedefaultendpunct}{\mcitedefaultseppunct}\relax
\EndOfBibitem
\bibitem[Harman and Kim(2019)]{harman2019low}
C.~J. Harman and M.~Kim, \emph{Hydrological Processes}, 2019, \textbf{33},
  466--475\relax
\mciteBstWouldAddEndPuncttrue
\mciteSetBstMidEndSepPunct{\mcitedefaultmidpunct}
{\mcitedefaultendpunct}{\mcitedefaultseppunct}\relax
\EndOfBibitem
\bibitem[Jules(2020)]{jules2020couplage}
V.~Jules, \emph{PhD thesis}, Universit{\'e} de Paris, 2020\relax
\mciteBstWouldAddEndPuncttrue
\mciteSetBstMidEndSepPunct{\mcitedefaultmidpunct}
{\mcitedefaultendpunct}{\mcitedefaultseppunct}\relax
\EndOfBibitem
\bibitem[Gu{\'{e}}rin \emph{et~al.}(2019)Gu{\'{e}}rin, Devauchelle, Robert,
  Kitou, Dessert, Quiquerez, Allemand, and Lajeunesse]{Guerin2019}
A.~Gu{\'{e}}rin, O.~Devauchelle, V.~Robert, T.~Kitou, C.~Dessert, A.~Quiquerez,
  P.~Allemand and E.~Lajeunesse, \emph{Geophysical Research Letters}, 2019,
  \textbf{46}, 7447--7455\relax
\mciteBstWouldAddEndPuncttrue
\mciteSetBstMidEndSepPunct{\mcitedefaultmidpunct}
{\mcitedefaultendpunct}{\mcitedefaultseppunct}\relax
\EndOfBibitem
\bibitem[Chandler \emph{et~al.}(2013)Chandler, Wadham, Lis, Cowton, Sole,
  Bartholomew, Telling, Nienow, Bagshaw,
  Mair,\emph{et~al.}]{chandler2013evolution}
D.~Chandler, J.~Wadham, G.~Lis, T.~Cowton, A.~Sole, I.~Bartholomew, J.~Telling,
  P.~Nienow, E.~Bagshaw, D.~Mair \emph{et~al.}, \emph{Nature Geoscience}, 2013,
  \textbf{6}, 195--198\relax
\mciteBstWouldAddEndPuncttrue
\mciteSetBstMidEndSepPunct{\mcitedefaultmidpunct}
{\mcitedefaultendpunct}{\mcitedefaultseppunct}\relax
\EndOfBibitem
\bibitem[Werder \emph{et~al.}(2013)Werder, Hewitt, Schoof, and
  Flowers]{werder2013modeling}
M.~A. Werder, I.~J. Hewitt, C.~G. Schoof and G.~E. Flowers, \emph{Journal of
  Geophysical Research: Earth Surface}, 2013, \textbf{118}, 2140--2158\relax
\mciteBstWouldAddEndPuncttrue
\mciteSetBstMidEndSepPunct{\mcitedefaultmidpunct}
{\mcitedefaultendpunct}{\mcitedefaultseppunct}\relax
\EndOfBibitem
\bibitem[Nye(1976)]{nye1976water}
J.~F. Nye, \emph{Journal of Glaciology}, 1976, \textbf{17}, 181--207\relax
\mciteBstWouldAddEndPuncttrue
\mciteSetBstMidEndSepPunct{\mcitedefaultmidpunct}
{\mcitedefaultendpunct}{\mcitedefaultseppunct}\relax
\EndOfBibitem
\bibitem[Damsgaard \emph{et~al.}(2017)Damsgaard, Suckale, Piotrowski, Houssais,
  Siegfried, and Fricker]{Damsgaard2017}
A.~Damsgaard, J.~Suckale, J.~A. Piotrowski, M.~Houssais, M.~R. Siegfried and
  H.~A. Fricker, \emph{Journal of Glaciology}, 2017, \textbf{63},
  1034--1048\relax
\mciteBstWouldAddEndPuncttrue
\mciteSetBstMidEndSepPunct{\mcitedefaultmidpunct}
{\mcitedefaultendpunct}{\mcitedefaultseppunct}\relax
\EndOfBibitem
\bibitem[Hewitt and Creyts(2019)]{hewitt2019model}
I.~J. Hewitt and T.~T. Creyts, \emph{Geophysical Research Letters}, 2019,
  \textbf{46}, 6673--6680\relax
\mciteBstWouldAddEndPuncttrue
\mciteSetBstMidEndSepPunct{\mcitedefaultmidpunct}
{\mcitedefaultendpunct}{\mcitedefaultseppunct}\relax
\EndOfBibitem
\bibitem[Lowenstein and Hardie(1985)]{Lowenstein1985}
T.~K. Lowenstein and L.~A. Hardie, \emph{Sedimentology}, 1985, \textbf{32},
  627--644\relax
\mciteBstWouldAddEndPuncttrue
\mciteSetBstMidEndSepPunct{\mcitedefaultmidpunct}
{\mcitedefaultendpunct}{\mcitedefaultseppunct}\relax
\EndOfBibitem
\bibitem[Dang \emph{et~al.}(2018)Dang, Xiao, Xu, Zhang, Huang, Wang, Zhao,
  Komatsu, and Yue]{Dang2018}
Y.~Dang, L.~Xiao, Y.~Xu, F.~Zhang, J.~Huang, J.~Wang, J.~Zhao, G.~Komatsu and
  Z.~Yue, \emph{Journal of Geophysical Research: Planets}, 2018, \textbf{123},
  1910--1933\relax
\mciteBstWouldAddEndPuncttrue
\mciteSetBstMidEndSepPunct{\mcitedefaultmidpunct}
{\mcitedefaultendpunct}{\mcitedefaultseppunct}\relax
\EndOfBibitem
\bibitem[Christiansen(1963)]{Christiansen1963}
F.~W. Christiansen, \emph{Science}, 1963, \textbf{139}, 607--609\relax
\mciteBstWouldAddEndPuncttrue
\mciteSetBstMidEndSepPunct{\mcitedefaultmidpunct}
{\mcitedefaultendpunct}{\mcitedefaultseppunct}\relax
\EndOfBibitem
\bibitem[Krinsley(1970)]{Krinsley1970}
D.~Krinsley, \emph{U.S. Geol. Survey}, 1970, \textbf{CP 70-800}, 356\relax
\mciteBstWouldAddEndPuncttrue
\mciteSetBstMidEndSepPunct{\mcitedefaultmidpunct}
{\mcitedefaultendpunct}{\mcitedefaultseppunct}\relax
\EndOfBibitem
\bibitem[Nield \emph{et~al.}(2015)Nield, Bryant, Wiggs, King, Thomas, Eckardt,
  and Washington]{Nield2015}
J.~M. Nield, R.~G. Bryant, G.~F. Wiggs, J.~King, D.~S. Thomas, F.~D. Eckardt
  and R.~Washington, \emph{Geology}, 2015, \textbf{43}, 31\relax
\mciteBstWouldAddEndPuncttrue
\mciteSetBstMidEndSepPunct{\mcitedefaultmidpunct}
{\mcitedefaultendpunct}{\mcitedefaultseppunct}\relax
\EndOfBibitem
\bibitem[Lasser \emph{et~al.}(2020)Lasser, Nield, and Goehring]{Lasser2020}
J.~Lasser, J.~M. Nield and L.~Goehring, \emph{Earth System Science Data}, 2020,
  \textbf{12}, 2881--2898\relax
\mciteBstWouldAddEndPuncttrue
\mciteSetBstMidEndSepPunct{\mcitedefaultmidpunct}
{\mcitedefaultendpunct}{\mcitedefaultseppunct}\relax
\EndOfBibitem
\bibitem[Lokier(2012)]{Lokier2012}
S.~Lokier, \emph{Journal of Arid Environments}, 2012, \textbf{79}, 32--47\relax
\mciteBstWouldAddEndPuncttrue
\mciteSetBstMidEndSepPunct{\mcitedefaultmidpunct}
{\mcitedefaultendpunct}{\mcitedefaultseppunct}\relax
\EndOfBibitem
\bibitem[Lasser \emph{et~al.}(2021)Lasser, Ernst, and Goehring]{Lasser2021}
J.~Lasser, M.~Ernst and L.~Goehring, \emph{Journal of Fluid Mechanics}, 2021,
  \textbf{917}, A14\relax
\mciteBstWouldAddEndPuncttrue
\mciteSetBstMidEndSepPunct{\mcitedefaultmidpunct}
{\mcitedefaultendpunct}{\mcitedefaultseppunct}\relax
\EndOfBibitem
\bibitem[Lasser \emph{et~al.}(2023)Lasser, Nield, Ernst, Karius, Wiggs,
  Threadgold, Beaume, and Goehring]{Lasser2022}
J.~Lasser, J.~M. Nield, M.~Ernst, V.~Karius, G.~F.~S. Wiggs, M.~R. Threadgold,
  C.~Beaume and L.~Goehring, \emph{Physical Review X}, 2023, \textbf{13},
  011025\relax
\mciteBstWouldAddEndPuncttrue
\mciteSetBstMidEndSepPunct{\mcitedefaultmidpunct}
{\mcitedefaultendpunct}{\mcitedefaultseppunct}\relax
\EndOfBibitem
\bibitem[Hewitt \emph{et~al.}(2020)Hewitt, Peng, and Lister]{Hewitt2020}
D.~R. Hewitt, G.~G. Peng and J.~R. Lister, \emph{Journal of Fluid Mechanics},
  2020, \textbf{883}, A37\relax
\mciteBstWouldAddEndPuncttrue
\mciteSetBstMidEndSepPunct{\mcitedefaultmidpunct}
{\mcitedefaultendpunct}{\mcitedefaultseppunct}\relax
\EndOfBibitem
\bibitem[Wooding(1960)]{Wooding1960}
R.~Wooding, \emph{Journal of Fluid Mechanics}, 1960, \textbf{9}, 183--192\relax
\mciteBstWouldAddEndPuncttrue
\mciteSetBstMidEndSepPunct{\mcitedefaultmidpunct}
{\mcitedefaultendpunct}{\mcitedefaultseppunct}\relax
\EndOfBibitem
\bibitem[Wooding \emph{et~al.}(1997)Wooding, Tyler, White, and
  Anderson]{Wooding1997}
R.~A. Wooding, S.~W. Tyler, I.~White and P.~A. Anderson, \emph{Water Resour.
  Res.}, 1997, \textbf{33}, 1199--1217\relax
\mciteBstWouldAddEndPuncttrue
\mciteSetBstMidEndSepPunct{\mcitedefaultmidpunct}
{\mcitedefaultendpunct}{\mcitedefaultseppunct}\relax
\EndOfBibitem
\bibitem[Stevens \emph{et~al.}(2009)Stevens, Jr., Simmons, and
  Fenstemaker]{Stevens2009}
J.~D. Stevens, J.~M.~S. Jr., C.~T. Simmons and T.~Fenstemaker, \emph{Journal of
  Hydrology}, 2009, \textbf{375}, 394--409\relax
\mciteBstWouldAddEndPuncttrue
\mciteSetBstMidEndSepPunct{\mcitedefaultmidpunct}
{\mcitedefaultendpunct}{\mcitedefaultseppunct}\relax
\EndOfBibitem
\bibitem[Groeneveld \emph{et~al.}(2010)Groeneveld, Huntington, and
  Barz]{Groeneveld2010}
D.~Groeneveld, J.~Huntington and D.~Barz, \emph{Journal of Hydrology}, 2010,
  \textbf{392}, 211--218\relax
\mciteBstWouldAddEndPuncttrue
\mciteSetBstMidEndSepPunct{\mcitedefaultmidpunct}
{\mcitedefaultendpunct}{\mcitedefaultseppunct}\relax
\EndOfBibitem
\bibitem[Eloukabi \emph{et~al.}(2013)Eloukabi, Sghaier, Nasrallah, and
  Prat]{Eloukabi2013}
H.~Eloukabi, N.~Sghaier, S.~B. Nasrallah and M.~Prat, \emph{International
  Journal of Heat and Mass Transfer}, 2013, \textbf{56}, 80--93\relax
\mciteBstWouldAddEndPuncttrue
\mciteSetBstMidEndSepPunct{\mcitedefaultmidpunct}
{\mcitedefaultendpunct}{\mcitedefaultseppunct}\relax
\EndOfBibitem
\bibitem[Huppert(1982)]{huppert1982flow}
H.~E. Huppert, \emph{Nature}, 1982, \textbf{300}, 427--429\relax
\mciteBstWouldAddEndPuncttrue
\mciteSetBstMidEndSepPunct{\mcitedefaultmidpunct}
{\mcitedefaultendpunct}{\mcitedefaultseppunct}\relax
\EndOfBibitem
\bibitem[Glade \emph{et~al.}(2021)Glade, Fratkin, Pouragha, Seiphoori, and
  Rowland]{glade2021arctic}
R.~C. Glade, M.~M. Fratkin, M.~Pouragha, A.~Seiphoori and J.~C. Rowland,
  \emph{Proceedings of the National Academy of Sciences}, 2021, \textbf{118},
  e2101255118\relax
\mciteBstWouldAddEndPuncttrue
\mciteSetBstMidEndSepPunct{\mcitedefaultmidpunct}
{\mcitedefaultendpunct}{\mcitedefaultseppunct}\relax
\EndOfBibitem
\bibitem[Furbish and Doane(2021)]{furbish2021rarefied}
D.~J. Furbish and T.~H. Doane, \emph{Earth Surface Dynamics}, 2021, \textbf{9},
  629--664\relax
\mciteBstWouldAddEndPuncttrue
\mciteSetBstMidEndSepPunct{\mcitedefaultmidpunct}
{\mcitedefaultendpunct}{\mcitedefaultseppunct}\relax
\EndOfBibitem
\bibitem[Haff(1996)]{haff199614}
P.~K. Haff, The Scientific Nature of Geomorphology: Proceedings of the 27th
  Binghamton Symposium in Geomorphology, Held 27-29 September, 1996, 1996, p.
  337\relax
\mciteBstWouldAddEndPuncttrue
\mciteSetBstMidEndSepPunct{\mcitedefaultmidpunct}
{\mcitedefaultendpunct}{\mcitedefaultseppunct}\relax
\EndOfBibitem
\bibitem[Matsuoka(2001)]{matsuoka2001solifluction}
N.~Matsuoka, \emph{Earth-Science Reviews}, 2001, \textbf{55}, 107--134\relax
\mciteBstWouldAddEndPuncttrue
\mciteSetBstMidEndSepPunct{\mcitedefaultmidpunct}
{\mcitedefaultendpunct}{\mcitedefaultseppunct}\relax
\EndOfBibitem
\bibitem[Deshpande \emph{et~al.}(2021)Deshpande, Furbish, Arratia, and
  Jerolmack]{deshpandePerpetualFragilityCreeping2021}
N.~S. Deshpande, D.~J. Furbish, P.~E. Arratia and D.~J. Jerolmack, \emph{Nature
  Communications}, 2021, \textbf{12}, 3909\relax
\mciteBstWouldAddEndPuncttrue
\mciteSetBstMidEndSepPunct{\mcitedefaultmidpunct}
{\mcitedefaultendpunct}{\mcitedefaultseppunct}\relax
\EndOfBibitem
\bibitem[Fazelpour \emph{et~al.}(2022)Fazelpour, Tang, and
  Daniels]{fazelpour2022effect}
F.~Fazelpour, Z.~Tang and K.~E. Daniels, \emph{Soft Matter}, 2022, \textbf{18},
  1435--1442\relax
\mciteBstWouldAddEndPuncttrue
\mciteSetBstMidEndSepPunct{\mcitedefaultmidpunct}
{\mcitedefaultendpunct}{\mcitedefaultseppunct}\relax
\EndOfBibitem
\bibitem[Mandal \emph{et~al.}(2020)Mandal, Nicolas, and
  Pouliquen]{mandal2020insights}
S.~Mandal, M.~Nicolas and O.~Pouliquen, \emph{Proceedings of the National
  Academy of Sciences}, 2020, \textbf{117}, 8366--8373\relax
\mciteBstWouldAddEndPuncttrue
\mciteSetBstMidEndSepPunct{\mcitedefaultmidpunct}
{\mcitedefaultendpunct}{\mcitedefaultseppunct}\relax
\EndOfBibitem
\bibitem[Harris \emph{et~al.}(2003)Harris, Davies, and
  Rea]{harris2003gelifluction}
C.~Harris, M.~C. Davies and B.~R. Rea, \emph{Earth Surface Processes and
  Landforms}, 2003, \textbf{28}, 1289--1301\relax
\mciteBstWouldAddEndPuncttrue
\mciteSetBstMidEndSepPunct{\mcitedefaultmidpunct}
{\mcitedefaultendpunct}{\mcitedefaultseppunct}\relax
\EndOfBibitem
\bibitem[Harkema \emph{et~al.}(2023)Harkema, Nijland, de~Jong, Kattenborn, and
  Eichel]{harkema2023monitoring}
M.~Harkema, W.~Nijland, S.~de~Jong, T.~Kattenborn and J.~Eichel,
  \emph{Geomorphology}, 2023, \textbf{433}, 108727\relax
\mciteBstWouldAddEndPuncttrue
\mciteSetBstMidEndSepPunct{\mcitedefaultmidpunct}
{\mcitedefaultendpunct}{\mcitedefaultseppunct}\relax
\EndOfBibitem
\bibitem[Bagnold(1941)]{Bagnold41}
R.~A. Bagnold, \emph{The {P}hysics of {B}lown {S}and and {D}esert {D}unes},
  Dover Publications Inc, 1941\relax
\mciteBstWouldAddEndPuncttrue
\mciteSetBstMidEndSepPunct{\mcitedefaultmidpunct}
{\mcitedefaultendpunct}{\mcitedefaultseppunct}\relax
\EndOfBibitem
\bibitem[Lorenz and Zimbelman(2014)]{Lorenz14}
R.~D. Lorenz and J.~R. Zimbelman, \emph{Dune {W}orlds, {H}ow {W}indblown {S}and
  {S}hapes {P}lanetary {L}andscapes}, Springer Praxis Books, 2014\relax
\mciteBstWouldAddEndPuncttrue
\mciteSetBstMidEndSepPunct{\mcitedefaultmidpunct}
{\mcitedefaultendpunct}{\mcitedefaultseppunct}\relax
\EndOfBibitem
\bibitem[Livingstone \emph{et~al.}(2007)Livingstone, Wiggs, and
  Weaver]{Livingstone07}
I.~Livingstone, G.~S.~S. Wiggs and C.~S. Weaver, \emph{Earth-Science Reviews},
  2007, \textbf{80}, 239--257\relax
\mciteBstWouldAddEndPuncttrue
\mciteSetBstMidEndSepPunct{\mcitedefaultmidpunct}
{\mcitedefaultendpunct}{\mcitedefaultseppunct}\relax
\EndOfBibitem
\bibitem[Berte(2010)]{Berte10}
C.~J. Berte, \emph{Fighting sand encroachment. {L}essons from {M}auritania},
  Food and Agriculture Organisation of the United Nations, Rome, 2010\relax
\mciteBstWouldAddEndPuncttrue
\mciteSetBstMidEndSepPunct{\mcitedefaultmidpunct}
{\mcitedefaultendpunct}{\mcitedefaultseppunct}\relax
\EndOfBibitem
\bibitem[Andreotti \emph{et~al.}(2013)Andreotti, Forterre, and
  Pouliquen]{Andreotti13}
B.~Andreotti, Y.~Forterre and O.~Pouliquen, \emph{{G}ranular {M}edia: {B}etween
  {F}luid and {S}olid}, Cambridge University Press, 2013\relax
\mciteBstWouldAddEndPuncttrue
\mciteSetBstMidEndSepPunct{\mcitedefaultmidpunct}
{\mcitedefaultendpunct}{\mcitedefaultseppunct}\relax
\EndOfBibitem
\bibitem[Kroy \emph{et~al.}(2002)Kroy, Sauermann, and Herrmann]{Kroy02}
K.~Kroy, G.~Sauermann and H.~J. Herrmann, \emph{Physical Review Letters}, 2002,
  \textbf{88}, 054301\relax
\mciteBstWouldAddEndPuncttrue
\mciteSetBstMidEndSepPunct{\mcitedefaultmidpunct}
{\mcitedefaultendpunct}{\mcitedefaultseppunct}\relax
\EndOfBibitem
\bibitem[Charru \emph{et~al.}(2013)Charru, Andreotti, and Claudin]{Charru2013}
F.~Charru, B.~Andreotti and P.~Claudin, \emph{Annual Review of Fluid
  Mechanics}, 2013, \textbf{45}, 469--493\relax
\mciteBstWouldAddEndPuncttrue
\mciteSetBstMidEndSepPunct{\mcitedefaultmidpunct}
{\mcitedefaultendpunct}{\mcitedefaultseppunct}\relax
\EndOfBibitem
\bibitem[Bacik \emph{et~al.}(2020)Bacik, Lovett, Caulfield, and
  Vriend]{Bacik2020}
K.~A. Bacik, S.~Lovett, C.~P. Caulfield and N.~M. Vriend, \emph{Physical Review
  Letters}, 2020, \textbf{124}, 054501\relax
\mciteBstWouldAddEndPuncttrue
\mciteSetBstMidEndSepPunct{\mcitedefaultmidpunct}
{\mcitedefaultendpunct}{\mcitedefaultseppunct}\relax
\EndOfBibitem
\bibitem[Jarvis \emph{et~al.}(2022)Jarvis, Bacik, Narteau, and
  Vriend]{Jarvis2022a}
P.~A. Jarvis, K.~A. Bacik, C.~Narteau and N.~M. Vriend, \emph{Journal of
  Geophysical Research: Earth Surface}, 2022, \textbf{127}, e2021JF006492\relax
\mciteBstWouldAddEndPuncttrue
\mciteSetBstMidEndSepPunct{\mcitedefaultmidpunct}
{\mcitedefaultendpunct}{\mcitedefaultseppunct}\relax
\EndOfBibitem
\bibitem[Jarvis \emph{et~al.}(2022)Jarvis, Narteau, Rozier, and
  Vriend]{Jarvis2022b}
P.~A. Jarvis, C.~Narteau, O.~Rozier and N.~M. Vriend, \emph{Earth Surface
  Dynamics Discussions}, 2022\relax
\mciteBstWouldAddEndPuncttrue
\mciteSetBstMidEndSepPunct{\mcitedefaultmidpunct}
{\mcitedefaultendpunct}{\mcitedefaultseppunct}\relax
\EndOfBibitem
\bibitem[Bacik \emph{et~al.}(2021)Bacik, Caulfield, and Vriend]{Bacik2021a}
K.~A. Bacik, C.~P. Caulfield and N.~M. Vriend, \emph{Physical Review Letters},
  2021, \textbf{127}, 154501\relax
\mciteBstWouldAddEndPuncttrue
\mciteSetBstMidEndSepPunct{\mcitedefaultmidpunct}
{\mcitedefaultendpunct}{\mcitedefaultseppunct}\relax
\EndOfBibitem
\bibitem[Bacik \emph{et~al.}(2021)Bacik, Canizares, , Caulfield, Williams, and
  Vriend]{Bacik2021b}
K.~A. Bacik, P.~Canizares, , C.~P. Caulfield, M.~J. Williams and N.~M. Vriend,
  \emph{Physical Review Fluids}, 2021, \textbf{6}, 104308\relax
\mciteBstWouldAddEndPuncttrue
\mciteSetBstMidEndSepPunct{\mcitedefaultmidpunct}
{\mcitedefaultendpunct}{\mcitedefaultseppunct}\relax
\EndOfBibitem
\bibitem[Hole(1981)]{Hole1981}
F.~D. Hole, \emph{Geoderma}, 1981, \textbf{25}, 75--112\relax
\mciteBstWouldAddEndPuncttrue
\mciteSetBstMidEndSepPunct{\mcitedefaultmidpunct}
{\mcitedefaultendpunct}{\mcitedefaultseppunct}\relax
\EndOfBibitem
\bibitem[Hosoi and Goldman(2015)]{Hosoi2015}
A.~Hosoi and D.~I. Goldman, \emph{Annual Review of Fluid Mechanics}, 2015,
  \textbf{47}, 431--453\relax
\mciteBstWouldAddEndPuncttrue
\mciteSetBstMidEndSepPunct{\mcitedefaultmidpunct}
{\mcitedefaultendpunct}{\mcitedefaultseppunct}\relax
\EndOfBibitem
\bibitem[Kudrolli and Ramirez(2019)]{Kudrolli2019}
A.~Kudrolli and B.~Ramirez, \emph{Proceedings of the National Academy of
  Sciences}, 2019, \textbf{116}, 25569--25574\relax
\mciteBstWouldAddEndPuncttrue
\mciteSetBstMidEndSepPunct{\mcitedefaultmidpunct}
{\mcitedefaultendpunct}{\mcitedefaultseppunct}\relax
\EndOfBibitem
\bibitem[Dorgan \emph{et~al.}(2013)Dorgan, Law, and Rouse]{Dorgan2013}
K.~M. Dorgan, C.~J. Law and G.~W. Rouse, \emph{Proceedings of the Royal Society
  B: Biological Sciences}, 2013, \textbf{280}, 20122948\relax
\mciteBstWouldAddEndPuncttrue
\mciteSetBstMidEndSepPunct{\mcitedefaultmidpunct}
{\mcitedefaultendpunct}{\mcitedefaultseppunct}\relax
\EndOfBibitem
\bibitem[Maladen \emph{et~al.}(2009)Maladen, Ding, Li, and
  Goldman]{Maladen2009}
R.~Maladen, Y.~Ding, C.~Li and D.~Goldman, \emph{Science (New York, N.Y.)},
  2009, \textbf{325}, 314--8\relax
\mciteBstWouldAddEndPuncttrue
\mciteSetBstMidEndSepPunct{\mcitedefaultmidpunct}
{\mcitedefaultendpunct}{\mcitedefaultseppunct}\relax
\EndOfBibitem
\bibitem[Hewitt and Balmforth(2022)]{Hewitt2022}
D.~Hewitt and N.~Balmforth, \emph{Journal of Fluid Mechanics}, 2022,
  \textbf{936}, A17\relax
\mciteBstWouldAddEndPuncttrue
\mciteSetBstMidEndSepPunct{\mcitedefaultmidpunct}
{\mcitedefaultendpunct}{\mcitedefaultseppunct}\relax
\EndOfBibitem
\bibitem[Panaitescu \emph{et~al.}(2017)Panaitescu, Clotet, and
  Kudrolli]{Panaitescu2017}
A.~Panaitescu, X.~Clotet and A.~Kudrolli, \emph{Physical Review E}, 2017,
  \textbf{95}, 032901\relax
\mciteBstWouldAddEndPuncttrue
\mciteSetBstMidEndSepPunct{\mcitedefaultmidpunct}
{\mcitedefaultendpunct}{\mcitedefaultseppunct}\relax
\EndOfBibitem
\bibitem[Jewel \emph{et~al.}(2018)Jewel, Panaitescu, and Kudrolli]{Jewel2018}
R.~Jewel, A.~Panaitescu and A.~Kudrolli, \emph{Physical Review Fluids}, 2018,
  \textbf{3}, 084303\relax
\mciteBstWouldAddEndPuncttrue
\mciteSetBstMidEndSepPunct{\mcitedefaultmidpunct}
{\mcitedefaultendpunct}{\mcitedefaultseppunct}\relax
\EndOfBibitem
\bibitem[Allen and Kudrolli(2019)]{Allen2019}
B.~Allen and A.~Kudrolli, \emph{Physical Review E}, 2019, \textbf{100},
  022901\relax
\mciteBstWouldAddEndPuncttrue
\mciteSetBstMidEndSepPunct{\mcitedefaultmidpunct}
{\mcitedefaultendpunct}{\mcitedefaultseppunct}\relax
\EndOfBibitem
\bibitem[Pal and Kudrolli(2021)]{Pal2021}
A.~Pal and A.~Kudrolli, \emph{Physical Review Fluids}, 2021, \textbf{6},
  124302\relax
\mciteBstWouldAddEndPuncttrue
\mciteSetBstMidEndSepPunct{\mcitedefaultmidpunct}
{\mcitedefaultendpunct}{\mcitedefaultseppunct}\relax
\EndOfBibitem
\bibitem[Chang and Kudrolli(2022)]{Chang2022}
B.~Chang and A.~Kudrolli, \emph{Physical Review E}, 2022, \textbf{105},
  034901\relax
\mciteBstWouldAddEndPuncttrue
\mciteSetBstMidEndSepPunct{\mcitedefaultmidpunct}
{\mcitedefaultendpunct}{\mcitedefaultseppunct}\relax
\EndOfBibitem
\bibitem[Park \emph{et~al.}(2003)Park, Wolanin, Yuzbashyan, Lin, Darnton,
  Stock, Silberzan, and Austin]{Park2003}
S.~Park, P.~M. Wolanin, E.~A. Yuzbashyan, H.~Lin, N.~C. Darnton, J.~B. Stock,
  P.~Silberzan and R.~Austin, \emph{Proceedings of the National Academy of
  Sciences}, 2003, \textbf{100}, 13910--13915\relax
\mciteBstWouldAddEndPuncttrue
\mciteSetBstMidEndSepPunct{\mcitedefaultmidpunct}
{\mcitedefaultendpunct}{\mcitedefaultseppunct}\relax
\EndOfBibitem
\bibitem[Biswas and Kudrolli(2023)]{Biswas2023}
A.~Biswas and A.~Kudrolli, \emph{Soft Matter}, 2023, \textbf{19},
  4376--4384\relax
\mciteBstWouldAddEndPuncttrue
\mciteSetBstMidEndSepPunct{\mcitedefaultmidpunct}
{\mcitedefaultendpunct}{\mcitedefaultseppunct}\relax
\EndOfBibitem
\bibitem[Bechinger \emph{et~al.}(2016)Bechinger, Di~Leonardo, L\"owen,
  Reichhardt, Volpe, and Volpe]{Bechinger2016}
C.~Bechinger, R.~Di~Leonardo, H.~L\"owen, C.~Reichhardt, G.~Volpe and G.~Volpe,
  \emph{Rev. Mod. Phys.}, 2016, \textbf{88}, 045006\relax
\mciteBstWouldAddEndPuncttrue
\mciteSetBstMidEndSepPunct{\mcitedefaultmidpunct}
{\mcitedefaultendpunct}{\mcitedefaultseppunct}\relax
\EndOfBibitem
\bibitem[Mokhtari and Zippelius(2019)]{Mokhtari2019}
Z.~Mokhtari and A.~Zippelius, \emph{Physical Review Letters}, 2019,
  \textbf{123}, 028001\relax
\mciteBstWouldAddEndPuncttrue
\mciteSetBstMidEndSepPunct{\mcitedefaultmidpunct}
{\mcitedefaultendpunct}{\mcitedefaultseppunct}\relax
\EndOfBibitem
\bibitem[Kurzthaler \emph{et~al.}(2021)Kurzthaler, Mandal, Bhattacharjee,
  Löwen, Datta, and Stone]{Kurzthaler2021}
C.~Kurzthaler, S.~Mandal, T.~Bhattacharjee, H.~Löwen, S.~Datta and H.~Stone,
  \emph{Nature Communications}, 2021, \textbf{12}, 7088\relax
\mciteBstWouldAddEndPuncttrue
\mciteSetBstMidEndSepPunct{\mcitedefaultmidpunct}
{\mcitedefaultendpunct}{\mcitedefaultseppunct}\relax
\EndOfBibitem
\bibitem[Melillo \emph{et~al.}(2017)Melillo, Frey, DeAngelis, Werner, Bernard,
  Bowles, Pold, Knorr, and Grandy]{melillo2017long}
J.~M. Melillo, S.~D. Frey, K.~M. DeAngelis, W.~J. Werner, M.~J. Bernard, F.~P.
  Bowles, G.~Pold, M.~A. Knorr and A.~S. Grandy, \emph{Science}, 2017,
  \textbf{358}, 101--105\relax
\mciteBstWouldAddEndPuncttrue
\mciteSetBstMidEndSepPunct{\mcitedefaultmidpunct}
{\mcitedefaultendpunct}{\mcitedefaultseppunct}\relax
\EndOfBibitem
\bibitem[Doetterl \emph{et~al.}(2015)Doetterl, Stevens, Six, Merckx, Van~Oost,
  Casanova~Pinto, Casanova-Katny, Muñoz, Boudin, Zagal~Venegas, and
  Boeckx]{doetterlSoilCarbon}
S.~Doetterl, A.~Stevens, J.~Six, R.~Merckx, K.~Van~Oost, M.~Casanova~Pinto,
  A.~Casanova-Katny, C.~Muñoz, M.~Boudin, E.~Zagal~Venegas and P.~Boeckx,
  \emph{Nature Geoscience}, 2015, \textbf{8}, 780--783\relax
\mciteBstWouldAddEndPuncttrue
\mciteSetBstMidEndSepPunct{\mcitedefaultmidpunct}
{\mcitedefaultendpunct}{\mcitedefaultseppunct}\relax
\EndOfBibitem
\bibitem[Yang \emph{et~al.}(2021)Yang, Zhang, Bourg, and
  Stone]{YangZhangBourgStone}
J.~Yang, X.~Zhang, I.~Bourg and H.~Stone, \emph{Nature Communications}, 2021,
  \textbf{12}, 622\relax
\mciteBstWouldAddEndPuncttrue
\mciteSetBstMidEndSepPunct{\mcitedefaultmidpunct}
{\mcitedefaultendpunct}{\mcitedefaultseppunct}\relax
\EndOfBibitem
\bibitem[Yang \emph{et~al.}(2021)Yang, Sanfilippo, Abbasi, Gitai, Bassler, and
  Stone]{YangSanfilippoAbbasiGitaiBasslerStone}
J.~Yang, J.~Sanfilippo, N.~Abbasi, Z.~Gitai, B.~Bassler and H.~Stone,
  \emph{Proceedings of the National Academy of Sciences}, 2021, \textbf{118},
  e2111060118\relax
\mciteBstWouldAddEndPuncttrue
\mciteSetBstMidEndSepPunct{\mcitedefaultmidpunct}
{\mcitedefaultendpunct}{\mcitedefaultseppunct}\relax
\EndOfBibitem
\bibitem[Stanley \emph{et~al.}(2016)Stanley, Grossmann, Solvas, and
  DeMello]{StanleyGrossmannSolvasDeMello}
C.~Stanley, G.~Grossmann, X.~C.~i. Solvas and A.~DeMello, \emph{Lab on a Chip},
  2016, \textbf{16}, 228--241\relax
\mciteBstWouldAddEndPuncttrue
\mciteSetBstMidEndSepPunct{\mcitedefaultmidpunct}
{\mcitedefaultendpunct}{\mcitedefaultseppunct}\relax
\EndOfBibitem
\bibitem[Aleklett \emph{et~al.}(2018)Aleklett, Kiers, Ohlsson, Shimizu, Caldas,
  and Hammer]{AleklettKiersOhlssonShimizuCaldasHammer}
K.~Aleklett, E.~Kiers, P.~Ohlsson, T.~Shimizu, V.~Caldas and E.~Hammer,
  \emph{The ISME Journal}, 2018, \textbf{12}, 312–319\relax
\mciteBstWouldAddEndPuncttrue
\mciteSetBstMidEndSepPunct{\mcitedefaultmidpunct}
{\mcitedefaultendpunct}{\mcitedefaultseppunct}\relax
\EndOfBibitem
\bibitem[Dragulet \emph{et~al.}(2022)Dragulet, Goyal, Ioannidou, Pellenq, and
  Del~Gado]{Dragulet2022ion}
F.~Dragulet, A.~Goyal, K.~Ioannidou, R.~J.-M. Pellenq and E.~Del~Gado,
  \emph{Journal of Physical Chemistry B}, 2022, \textbf{126}, 4977\relax
\mciteBstWouldAddEndPuncttrue
\mciteSetBstMidEndSepPunct{\mcitedefaultmidpunct}
{\mcitedefaultendpunct}{\mcitedefaultseppunct}\relax
\EndOfBibitem
\bibitem[Goyal \emph{et~al.}(2020)Goyal, Ioannidou, Tiede, Levitz, Pellenq, and
  {Del Gado}]{Goyal2020}
A.~Goyal, K.~Ioannidou, C.~Tiede, P.~Levitz, R.~J.-M. Pellenq and E.~{Del
  Gado}, \emph{The Journal of Physical Chemistry C}, 2020, \textbf{124},
  15500--15510\relax
\mciteBstWouldAddEndPuncttrue
\mciteSetBstMidEndSepPunct{\mcitedefaultmidpunct}
{\mcitedefaultendpunct}{\mcitedefaultseppunct}\relax
\EndOfBibitem
\bibitem[Fontaine and Auger(2009)]{BatirEnTerre}
L.~Fontaine and R.~Auger, \emph{Batir en terre}, Edition B\'elin, 2009\relax
\mciteBstWouldAddEndPuncttrue
\mciteSetBstMidEndSepPunct{\mcitedefaultmidpunct}
{\mcitedefaultendpunct}{\mcitedefaultseppunct}\relax
\EndOfBibitem
\bibitem[Guillaud(2009)]{Guillaud2009}
H.~Guillaud, in \emph{Terra Literature Review-An Overview of Research in
  Earthen Architecture Conservation}, ed. E.~Avrami, H.~Guillaud and M.~Hardy,
  The Getty Conservation Institute, Los Angeles, 2009\relax
\mciteBstWouldAddEndPuncttrue
\mciteSetBstMidEndSepPunct{\mcitedefaultmidpunct}
{\mcitedefaultendpunct}{\mcitedefaultseppunct}\relax
\EndOfBibitem
\bibitem[Abergel \emph{et~al.}(2017)Abergel, Dean, and Dulac]{globalstatus2017}
T.~Abergel, B.~D. Dean and J.~Dulac, \emph{Global Alliance for Buildings and
  Construction/International Energy Agency: Paris, France}, 2017\relax
\mciteBstWouldAddEndPuncttrue
\mciteSetBstMidEndSepPunct{\mcitedefaultmidpunct}
{\mcitedefaultendpunct}{\mcitedefaultseppunct}\relax
\EndOfBibitem
\bibitem[Habert \emph{et~al.}(2020)Habert, Miller, John, Provis, Favier,
  Horvath, and Scrivener]{Habert2020environmental}
G.~Habert, S.~A. Miller, V.~M. John, J.~L. Provis, A.~Favier, A.~Horvath and
  K.~L. Scrivener, \emph{Nature Reviews Earth \& Environment}, 2020,
  \textbf{1}, 559--573\relax
\mciteBstWouldAddEndPuncttrue
\mciteSetBstMidEndSepPunct{\mcitedefaultmidpunct}
{\mcitedefaultendpunct}{\mcitedefaultseppunct}\relax
\EndOfBibitem
\bibitem[Gangotra \emph{et~al.}(2023)Gangotra, Del~Gado, and
  Lewis]{Gangotra2023}
A.~Gangotra, E.~Del~Gado and J.~Lewis, \emph{Technology \& Policy Opportunities
  for Upgrading Low-Carbon Cement Production in the United States}, 2023\relax
\mciteBstWouldAddEndPuncttrue
\mciteSetBstMidEndSepPunct{\mcitedefaultmidpunct}
{\mcitedefaultendpunct}{\mcitedefaultseppunct}\relax
\EndOfBibitem
\bibitem[Malakoff(2020)]{mud-2020}
D.~Malakoff, \emph{Science}, 2020, \textbf{369}, 894--895\relax
\mciteBstWouldAddEndPuncttrue
\mciteSetBstMidEndSepPunct{\mcitedefaultmidpunct}
{\mcitedefaultendpunct}{\mcitedefaultseppunct}\relax
\EndOfBibitem
\bibitem[Israelachvili(2015)]{israelachvili2015intermolecular}
J.~N. Israelachvili, \emph{Intermolecular and surface forces}, Academic press,
  2015\relax
\mciteBstWouldAddEndPuncttrue
\mciteSetBstMidEndSepPunct{\mcitedefaultmidpunct}
{\mcitedefaultendpunct}{\mcitedefaultseppunct}\relax
\EndOfBibitem
\bibitem[Goyal \emph{et~al.}(2021)Goyal, Palaia, Ioannidou, Ulm, van Damme,
  Pellenq, Trizac, and Del~Gado]{Goyal2021physics}
A.~Goyal, I.~Palaia, K.~Ioannidou, F.-J. Ulm, H.~van Damme, R.~J.-M. Pellenq,
  E.~Trizac and E.~Del~Gado, \emph{Science Advances}, 2021, \textbf{7},
  eabg5882\relax
\mciteBstWouldAddEndPuncttrue
\mciteSetBstMidEndSepPunct{\mcitedefaultmidpunct}
{\mcitedefaultendpunct}{\mcitedefaultseppunct}\relax
\EndOfBibitem
\bibitem[Palaia \emph{et~al.}(2022)Palaia, Goyal, Del~Gado, Samaj, and
  Trizac]{Palaia2022likecharge}
I.~Palaia, A.~Goyal, E.~Del~Gado, L.~Samaj and E.~Trizac, \emph{Journal of
  Physical Chemistry B}, 2022, \textbf{126}, 3143\relax
\mciteBstWouldAddEndPuncttrue
\mciteSetBstMidEndSepPunct{\mcitedefaultmidpunct}
{\mcitedefaultendpunct}{\mcitedefaultseppunct}\relax
\EndOfBibitem
\bibitem[Ioannidou \emph{et~al.}(2016)Ioannidou, Kandu{\v{c}}, Li, Frenkel,
  Dobnikar, and Del~Gado]{ioannidou2016crucial}
K.~Ioannidou, M.~Kandu{\v{c}}, L.~Li, D.~Frenkel, J.~Dobnikar and E.~Del~Gado,
  \emph{Nature Communications}, 2016, \textbf{7}, 1--9\relax
\mciteBstWouldAddEndPuncttrue
\mciteSetBstMidEndSepPunct{\mcitedefaultmidpunct}
{\mcitedefaultendpunct}{\mcitedefaultseppunct}\relax
\EndOfBibitem
\bibitem[Ioannidou \emph{et~al.}(2016)Ioannidou, Krakowiak, Bauchy, Hoover,
  Masoero, Yip, Ulm, Levitz, Pellenq, and Del~Gado]{ioannidou2016mesoscale}
K.~Ioannidou, K.~J. Krakowiak, M.~Bauchy, C.~G. Hoover, E.~Masoero, S.~Yip,
  F.-J. Ulm, P.~Levitz, R.~J.-M. Pellenq and E.~Del~Gado, \emph{Proceedings of
  the National Academy of Sciences}, 2016, \textbf{113}, 2029--2034\relax
\mciteBstWouldAddEndPuncttrue
\mciteSetBstMidEndSepPunct{\mcitedefaultmidpunct}
{\mcitedefaultendpunct}{\mcitedefaultseppunct}\relax
\EndOfBibitem
\bibitem[Roth \emph{et~al.}(2017)Roth, Finnegan, Brodsky, Rickenmann, Turowski,
  Badoux, and Gimbert]{Roth2017}
D.~L. Roth, N.~J. Finnegan, E.~E. Brodsky, D.~Rickenmann, J.~M. Turowski,
  A.~Badoux and F.~Gimbert, \emph{Journal of Geophysical Research: Earth
  Surface}, 2017, \textbf{122}, 1182--1200\relax
\mciteBstWouldAddEndPuncttrue
\mciteSetBstMidEndSepPunct{\mcitedefaultmidpunct}
{\mcitedefaultendpunct}{\mcitedefaultseppunct}\relax
\EndOfBibitem
\bibitem[Lovejoy \emph{et~al.}(2008)Lovejoy, Gaonac'h, and
  Schertzer]{Lovejoy2008}
S.~Lovejoy, H.~Gaonac'h and D.~Schertzer, \emph{Mathematical Geosciences},
  2008, \textbf{40}, 533--573\relax
\mciteBstWouldAddEndPuncttrue
\mciteSetBstMidEndSepPunct{\mcitedefaultmidpunct}
{\mcitedefaultendpunct}{\mcitedefaultseppunct}\relax
\EndOfBibitem
\bibitem[Sadler and Jerolmack(2015)]{Sadler2015}
P.~M. Sadler and D.~J. Jerolmack, \emph{Geological Society, London, Special
  Publications}, 2015, \textbf{404}, 69--88\relax
\mciteBstWouldAddEndPuncttrue
\mciteSetBstMidEndSepPunct{\mcitedefaultmidpunct}
{\mcitedefaultendpunct}{\mcitedefaultseppunct}\relax
\EndOfBibitem
\bibitem[H{\'{e}}bert \emph{et~al.}(2022)H{\'{e}}bert, Herzschuh, and
  Laepple]{Hebert2022}
R.~H{\'{e}}bert, U.~Herzschuh and T.~Laepple, \emph{Nature Geoscience}, 2022,
  \textbf{15}, 899--905\relax
\mciteBstWouldAddEndPuncttrue
\mciteSetBstMidEndSepPunct{\mcitedefaultmidpunct}
{\mcitedefaultendpunct}{\mcitedefaultseppunct}\relax
\EndOfBibitem
\bibitem[Franzke \emph{et~al.}(2020)Franzke, Barbosa, Blender, Fredriksen,
  Laepple, Lambert, Nilsen, Rypdal, Rypdal, Scotto, Vannitsem, Watkins, Yang,
  and Yuan]{Franzke2020}
C.~L. Franzke, S.~Barbosa, R.~Blender, H.~B. Fredriksen, T.~Laepple,
  F.~Lambert, T.~Nilsen, K.~Rypdal, M.~Rypdal, M.~G. Scotto, S.~Vannitsem,
  N.~W. Watkins, L.~Yang and N.~Yuan, \emph{Reviews of Geophysics}, 2020,
  \textbf{58}, e2019RG000657\relax
\mciteBstWouldAddEndPuncttrue
\mciteSetBstMidEndSepPunct{\mcitedefaultmidpunct}
{\mcitedefaultendpunct}{\mcitedefaultseppunct}\relax
\EndOfBibitem
\bibitem[Ha and Kim(2020)]{Ha2020capillarity}
J.~Ha and H.-Y. Kim, \emph{Annual Review of Fluid Mechanics}, 2020,
  \textbf{52}, 263--284\relax
\mciteBstWouldAddEndPuncttrue
\mciteSetBstMidEndSepPunct{\mcitedefaultmidpunct}
{\mcitedefaultendpunct}{\mcitedefaultseppunct}\relax
\EndOfBibitem
\end{mcitethebibliography}
\bibliographystyle{rsc} 

\end{document}